\documentclass[12pt]{article}
\usepackage{amssymb,amsmath}
\usepackage{graphicx}
\usepackage{ytableau}
\usepackage{comment}
\usepackage[vcentermath]{youngtab}
\usepackage{hyperref}
\usepackage{tikz}

\DeclareMathOperator*{\Tr}{{\rm Tr}}

\numberwithin{equation}{section}

\begin{document}

\thispagestyle{empty}
\begin{flushright}
\end{flushright}
\vskip1cm
\begin{center}
{\bf {\LARGE Orbifold ETW brane and half-indices}}

\vskip1.5cm

Yasuyuki Hatsuda$^a$\footnote{yhatsuda@rikkyo.ac.jp}, 

\bigskip
$^a$
{\it Department of Physics, Rikkyo University, \\
Toshima, Tokyo 171-8501, Japan
}

\bigskip
\bigskip

Hai Lin$^b$\footnote{hailinhl@seu.edu.cn}

\bigskip
\bigskip

and 

\bigskip
\bigskip

Tadashi Okazaki$^b$\footnote{tokazaki@seu.edu.cn}

\bigskip
$^b$
{\it Shing-Tung Yau Center of Southeast University,\\
Yifu Architecture Building, No.2 Sipailou, Xuanwu district, \\
Nanjing, Jiangsu, 210096, China
}

\end{center}

\vskip1.25cm
\begin{abstract}
We study the half-indices of $\mathcal{N}=4$ super Yang-Mills theories of orthogonal and symplectic gauge groups. 
We find precise matching pairs of the half-indices as strong evidence of dualities of the half-BPS boundary conditions and interfaces. 
The half-indices encode the spectra of excitations on the holographically dual orbifold ETW brane and giant gravitons. 
The exact form of the index for the orbifold ETW giant gravitons is obtained from the giant graviton expansion of the half-index. 
\end{abstract}

\newpage
\setcounter{tocdepth}{3}
\tableofcontents

\section{Introduction and conclusion}

S-duality conjectures \cite{Montonen:1977sn,Osborn:1979tq} that 
4d $\mathcal{N}=4$ super Yang-Mills (SYM) theory with gauge group $G$ has a dual description at strong coupling 
as a weakly coupled theory with the Langlands dual gauge group $G^{\vee}$. 
The Lie algebra $\mathfrak{g}^{\vee}$ for $G^{\vee}$ has a root 
$\alpha^{\vee}$ $=$ $2\alpha/\alpha^2$ with $\alpha$ being a root of the Lie algebra $\mathfrak{g}$ for $G$. 
For gauge groups with simply-laced Lie algebras $\mathfrak{su}(N+1)$, $\mathfrak{so}(2N)$, $E_6$, $E_7$ and $E_8$ 
the Langlands duals are the same, i.e. $\mathfrak{g}^{\vee}$ $=$ $\mathfrak{g}$. 
For gauge groups with non-simply-laced Lie algebras $\mathfrak{so}(2N+1)$, $\mathfrak{usp}(2N)$, $G_2$ and $F_4$, 
the Langlands duals are $\mathfrak{so}(2N+1)^{\vee}$ $=$ $\mathfrak{usp}(2N)$, 
$\mathfrak{usp}(2N)^{\vee}$ $=$ $\mathfrak{so}(2N+1)$, 
$G_2^{\vee}$ $=$ $G_2'$ and $F_4^{\vee}$ $=$ $F_4'$ 
where the primes stand for rotations of their root systems \cite{Goddard:1976qe}. 

One of the most remarkable features of 4d $\mathcal{N}=4$ SYM theory 
is that it can be realized as a low-energy effective theory of a stack of D3-branes in Type IIB string theory. 
In the presence of additional $5$-branes, one finds boundaries and interfaces 
\cite{Hanany:1996ie,Gaiotto:2008sa,Gaiotto:2008sd,Gaiotto:2008ak,Mikhaylov:2014aoa} as well as corners \cite{Chung:2016pgt,Gaiotto:2017euk,Hanany:2018hlz,Gaiotto:2019jvo,Okazaki:2019bok} in $\mathcal{N}=4$ SYM theories. 
Such brane setups of defects in $\mathcal{N}=4$ SYM theories are useful to study more sophisticated dualities in gauge theories across the dimensions of spacetime 
and to construct natural frameworks of double holography \cite{Karch:2000ct,Karch:2000gx,Takayanagi:2011zk,Karch:2022rvr} in string theory.

In this paper we investigate the half-BPS boundary conditions and interfaces in $\mathcal{N}=4$ supersymmetric gauge theories of orthogonal and symplectic gauge groups 
by analyzing the half-indices \cite{Dimofte:2011py,Gang:2012yr,Gaiotto:2019jvo} that encode the spectra of the BPS local operators in the presence of boundary. 
We obtain exact forms of the half-indices of Neumann and Nahm pole boundary conditions by applying the Higgsing procedure \cite{Gaiotto:2012xa}. 
As these setups can be realized as configurations of D3-branes, $5$-branes as well as an O3-plane, 
dualities of the half-BPS boundary conditions and interfaces are conjectured by Gaiotto and Witten \cite{Gaiotto:2008ak} upon S-duality in Type IIB string theory. 
We find precise matching pairs of half-indices for the S-dual half-BPS boundary conditions and interfaces as powerful evidence of these dualities. 
For the case with $\mathfrak{so}(2N)$ gauge algebra, the dualities can be further generalized to disconnected $O(2N)$ gauge groups. 
Our results generalize the matching of half-indices in \cite{Gaiotto:2019jvo} for the cases with unitary gauge groups. 

In addition, the half-BPS boundary conditions and interfaces are holographically dual to the \textit{orbifold bagpipe geometries} in Type IIB supergravity, 
which can be obtained by taking the orbifold of the $AdS_4$ $\times$ $S^2$ $\times$ $S^2$ warped over the Riemann surface $\Sigma$ 
constructed by D'Hoker, Estes and Gutperle \cite{DHoker:2007zhm,DHoker:2007hhe}. 
The latter supergravity solutions are also referred to as the ``bagpipe geometries'' \cite{Bachas:2018zmb} 
which are composed of the bag corresponding to end-of-the-world (ETW) brane with $AdS_4$ factor and the pipes to semi-infinite Janus throats. 
We examine the large $N$ limits of the half-indices which capture the Kaluza-Klein (KK) excitations on the orbifold bagpipe geometries 
and analyze the giant graviton expansion \cite{Arai:2019xmp,Arai:2020qaj,Gaiotto:2021xce} 
of the Neumann (or equivalently Nahm) half-index for the orthogonal or symplectic gauge theory. 
\footnote{See \cite{Hatsuda:2024uwt} for the analysis of the giant graviton expansions of the half-indices of unitary gauge theories. }
We derive the exact form of the index for the orbifold ETW giant gravitons, wrapped D3-branes in the $AdS_4$ ETW region of the orbifold bagpipe geometry. 

\subsection{Structure}
The paper is organized as follows. 
In section \ref{sec_brane_conf} we review the brane construction  
of $\mathcal{N}=4$ SYM theories with orthogonal and symplectic gauge groups and their half-BPS boundary conditions and interfaces in Type IIB string theory 
involving D3-branes, $5$-branes and O3-planes. The holographically dual geometries are identified with the orbifold bagpipe geometries. 
In section \ref{sec_hindex} we compute the half-indices of half-BPS boundary conditions and interfaces in orthogonal and symplectic gauge theories 
to find the matching pairs for the S-dual configurations. 
In section \ref{sec_ggexp} we analyze the large $N$ limits of the half-indices and the giant graviton expansions of the half-indices. 
In Appendix \ref{app_expansion} we show the $q$-series expansions of the half-indices. 
We have confirmed the matching pairs of half-indices at least up to the terms with $q^5$ apart from the cases indicated in the tables. 

\subsection{Future works}
There are interesting open problems which we hope to address in future works. 

\begin{itemize}

\item There exist more general half-BPS boundary conditions and interfaces 
in $\mathcal{N}=4$ SYM theories with orthogonal and symplectic gauge groups, including 
more general successive Nahm pole boundary conditions and enriched Neumann boundary conditions which couple to the 3d gauge theories. 
It would be interesting to examine the half-indices to test the dualities as worked out in \cite{Okazaki:2019ony} for unitary gauge groups. 

\item While we have checked the matching pairs of half-indices, 
we have not yet obtained the closed-form expressions for the interface half-indices as well as for the full-indices of the orthogonal/symplectic gauge theories. 
It is tempting to generalize the ``vortex expansions'' for unitary gauge groups in \cite{Gaiotto:2019jvo} by summing over residues. 

\item While we have examined the giant graviton expansions of the half-indices of basic Neumann or Nahm pole boundary conditions, 
it would be intriguing to figure out those for general half-indices for orthogonal and symplectic gauge theories. 
As they are captured by the line defect indices for unitary gauge theories \cite{Hatsuda:2024uwt} (also see \cite{Imamura:2024lkw,Imamura:2024pgp,Beccaria:2024oif,Beccaria:2024dxi,Imamura:2024zvw,Beccaria:2024lbt} for the inverse relation), 
we hope to report detailed analysis of the line defect indices for orthogonal and symplectic gauge theories as performed in 
\cite{Hatsuda:2023iwi,Hatsuda:2023imp,Hatsuda:2023iof}. 

\item 
The orbifold giant graviton expansions of the full indices for $\mathcal{N}=4$ SYM theories 
of orthogonal and symplectic gauge groups were numerically investigated in \cite{Fujiwara:2023bdc}. 
While we have presented the exact forms of the half-BPS indices and their giant graviton expansions for the half-BPS states, 
it would be a good future direction to give a detailed analysis of the $1/4$-BPS states, which have also been discussed in e.g. \cite{Lewis-Brown:2018dje,Kemp:2014apa} and related contexts. 
These would also play important roles for orbifold giant graviton cases. 

\end{itemize}

\section{Brane configuration}
\label{sec_brane_conf}
We begin by reviewing the brane construction in Type IIB string theory 
of half-BPS boundary conditions and interfaces in $\mathcal{N}=4$ supersymmetric gauge theories of orthogonal and symplectic gauge groups. 

\subsection{O3-planes}
4d $\mathcal{N}=4$ SYM theories with orthogonal or symplectic gauge group can be realized in Type IIB string theory 
by introducing D3-branes in the background of an O3-plane \cite{Witten:1998xy,Feng:2000eq}. 
There are four kinds of O3-planes characterized by $\mathbb{Z}_2$-valued discrete fluxes, 
or discrete torsions $\theta_{RR}$ and $\theta_{NS}$ for RR and NSNS 2-forms (see also section \ref{sec_orbETW}). 

These O3-planes are related with each other by the $SL(2,\mathbb{Z})$ S-duality transformation in Type IIB string theory as it acts on the discrete fluxes. 
The $SL(2,\mathbb{Z})$ transformation is generated by
\begin{align}
\label{S_transf}
S: \left(
\begin{matrix}
0&1\\
-1&0\\
\end{matrix}
\right)
\end{align}
and 
\begin{align}
\label{T_transf}
T: \left(
\begin{matrix}
1&1\\
0&1\\
\end{matrix}
\right). 
\end{align}

With the non-trivial RR flux, we have $\widetilde{\textrm{O3}}^-$-plane. 
It has $1/4$ unit of D3-brane charge and the effective theory on $N$ D3-branes with a parallel $\widetilde{\textrm{O3}}^-$-plane has $SO(2N+1)$ gauge group. 
While the $\widetilde{\textrm{O3}}^-$-plaine is invariant under the $T$ transformation (\ref{T_transf}) of $SL(2,\mathbb{Z})$, 
it transforms into O3$^+$-plane with the non-trivial NS flux under the $S$ transformation (\ref{S_transf}). 

The O3$^+$-plane has $1/4$ unit of D3-brane charge. 
The theory on $N$ D3-branes in the background of O3$^+$-plane has $USp(2N)$ gauge group. 
Thus $\mathcal{N}=4$ $SO(2N+1)$ SYM theory is dual to $\mathcal{N}=4$ $USp(2N)$ SYM theory. 

The O3$^+$-plane transforms under the $T$ operation into $\widetilde{\textrm{O3}}^+$-plane that carries $1/4$ unit of D3-brane charge. 
While the theory on $N$ D3-branes in the presence of $\widetilde{\textrm{O3}}^+$ still supports $USp(2N)$ gauge group, 
it has a unit of theta-angle, which we call $USp(2N)'$ theory. 
The $\widetilde{\textrm{O3}}^+$-plane is invariant under the $S$ transformation. 

For zero fluxes one has O3$^-$-plane. 
It has $-1/4$ unit of D3-brane charge unlike the previous three cases. 
The gauge group of the world-volume theory on $N$ D3-branes in the presence of O3$^-$-plane is $O(2N)$ \cite{Garcia-Etxebarria:2015wns} (also see \cite{Aharony:2016kai}). 
Since it is invariant under S-duality, $\mathcal{N}=4$ $O(2N)$ SYM theory is conjecturally self S-dual theory. 

The four possible choices of O3-planes 
and SYM theories with orthogonal or symplectic gauge groups as the low-energy effective theories of the D3-branes are summarized as
\begin{align}
\label{O3_SYM}
\begin{array}{c|c|c|c|c}
&SO(2N+1)&USp(2N)&O(2N)&USp(2N)'\\ \hline 
\theta_{RR}&1/2&0&0&1/2 \\
\theta_{NS}&0&1/2&0&1/2 \\
\textrm{D3-brane charge}&1/4&1/4&-1/4&1/4 \\
\textrm{orientifold}&\widetilde{\textrm{O3}}^{-}&\textrm{O3}^+&\textrm{O3}^{-}&\widetilde{\textrm{O3}}^+ \\
\textrm{$S$ operation}&\textrm{O3}^+&\widetilde{\textrm{O3}}^{-}&\textrm{O3}^{-}&\widetilde{\textrm{O3}}^+ \\
\end{array}.
\end{align}

\subsection{Brane setup}
We consider the D3-branes which fill the $0126$ directions 
in the presence of one kind of O3-planes parallel to the D3-branes 
to realize $\mathcal{N}=4$ SYM theory of orthogonal or symplectic gauge group. 

The half-BPS boundary conditions are generated by $5$-branes localized at $x_6=0$, 
D5-branes extending along the $012789$ directions and NS5-branes along the $012345$ directions \cite{Hanany:1996ie,Gaiotto:2008sa}. 
As the configuration is invariant under reflection of directions $345789$, 
these $5$-branes interacting with the O3-plane are identified with half D5- and NS5-branes. 
The half NS5-brane realizes the Neumann boundary condition $\mathcal{N}$ in $\mathcal{N}=4$ SYM theory, 
which is compatible with the 3d $\mathcal{N}=4$ vector multiplet. 
The half D5-brane gives rise to the Nahm pole boundary condition or Dirichlet boundary condition 
(which we denote by Nahm and $\mathcal{D}$) in $\mathcal{N}=4$ SYM theory. 
The fluctuations of open strings between the half D5-branes and D3-branes are described by the 3d $\mathcal{N}=4$ (half-)hypermultiplets 
transforming in the fundamental representation of the gauge group. 

Alternatively, we can consider the boundary conditions at $x^2=0$ that are produced by half D5-branes along the $013456$ directions 
and half NS5-branes filling the $016789$ directions, 
which we call half D5$'$-branes and half NS5$'$-branes respectively. 
Upon the $S$ transformation, the half NS5-branes map to the half D5$'$-brane and the half D5-brane to the half NS5$'$-brane 
so that the dual descriptions of the boundary conditions at $x^6=0$ can be obtained by reading off the resulting boundary conditions at $x^2=0$ 
which are constructed by the half NS5$'$- and the half D5$'$-branes. 
Correspondingly, we denote the Neumann-type (resp. Dirichlet-type) boundary condition at $x_2=0$ 
by $\mathcal{N}'$ (resp. Nahm$'$ or $\mathcal{D}'$), 
which can be decorated by the 3d twisted vector multiplets and twisted (half-)hypermultiplets at the boundary. 
In this paper, we consider the configuration involving either of $5$-branes defining a boundary at $x^6=0$ 
or those associated with a boundary at $x^2=0$. 
\footnote{More generally, one can find dual descriptions of corner configurations from the junctions of 5-branes preserving $\mathcal{N}=(0,4)$ supersymmetry  \cite{Hanany:2018hlz,Gaiotto:2019jvo,Okazaki:2019bok}. } 
The brane configuration is summarized as follows: 
\begin{align}
\begin{array}{c|cccccccccc}
&0&1&2&3&4&5&6&7&8&9\\ \hline
\textrm{D3}&\circ&\circ&\circ&&&&\circ&&&\\ 
\textrm{O3}&\circ&\circ&\circ&&&&\circ&&&\\ 
\textrm{NS5}&\circ&\circ&\circ&\circ&\circ&\circ&&&&\\ 
\textrm{D5}&\circ&\circ&\circ&&&&&\circ&\circ&\circ \\ 
\textrm{NS5$'$}&\circ&\circ&&&&&&\circ&\circ&\circ \\ 
\textrm{D5$'$}&\circ&\circ&&\circ&\circ&\circ&&&&\\ 
\end{array}
\end{align}
where $\circ$ indicates the directions in which the extended objects are supported. 

\subsection{Orbifold ETW giant gravitons}
\label{sec_orbETW}
The near-horizon limit of D3-branes on an O3-plane is 
Type IIB string theory on $AdS_5\times \mathbb{RP}^5$ where $\mathbb{RP}^5$ $\cong$ $S^5/\mathbb{Z}_2$ is the five-dimensional projective plane 
defined by the five-sphere with identification of antipodal points. 
So it is holographically dual to $\mathcal{N}=4$ SYM theory of orthogonal or symplectic gauge group \cite{Witten:1998xy}. 
The $AdS_5\times \mathbb{RP}^5$ background is characterized by the discrete torsions $\theta_{RR}$ and $\theta_{NS}$ 
for RR $2$-form $B_{RR}$ and NSNS $2$-form $B_{NS}$. 
The discrete torsion takes values in $H^3(\mathbb{RP}^5,\tilde{\mathbb{Z}})$, 
where $\tilde{\mathbb{Z}}$ stands for the sheaf of integers twisted by the unorientable real line bundle over $\mathbb{RP}^5$ 
in such a way that as one goes around the $\mathbb{Z}_2$ torsion $3$-cycle in $\mathbb{RP}^5$, 
a section of $\tilde{\mathbb{Z}}$ is multiplied by $(-1)$. 
They are determined by the holonomy on a $\mathbb{RP}^2$ $\subset$ $\mathbb{RP}^5$, 
\begin{align}
e^{i\int_{\mathbb{RP}^2} B_{RR}}=e^{2\pi i\theta_{RR}}=\pm 1, \qquad
e^{i\int_{\mathbb{RP}^2} B_{NS}}=e^{2\pi i\theta_{NS}}=\pm 1. 
\end{align}
Hence $H^3(\mathbb{RP}^5,\tilde{\mathbb{Z}})$ $=$ $\mathbb{Z}_2\oplus \mathbb{Z}_2$ 
and four possible choices correspond to four kinds of O3-planes and four types of $\mathcal{N}=4$ SYM theories in (\ref{O3_SYM}). 

Continuous global symmetries in gauge theories correspond to the bulk gauge fields whose field strength is set to zero at the boundary. 
When one chooses other boundary conditions with non-vanishing field strength, 
its boundary value can be identified with a gauge field in the dual gauge theory (see e.g. \cite{Witten:2003ya}). 
Similar phenomena occur for discrete global symmetries in gauge theories \cite{Aharony:2016kai} (also see \cite{Harlow:2018tng}). 
In our setup, the gauge theories based on the Lie algebra $\mathfrak{so}(2N)$ can realize $SO(2N)$ gauge group and $O(2N)$ gauge group 
by (un)gauging the discrete $\mathbb{Z}_2$ symmetry. 
The full quantum gravity theory on the bulk $AdS_5\times \mathbb{RP}^5$ 
includes a topological sector corresponding to a topological theory with $\mathbb{Z}_2$ gauge symmetry on $AdS_5$ described by an action of the form
\begin{align}
\label{AdS5_top_action}
\frac{i}{\pi}A\wedge dC, 
\end{align}
where $A$ is the $1$-form $\mathbb{Z}_2$ gauge field and $C$ is the $3$-form gauge field. 
They obey the equations of motion as the flatness condition on both $A$ and $C$, $dA=dC=0$. 
When we introduce the action (\ref{AdS5_top_action}), we need to impose boundary conditions. 
According to the variational principle, $A\wedge C$ must vanish at the boundary. 
When one chooses vanishing $C$, the boundary value of $A$ can be identified with the discrete $\mathbb{Z}_2$ gauge field 
so that the resulting gauge theory has $O(2N)$ gauge group. 
On the other hand, when $A$ is set to zero at the boundary, 
$\mathbb{Z}_2$ symmetry is identified with a global symmetry in $SO(2N)$ gauge theory. 
In the following discussion, we consider both $SO(2N)$ and $O(2N)$ gauge theories 
which arise from the same holographic dual geometry but with different boundary conditions for bulk $\mathbb{Z}_2$ gauge fields. 

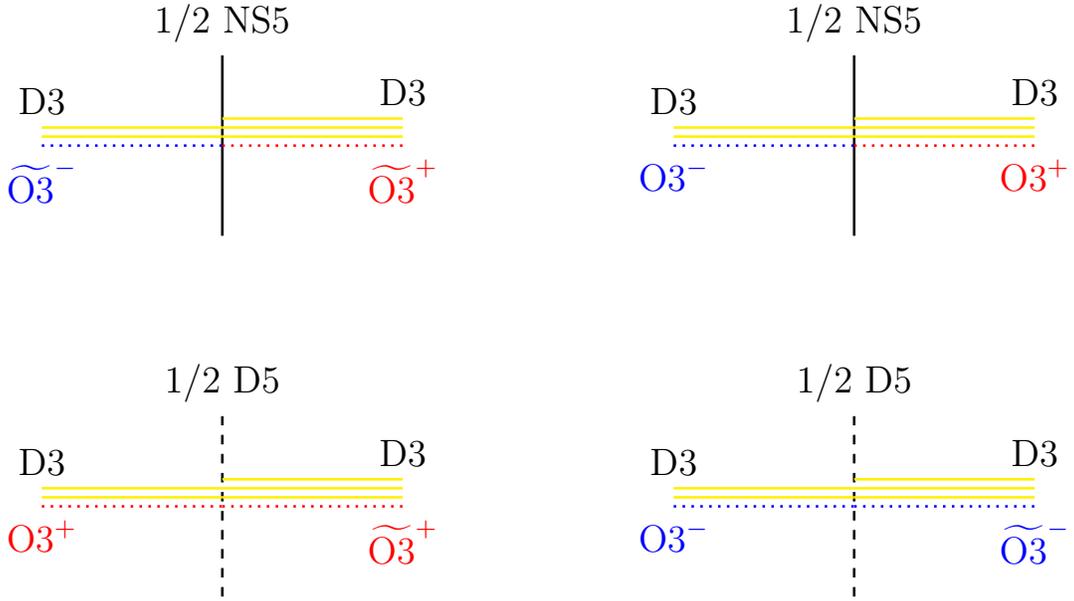
\begin{figure}
\centering
\scalebox{1.2}{
\begin{tikzpicture}
\draw[thick,solid,black] (-3.5,1) -- (-3.5,3) node[above] {\textrm{1/2 NS5}};
\draw[thick,solid,black] (3.5,1) -- (3.5,3) node[above] {\textrm{1/2 NS5}};
\draw[thick,dashed,black] (-3.5,-3) -- (-3.5,-1) node[above] {\textrm{1/2 D5}};
\draw[thick,dashed,black] (3.5,-3) -- (3.5,-1) node[above] {\textrm{1/2 D5}};
\draw[thick,dotted,blue] (-5.5,2) node[below] {$\widetilde{\textrm{O3}}^-$} -- (-3.5,2);
\draw[thick,dotted,red] (-3.5,2) -- (-1.5,2) node[below] {$\widetilde{\textrm{O3}}^+$};
\draw[thick,dotted,blue] (1.5,2) node[below] {$\textrm{O3}^-$} -- (3.5,2);
\draw[thick,dotted,red] (3.5,2) -- (5.5,2) node[below] {$\textrm{O3}^+$};
\draw[thick,dotted,red] (-5.5,-2) node[below] {$\textrm{O3}^+$} -- (-3.5,-2);
\draw[thick,dotted,red] (-3.5,-2) -- (-1.5,-2) node[below] {$\widetilde{\textrm{O3}}^+$};
\draw[thick,dotted,blue] (1.5,-2) node[below] {$\textrm{O3}^-$} -- (3.5,-2);
\draw[thick,dotted,blue] (3.5,-2) -- (5.5,-2) node[below] {$\widetilde{\textrm{O3}}^-$};
\draw[thick,solid,yellow] (-5.5,2.2) node[above] {\textcolor{black}{\textrm{D3}}} -- (-3.5,2.2);
\draw[thick,solid,yellow] (-5.5,2.1)  -- (-3.5,2.1);
\draw[thick,solid,yellow] (-3.5,2.3)  -- (-1.5,2.3) node[above] {\textcolor{black}{\textrm{D3}}};
\draw[thick,solid,yellow] (-3.5,2.2)  -- (-1.5,2.2);
\draw[thick,solid,yellow] (-3.5,2.1)  -- (-1.5,2.1);
\draw[thick,solid,yellow] (1.5,2.2) node[above] {\textcolor{black}{\textrm{D3}}} -- (3.5,2.2);
\draw[thick,solid,yellow] (1.5,2.1)  -- (3.5,2.1);
\draw[thick,solid,yellow] (3.5,2.3)  -- (5.5,2.3) node[above] {\textcolor{black}{\textrm{D3}}};
\draw[thick,solid,yellow] (3.5,2.2)  -- (5.5,2.2);
\draw[thick,solid,yellow] (3.5,2.1)  -- (5.5,2.1);
\draw[thick,solid,yellow] (-5.5,-1.8) node[above] {\textcolor{black}{\textrm{D3}}} -- (-3.5,-1.8);
\draw[thick,solid,yellow] (-5.5,-1.9)  -- (-3.5,-1.9);
\draw[thick,solid,yellow] (-3.5,-1.7)  -- (-1.5,-1.7) node[above] {\textcolor{black}{\textrm{D3}}};
\draw[thick,solid,yellow] (-3.5,-1.8)  -- (-1.5,-1.8);
\draw[thick,solid,yellow] (-3.5,-1.9)  -- (-1.5,-1.9);
\draw[thick,solid,yellow] (1.5,-1.8) node[above] {\textcolor{black}{\textrm{D3}}} -- (3.5,-1.8);
\draw[thick,solid,yellow] (1.5,-1.9)  -- (3.5,-1.9);
\draw[thick,solid,yellow] (3.5,-1.7)  -- (5.5,-1.7) node[above] {\textcolor{black}{\textrm{D3}}};
\draw[thick,solid,yellow] (3.5,-1.8)  -- (5.5,-1.8);
\draw[thick,solid,yellow] (3.5,-1.9)  -- (5.5,-1.9);
\end{tikzpicture}
}
\caption{\textrm{The configurations of O3-planes, half $5$-branes and D3-branes 
where the horizontal direction is $x^6$. 
Four types of O3-planes (blue and red horizontal lines) 
which are parallel to the D3-branes (yellow horizontal lines) 
change when they cross a half $5$-branes. 
The configuration with a half NS5-brane (vertical solid line) realizes 
the interface between an orthogonal gauge theory and a symplectic gauge theory 
while that with a half D5-brane (vertical dashed line) corresponds to the domain wall between orthogonal gauge theories 
(or between symplectic gauge theories). 
}}
\label{fig_o3-5branes}
\end{figure}

Our interest is the half-BPS boundary conditions and domain walls in $\mathcal{N}=4$ SYM theories 
of orthogonal and symplectic gauge groups made by additional half $5$-branes. 
Such half $5$-branes should wrap $\mathbb{RP}^2$ $\subset$ $\mathbb{RP}^5$ 
so that the world-volume of $5$-branes are given by $AdS_4$ $\times$ $\mathbb{RP}^2$. 
In fact, there is no obstruction on wrapping of $5$-branes on $\mathbb{RP}^2$ $\subset$ $\mathbb{RP}^5$ \cite{Witten:1998xy}. 
For a half D5-brane wrapped on $\mathbb{RP}^2$, 
we obtain a domain wall across which $\theta_{RR}$ jumps. 
On the other hand, when we have a half NS5-brane wrapped on $\mathbb{RP}^2$, 
we get an interface across which $\theta_{NS}$ jumps \cite{Evans:1997hk,Witten:1998xy}. 
This implies that the configuration with a half NS5-brane realizes 
the interface between an orthogonal gauge theory and a symplectic gauge theory. 
Besides, since a $5$-brane carries D3-brane charge
and $H^2(\mathbb{RP}^2,\tilde{\mathbb{Z}})$ $=$ $\mathbb{Z}$, arbitrary number of D3-branes can appear in both sides of the $5$-branes (see Figure \ref{fig_o3-5branes}). 
%
%
%
%
%

The holographic dual geometries in Type IIB supergravity can be considered as an orbifold 
\footnote{See \cite{Huertas:2023syg} for more general discussion. }
of the geometry in Type IIB supergravity constructed by D'Hoker, Estes and Gutperle \cite{DHoker:2007zhm,DHoker:2007hhe} 
with the form 
\begin{align}
AdS_4\times S_{(1)}^2\times S_{(2)}^2\times \Sigma
\end{align}
as the $AdS_4\times S^2\times S^2$ fibration over the Riemann surface $\Sigma$ with the topology of a disk. 
When we parameterize $S_{(1)}^2$ by 
\begin{align}
x_3&=R_1\cos\theta_1\sin\varphi_1, \\
x_4&=R_1\cos\theta_1\cos\varphi_1, \\
x_5&=R_1\sin\theta_1
\end{align}
and $S_{(2)}^2$ by 
\begin{align}
x_7&=R_2\cos\theta_2\sin\varphi_2, \\
x_8&=R_2\cos\theta_2\cos\varphi_2, \\
x_9&=R_2\sin\theta_2
\end{align}
in terms of spherical coordinates, we can consider the $\mathbb{Z}_2$ action as
\begin{align}
\label{orb_act1}
\varphi_1&\rightarrow e^{\pi i}\varphi_1, 
\qquad
\varphi_2\rightarrow e^{\pi i}\varphi_2, 
\end{align}
while the coordinates on $\Sigma$, $\theta_1$ and $\theta_2$ are invariant. 
At a generic point in $\Sigma$, the action (\ref{orb_act1}) has four fixed points 
corresponding to the two poles of $S_{(1)}^2$ and those of $S_{(2)}^2$. 
However, one of the two-spheres $S_{(1)}^2$ and $S_{(2)}^2$ shrinks to zero size at the boundary of $\Sigma$ 
in the construction of \cite{DHoker:2007zhm,DHoker:2007hhe}. 
So the four different fixed points collapse pairwise at the boundary of $\Sigma$ 
and the fixed point set has no non-trivial topology. 
Such gravity dual geometries give rise to the Karch-Randall models \cite{Karch:2000ct,Karch:2000gx} 
which contain $AdS_5$ factors cut off by end-of-the-world (ETW) brane with $AdS_4$ factor, 
a configuration ending spacetime geometry in quantum gravity. 
They have a structure of ``bagpipes'' in which 
the bag corresponds to a small perturbation to $AdS_4$ geometry 
and the pipes to semi-infinite Janus throats \cite{Bachas:2018zmb}. 

For the $AdS_5$ $\times$ $S^5$ geometry 
which is holographically dual to $\mathcal{N}=4$ $U(N)$ SYM theory, 
the spectrum of the KK excitations on the geometry correspond to that of the primary operators obtained from $\mathcal{N}=4$ SYM theory in the large $N$ limit. 
However, when the mass becomes large, the excitations are considered as the giant gravitons \cite{McGreevy:2000cw,Grisaru:2000zn,Hashimoto:2000zp} 
which carry large angular momenta and behave as wrapped branes. 
In \cite{Hatsuda:2024uwt} we geometrically constructed the giant gravitons in the $AdS_4$ ETW brane region for the supergravity background 
involving asymptotic $AdS_5$ $\times$ $S^5$ regions, which we refer to as the \textit{ETW giant gravitons}. 
They are the D3-branes wrapped on the $3$-cycle as a fibration of $S_{(i)}^2$, $i=1,2$ over the segment on $\Sigma$. 
We showed that they are the BPS configurations and their energy is bounded by the background $5$-form flux associated with the $5$-brane singularity. 
The construction can be generalized, as follows, to the \textit{orbifold ETW giant gravitons} as the D3-branes wrapping the $3$-cycle in the bagpipe geometries 
with the asymptotic $AdS_5$ $\times$ $\mathbb{RP}^5$ regions 
by quotienting the ordinary ETW giant gravitons by the orbifold action (\ref{orb_act1}). 
Since the $5$-form flux in the covering spacetime with the $\mathbb{Z}_2$ orbifold action is twice as the ordinary flux, 
the effective topological wrapping numbers carried by the orbifold ETW giant gravitons are even integers. 
\footnote{More generally, for the $\mathbb{Z}_k$ orbifold action, 
the orbifold ETW giant gravitons have the topological wrapping numbers $m$ $\in$ $\mathbb{Z}/k \mathbb{Z}$. }
In view of orthogonal and symplectic gauge theories, 
this means that the baryonic charges carried by the BPS local operators, which can be identified with the wrapping numbers \cite{Arai:2019wgv}, 
are twice as those in unitary gauge theories. 
In fact, in section \ref{sec_ggexp}, we find that 
the half-indices of boundary conditions and interfaces in orthogonal and symplectic gauge theories 
which are holographically dual to the orbifold bagpipe geometries 
admit the giant graviton expansions associated with the orbifold ETW giant gravitons 
with wrapping numbers being even integers. 

\section{Half-indices}
\label{sec_hindex}
In this section, we present the half-indices that can count the boundary BPS local operators 
for the half-BPS boundary conditions $\mathcal{B}$ in 4d $\mathcal{N}=4$ SYM theories with orthogonal and symplectic gauge groups. 
It is defined as \cite{Gaiotto:2019jvo} \footnote{See \cite{Gaiotto:2019jvo,Okazaki:2019ony} for convention and definition of the half-indices for various boundary conditions. 
Also see \cite{Dimofte:2011py,Gang:2012yr} for the half-indices involving the Neumann boundary conditions. }
\begin{align}
\label{hindex_def}
\mathbb{II}_{\mathcal{B}}^{\textrm{4d $G$}}(t,z;q)
&={\Tr}_{\mathcal{H}}(-1)^{F} q^{J+\frac{H+C}{4}}t^{H-C}z^f. 
\end{align}
Here $F$ is the Fermion number operator, $J$ is the spin, 
$H,C$ are the Cartans of the two $SU(2)_H$ and $SU(2)_C$ factors of the R-symmetry group 
and $f$ are the Cartans of the other global symmetries. 
The trace is taken over the cohomology $\mathcal{H}$ of the chosen supercharges 
which belong to the subalgebra of the 3d $\mathcal{N}=4$ superalgebra. 

To express the half-indices, it is convenient to introduce the $q$-shifted factorial. 
We define 
\begin{align}
\label{qpoch_def}
(a;q)_{0}&:=1,\qquad
(a;q)_{n}:=\prod_{k=0}^{n-1}(1-aq^{k}),\qquad 
(q)_{n}:=\prod_{k=1}^{n}(1-q^{k}),
\nonumber \\
(a;q)_{\infty}&:=\prod_{k=0}^{\infty}(1-aq^{k}),\qquad 
(q)_{\infty}:=\prod_{k=1}^{\infty} (1-q^k), 
\end{align}
with $a, q \in \mathbb{C}$ and $|q|<1$. 
For simplicity we also use the following notation: 
\begin{align}
(x^{\pm};q)_{n}:=(x;q)_{n}(x^{-1};q)_{n}. 
\end{align}

In the presence of boundary, 4d $\mathcal{N}=4$ SYM theory can preserve half of supersymmetry and the $SU(2)_C\times SU(2)_H$ R-symmetry so that the adjoint scalar fields transforming as ${\bf 6}$ under the $SU(4)_{R}$ split into two scalar fields $X$ and $Y$ transforming as $({\bf 1},{\bf 3})$ and $({\bf 3},{\bf 1})$ respectively. Under the $SU(2)_{C}$ $\times$ $SU(2)_{H}$ the 4d gauginos $\lambda$ transform as $({\bf 2},{\bf 2})$. 
Accordingly, the local operators in 4d $\mathcal{N}=4$ SYM theory of gauge group $G$ which contribute to index are as follows:  
\begin{align}
\label{4dn4_ch}
\begin{array}{c|cccc}
&\partial_{z}^{n}X&\partial_{z}^{n}Y&\partial_{z}^{n}\lambda&\partial_{z}^{n}\overline{\lambda} \\ \hline
G&\textrm{adj}&\textrm{adj}&\textrm{adj}&\textrm{adj} \\
U(1)_{J}&n&n&n+\frac12&n+\frac12 \\
U(1)_{C}&0&2&+&+ \\
U(1)_{H}&2&0&+&+  \\
\textrm{fugacity}
&q^{n+\frac12}t^{2}s_{\alpha}
&q^{n+\frac12}t^{-2}s_{\alpha}
&-q^{n+1}s_{\alpha}
&-q^{n+1}s_{\alpha} \\
\end{array}
\end{align}
Here $s$ are the fugacities for gauge group $G$ and $\alpha$ are the associated weights of the adjoint representation of gauge group $G$. We consider the half-BPS boundary conditions or interfaces in 4d $\mathcal{N}=4$ gauge theories that are realized by a single $5$-brane. The half-BPS boundary condition which is realized by the NS5-brane on which a stack of D3-branes terminate, one finds the Neumann boundary condition where the gauge group is preserved \cite{Gaiotto:2008sa}. By collecting the contributions to the index and projecting these to $G$-invariants by integrating over the gauge fugacities $s$, we then obtain the half-index for the Neumann boundary condition \cite{Dimofte:2011py}. On the other hand, when D3-branes end on a single D5-brane, one finds the singular Nahm pole boundary condition \cite{Gaiotto:2008sa}. The half-index for the Nahm pole half-index can be equivalent to the Neumann half-index as a consequence of S-duality, as discussed for unitary gauge theories in \cite{Gaiotto:2019jvo} and in the subsections below for orthogonal or symplectic gauge theories. One can derive it by applying the Higgsing procedure to the Dirichlet half-index that results from the full-index by projecting out the contributions which cannot freely fluctuate at the boundary as discussed for the unitary gauge theories in  \cite{Gaiotto:2019jvo}. 

The half-index can be thought of as a generalized supersymmetric index of 3d $\mathcal{N}=4$ supersymmetric index. 
The indices admit two types of special fugacity limits, the Higgs and Coulomb limits \cite{Razamat:2014pta}
\begin{align}
\label{H_lim}
{\mathbb{II}_{\mathcal{B}}^{\textrm{4d $G$}}}^{(H)}(z;\mathfrak{q})
&=\lim_{
\begin{smallmatrix}
\textrm{$\mathfrak{q}:=q^{1/4}t$: fixed}\\
q\rightarrow 0\\
\end{smallmatrix}
}
\mathbb{II}_{\mathcal{B}}^{\textrm{4d $G$}}(t,z;q),\\
\label{C_lim}
{\mathbb{II}_{\mathcal{B}}^{\textrm{4d $G$}}}^{(C)}(z;\mathfrak{q})
&=\lim_{
\begin{smallmatrix}
\textrm{$\mathfrak{q}:=q^{1/4}t^{-1}$: fixed}\\
q\rightarrow 0\\
\end{smallmatrix}
}
\mathbb{II}_{\mathcal{B}}^{\textrm{4d $G$}}(t,z;q). 
\end{align}

The half-index is protected in the infrared so that it is a powerful tool that 
allows us to test the dualities of the 4d/3d systems and to the double holography \cite{Karch:2000ct,Karch:2000gx,Takayanagi:2011zk,Karch:2022rvr}. 
In the following we find strong evidence of S-duality of half-BPS boundary conditions and interfaces 
proposed by Gaiotto and Witten \cite{Gaiotto:2008ak} as precise matching of pairs of the half-indices. 
The results generalize those worked out in \cite{Gaiotto:2019jvo} for unitary gauge groups.  
For orthogonal gauge groups associated with Lie algebra $\mathfrak{so}(2N)$, 
we also propose S-duality of boundary conditions as well as of interfaces involving disconnected $O(2N)$ gauge theories. 

\subsection{$SO(2N+1)$ and $USp(2N)$}
When $N$ D3-branes terminate on a single NS5-brane, one finds Neumann boundary condition in $\mathcal{N}=4$ $U(N)$ SYM theory. 
In the presence of an O3-plane, when $N$ D3-branes end on a single half NS5-brane, 
we get the Neumann boundary condition in $\mathcal{N}=4$ orthogonal or symplectic gauge theory \cite{Gaiotto:2008ak}. 

\begin{figure}
\centering
\scalebox{1.2}{
\begin{tikzpicture}
\draw[thick,solid,black] (-3.5,1) -- (-3.5,3) node[above] {\textrm{1/2 NS5}};
\draw[thick,dashed,black] (3.5,1) -- (3.5,3) node[above] {\textrm{1/2 D5$'$}};
\draw[thick,dotted,blue] (-3.5,2) -- (-1.5,2) node[below] {$\widetilde{\textrm{O3}}^-$};
\draw[thick,dotted,red] (3.5,2) -- (5.5,2) node[below] {$\textrm{O3}^+$};
\draw[thick,solid,yellow] (-3.5,2.3)  -- (-1.5,2.3) node[above] {\textcolor{black}{\textrm{D3}}};
\draw[thick,solid,yellow] (-3.5,2.2)  -- (-1.5,2.2);
\draw[thick,solid,yellow] (-3.5,2.1)  -- (-1.5,2.1);
\draw[thick,solid,yellow] (3.5,2.3)  -- (5.5,2.3) node[above] {\textcolor{black}{\textrm{D3}}};
\draw[thick,solid,yellow] (3.5,2.2)  -- (5.5,2.2);
\draw[thick,solid,yellow] (3.5,2.1)  -- (5.5,2.1);
\end{tikzpicture}
}
\caption{\textrm{
The S-dual brane configurations for the Neumann boundary condition in $SO(2N+1)$ SYM theory 
and the Nahm pole boundary condition in $USp(2N)$ SYM theory. 
}}
\label{fig_bcNeuNahm1}
\end{figure}
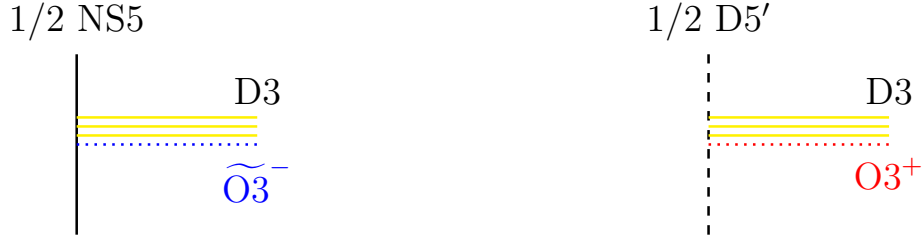

When $N$ D3-branes in the background of $\widetilde{\textrm{O3}}^-$ end on a single half NS5-brane, 
we have the Neumann boundary condition in $\mathcal{N}=4$ $SO(2N+1)$ gauge theory (see Figure \ref{fig_bcNeuNahm1}). 

The half-index is given by the matrix integral 
\begin{align}
\label{h_B_Neu}
\mathbb{II}_{\mathcal{N}}^{\textrm{4d $SO(2N+1)$}}(t;q)
&=\frac{1}{2^N N!}
\frac{(q)_{\infty}^N}{(q^{\frac12}t^{-2};q)_{\infty}^N}
\oint \prod_{i=1}^N \frac{ds_i}{2\pi is_i}
\frac{(s_i^{\pm};q)_{\infty}}{(q^{\frac12}t^{-2}s_i^{\pm};q)_{\infty}}
\nonumber\\
&\times 
\prod_{1\le i<j\le N}
\frac{(s_i^{\pm} s_j^{\mp};q)_{\infty} (s_{i}^{\pm}s_{j}^{\pm};q)_{\infty}}
{(q^{\frac12}t^{-2}s_i^{\pm}s_j^{\mp};q)_{\infty} (q^{\frac12}t^{-2}s_i^{\pm}s_j^{\pm};q)_{\infty}}. 
\end{align}
Under the $S$ transformation, $\widetilde{\textrm{O3}}^-$ and the NS5-brane become $O3^+$ and a half D5$'$-brane respectively (see Figure \ref{fig_bcNeuNahm1}). 
The half D5$'$-brane creates the Nahm pole so that 
one finds the dual boundary condition as the regular Nahm pole boundary condition for $\mathcal{N}=4$ $USp(2N)$ gauge theory 
whose regular $\mathfrak{su}(2)$ embedding corresponds to the $2N$-dimensional irreducible representation of $SU(2)$ \cite{Gaiotto:2008ak}. 

The Nahm pole boundary condition can be obtained from the deformed Dirichlet boundary condition in the RG flow 
by turning on a regular nilpotent vev for the adjoint scalar field. 
In the RG flow there are extra decoupled three-dimensional degrees of freedom. 
By means of the Higgsing procedure in \cite{Gaiotto:2012xa} the Nahm pole half-index for $U(N)$ gauge theory is obtained   
from the Dirichlet half-index for $U(N)$ gauge theory in \cite{Gaiotto:2019jvo}. 
Let us derive the closed-form formula for (\ref{h_B_Neu}) as the Nahm pole half-index for $USp(2N)$ gauge theory by applying the Higgsing method. 

The half-index of Dirichlet boundary condition $\mathcal{D}'$ for $\mathcal{N}=4$ $USp(2N)$ gauge theory is given by 
\begin{align}
\label{h_C_Dir}
\mathbb{II}_{\mathcal{D}'}^{\textrm{4d $USp(2N)$}}(t,x_i;q)
&=
\frac{(q)_{\infty}^{N}}{(q^{\frac12}t^{-2};q)_{\infty}^{N}}
\prod_{i=1}^{N}
\frac{(qx_{i}^{\pm2};q)_{\infty}}
{(q^{\frac12}t^{-2} x_{i}^{\pm2};q)_{\infty}}
\nonumber\\
&\times 
\prod_{1\le i<j\le N}
\frac{
(qx_{i}^{\pm}x_{j}^{\mp};q)_{\infty}
(qx_{i}^{\pm}x_{j}^{\pm};q)_{\infty}
}
{
(q^{\frac12}t^{-2} x_{i}^{\pm}x_{j}^{\mp};q)_{\infty}
(q^{\frac12}t^{-2} x_{i}^{\pm}x_{j}^{\pm};q)_{\infty}
}, 
\end{align}
where $x_i$ are the fugacities for the bulk gauge group 
which breaks down to the global symmetry $USp(2N)$ at a boundary. 
Turning on a regular nilpotent vev for the adjoint
scalar field, the Dirichlet boundary condition is deformed to the Nahm pole boundary condition. 
Correspondingly, let us deform the Dirichlet half-index (\ref{h_C_Dir}) by identifying the fugacities $x_{i}$ with $q^{\frac{2i-1}{4}} t^{-2i+1}$. 
We find that it is factorized as  
\begin{align}
\label{h_C_Higgs}
&\mathbb{II}_{\mathcal{D}'}^{\textrm{4d $USp(2N)$}}
(t,x_{i}=q^{\frac{2i-1}{4}}t^{-2i+1})
\nonumber\\
&=\prod_{k=1}^{N}
\frac{(q^{k+\frac12}t^{-4k+2};q)_{\infty}}
{(q^{k}t^{-4k};q)_{\infty}}
\times 
\mathcal{I}^{\textrm{3d tHM}}(x=q^{\frac14}t^{-1})^{N}
\nonumber\\
&\times 
\prod_{l=1}^{N-1}\mathcal{I}^{\textrm{3d tHM}}(x=q^{\frac{4l-1}{4}}t^{-4l+1})^{N-l}
\times 
\prod_{m=1}^{N-1}\mathcal{I}^{\textrm{3d tHM}}(x=q^{\frac{4m+1}{4}}t^{-4m-1})^{N-m}, 
\end{align}
where 
\begin{align}
\mathcal{I}^{\textrm{3d tHM}}(x)
&=\frac{(q^{\frac34}tx;q)_{\infty}(q^{\frac34}tx^{-1};q)_{\infty}}
{(q^{\frac14}t^{-1}x;q)_{\infty}(q^{\frac14}t^{-1}x^{-1};q)_{\infty}}
\end{align}
is the full-index of the 3d twisted hypermultiplet. 
By stripping off the full-indices of the 3d twisted hypermultiplets in (\ref{h_C_Higgs}), 
we find the half-index of Nahm pole boundary condition for $USp(2N)$ gauge theory
\begin{align}
\label{h_C_Nahm}
\mathbb{II}_{\textrm{Nahm}'}^{\textrm{4d $USp(2N)$}}(t;q)
&=\prod_{k=1}^N \frac{(q^{k+\frac12}t^{-4k+2};q)_{\infty}}{(q^k t^{-4k};q)_{\infty}}. 
\end{align}
In fact, it follows that the half-indices (\ref{h_B_Neu}) and (\ref{h_C_Nahm}) exactly coincide with one another. 
\footnote{
The equality between (\ref{h_B_Neu}) and (\ref{h_C_Nahm}) can be proven by means of the inner product of the Macdonald polynomials \cite{MR1354144,MR1314036,MR1354956}. 
}
It supports the conjectural duality: 
\begin{align}
&\textrm{Neumann b.c. for $\mathcal{N}=4$ $SO(2N+1)$ SYM}
\nonumber\\
&\Leftrightarrow
\textrm{Nahm pole b.c. for $\mathcal{N}=4$ $USp(2N)$ SYM}. 
\end{align}

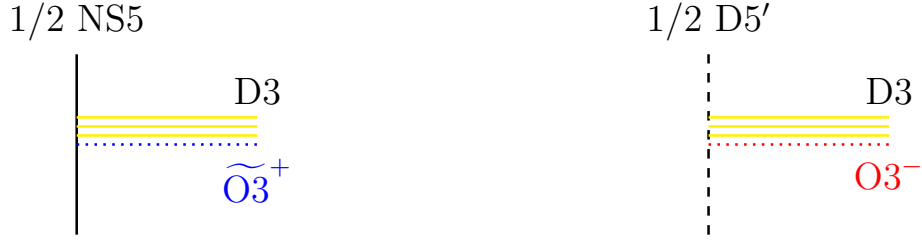
\begin{figure}
\centering
\scalebox{1.2}{
\begin{tikzpicture}
\draw[thick,solid,black] (-3.5,1) -- (-3.5,3) node[above] {\textrm{1/2 NS5}};
\draw[thick,dashed,black] (3.5,1) -- (3.5,3) node[above] {\textrm{1/2 D5$'$}};
\draw[thick,dotted,blue] (-3.5,2) -- (-1.5,2) node[below] {$\widetilde{\textrm{O3}}^+$};
\draw[thick,dotted,red] (3.5,2) -- (5.5,2) node[below] {$\textrm{O3}^-$};
\draw[thick,solid,yellow] (-3.5,2.3)  -- (-1.5,2.3) node[above] {\textcolor{black}{\textrm{D3}}};
\draw[thick,solid,yellow] (-3.5,2.2)  -- (-1.5,2.2);
\draw[thick,solid,yellow] (-3.5,2.1)  -- (-1.5,2.1);
\draw[thick,solid,yellow] (3.5,2.3)  -- (5.5,2.3) node[above] {\textcolor{black}{\textrm{D3}}};
\draw[thick,solid,yellow] (3.5,2.2)  -- (5.5,2.2);
\draw[thick,solid,yellow] (3.5,2.1)  -- (5.5,2.1);
\end{tikzpicture}
}
\caption{\textrm{
The S-dual brane configurations for the Neumann boundary condition in $USp(2N)$ SYM theory 
and the Nahm pole boundary condition in $SO(2N+1)$ SYM theory. 
}}
\label{fig_bcNeuNahm2}
\end{figure}

Next consider the case with $N$ D3-branes in the presence of $O3^+$ terminating on a single half NS5-brane (see Figure \ref{fig_bcNeuNahm2}). 
We have the Neumann boundary condition in $\mathcal{N}=4$ $USp(2N)$ gauge theory. 

The half-index reads
\begin{align}
\label{h_C_Neu}
&\mathbb{II}_{\mathcal{N}}^{\textrm{4d $USp(2N)$}}(t;q)
\nonumber\\
&=\frac{1}{2^N N!}
\frac{(q)_{\infty}^N}{(q^{\frac12}t^{-2};q)_{\infty}^N}
\oint \prod_{i=1}^N \frac{ds_i}{2\pi is_i}
\frac{(s_i^{\pm 2};q)_{\infty}}{(q^{\frac12}t^{-2}s_i^{\pm 2};q)_{\infty}}
\nonumber\\
&\times 
\prod_{1\le i<j\le N}
\frac{(s_i^{\pm} s_j^{\mp};q)_{\infty} (s_{i}^{\pm}s_{j}^{\pm};q)_{\infty}}
{(q^{\frac12}t^{-2}s_i^{\pm}s_j^{\mp};q)_{\infty} (q^{\frac12}t^{-2}s_i^{\pm}s_j^{\pm};q)_{\infty}}. 
\end{align}

The S-dual configuration contains 
$N$ D3-branes in the background of $\widetilde{\textrm{O3}}^-$ terminating on a single half D5$'$-brane (see Figure \ref{fig_bcNeuNahm2}). 
It realizes the Nahm pole boundary condition for $SO(2N+1)$ gauge theory 
that corresponds to the $2N+1$-dimensional irreducible representation of $SU(2)$. 
It is therefore conjectured \cite{Gaiotto:2008ak} that 
the $USp(2N)$ Neumann boundary condition is dual to the regular Nahm pole boundary condition for $SO(2N+1)$ gauge theory. 

Let us derive the dual Nahm pole half-index by performing the Higgsing prescription in a similar manner as performed in \cite{Gaiotto:2019jvo}. 
We begin with the half-index of the Dirichlet boundary condition $\mathcal{D}'$ for 4d $\mathcal{N}=4$ $SO(2N+1)$ gauge theory
\begin{align}
\label{h_B_Dir}
\mathbb{II}_{\mathcal{D}'}^{\textrm{4d $SO(2N+1)$}}(t,x_{i};q)
&=
\frac{(q)_{\infty}^{N}}{(q^{\frac12}t^{-2};q)_{\infty}^{N}}
\prod_{i=1}^{N}
\frac{(qx_{i}^{\pm};q)_{\infty}}
{(q^{\frac12}t^{-2} x_{i}^{\pm};q)_{\infty}}
\nonumber\\
&\times 
\prod_{1\le i<j\le N}
\frac{
(qx_{i}^{\pm}x_{j}^{\mp};q)_{\infty}
(qx_{i}^{\pm}x_{j}^{\pm};q)_{\infty}
}
{
(q^{\frac12}t^{-2} x_{i}^{\pm}x_{j}^{\mp};q)_{\infty}
(q^{\frac12}t^{-2} x_{i}^{\pm}x_{j}^{\pm};q)_{\infty}
}. 
\end{align}
We propose that the Nahm pole boundary condition for $SO(2N+1)$ gauge theory can be obtained by deforming the Dirichlet boundary condition 
in such a way that the Dirichlet half-index is specialized with the fugacities $x_{i}$ being replaced by $q^{\frac{i}{2}}t^{-2i}$. 
We find the following factorized form 
\begin{align}
\label{h_B_Higgs}
&
\mathbb{II}_{\mathcal{D}'}^{\textrm{4d $SO(2N+1)$}}
(t,x_{i}=q^{\frac{i}{2}}t^{-2i};q)
\nonumber\\
&=\prod_{k=1}^{N}
\frac{(q^{k+\frac12}t^{-4k+2};q)_{\infty}}
{(q^{k}t^{-4k};q)_{\infty}}
\times 
\mathcal{I}^{\textrm{3d tHM}}(x=q^{\frac14}t^{-1})^{N}
\nonumber\\
&\times 
\prod_{l=1}^{N-1}\mathcal{I}^{\textrm{3d tHM}}(x=q^{\frac{4l-1}{4}}t^{-4l+1})^{N-l}
\times 
\prod_{m=1}^{N-1}\mathcal{I}^{\textrm{3d tHM}}(x=q^{\frac{4m+1}{4}}t^{-4m-1})^{N-m}. 
\end{align}
By stripping off the full-indices of the 3d twisted hypermultiplets which will be decoupled along the RG-flow, 
we get the half-index of the regular Nahm pole boundary condition for $\mathcal{N}=4$ $SO(2N+1)$ gauge theory
\begin{align}
\label{h_B_Nahm}
\mathbb{II}_{\textrm{Nahm}'}^{\textrm{4d $SO(2N+1)$}}(t;q)
&=\prod_{k=1}^N \frac{(q^{k+\frac12}t^{-4k+2};q)_{\infty}}{(q^k t^{-4k};q)_{\infty}}. 
\end{align}
The half-indices (\ref{h_C_Neu}) and (\ref{h_B_Nahm}) agree with each other. 
The equality of these half-indices confirms the duality
\begin{align}
&\textrm{Neumann b.c. for $\mathcal{N}=4$ $USp(2N)$ SYM}
\nonumber\\
&\Leftrightarrow
\textrm{Nahm pole b.c. for $\mathcal{N}=4$ $SO(2N+1)$ SYM}
\end{align}
In addition, we observe that the Nahm pole half-indices (\ref{h_C_Nahm}) and (\ref{h_B_Nahm}) are equal, 
which implies the equality between the Neumann half-indices (\ref{h_B_Neu}) and (\ref{h_C_Neu}). 

We find that in the Coulomb limit, the half-indices (\ref{h_B_Neu}), (\ref{h_C_Nahm}), (\ref{h_C_Neu}) and (\ref{h_B_Nahm}) reduce to the half-BPS index
\begin{align}
\label{1/2BPSindex_BC}
\mathcal{I}^{SO(2N+1)}_{\textrm{$\frac12$BPS}}(\mathfrak{q})
&=\mathcal{I}^{USp(2N)}_{\textrm{$\frac12$BPS}}(\mathfrak{q})
=\prod_{n=1}^{N}\frac{1}{1-\mathfrak{q}^{4n}}. 
\end{align}
Note that the index (\ref{1/2BPSindex_BC}) is simply obtained from the half-BPS index of the $U(N)$ theory 
by replacing the fuagacity $\mathfrak{q}$ with its square. 
In view of $SO(2N+1)$ gauge theory, this follows from the fact that in $SO(2N+1)$ gauge theory the adjoint scalar field $X$ is antisymmetric and therefore
\begin{align}
\Tr X&=\Tr X^3=\Tr X^5=\cdots=0. 
\end{align}
Accordingly, the non-zero multi-trace half-BPS operators take the form 
\begin{align}
(\Tr X^{2})^{\lambda_1}(\Tr X^{4})^{\lambda_2}\cdots, 
\end{align}
which can be labeled by a partition $\lambda$. 
In fact, (\ref{1/2BPSindex_BC}) precisely agrees with the expression conjectured in \cite{Caputa:2013vla}. 

\subsection{$O(2N)$}
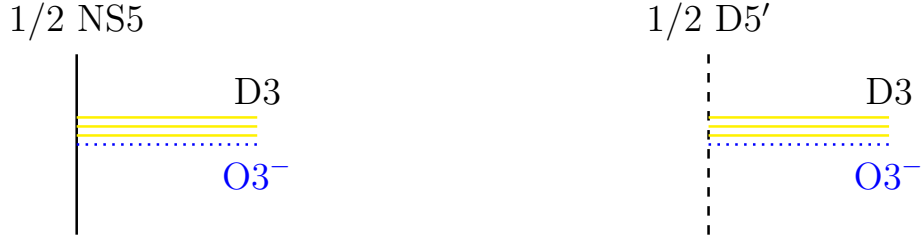
\begin{figure}
\centering
\scalebox{1.2}{
\begin{tikzpicture}
\draw[thick,solid,black] (-3.5,1) -- (-3.5,3) node[above] {\textrm{1/2 NS5}};
\draw[thick,dashed,black] (3.5,1) -- (3.5,3) node[above] {\textrm{1/2 D5$'$}};
\draw[thick,dotted,blue] (-3.5,2) -- (-1.5,2) node[below] {$\textrm{O3}^-$};
\draw[thick,dotted,blue] (3.5,2) -- (5.5,2) node[below] {$\textrm{O3}^-$};
\draw[thick,solid,yellow] (-3.5,2.3)  -- (-1.5,2.3) node[above] {\textcolor{black}{\textrm{D3}}};
\draw[thick,solid,yellow] (-3.5,2.2)  -- (-1.5,2.2);
\draw[thick,solid,yellow] (-3.5,2.1)  -- (-1.5,2.1);
\draw[thick,solid,yellow] (3.5,2.3)  -- (5.5,2.3) node[above] {\textcolor{black}{\textrm{D3}}};
\draw[thick,solid,yellow] (3.5,2.2)  -- (5.5,2.2);
\draw[thick,solid,yellow] (3.5,2.1)  -- (5.5,2.1);
\end{tikzpicture}
}
\caption{\textrm{
The S-dual brane configurations for the Neumann and Nahm pole boundary conditions in $O(2N)$ SYM theory. 
}}
\label{fig_o2NNeuNahm}
\end{figure}

When we have a single half NS5-brane on which $N$ D3-branes in the background of O3$^-$ end, 
we find the Neumann boundary condition in $\mathcal{N}=4$ $O(2N)$ gauge theory (see Figure \ref{fig_o2NNeuNahm}). 

The $O(2N)$ gauge theory can be obtained by gauging the $\mathbb{Z}_2$ global symmetry of the $SO(2N)$ gauge theory. 
The half-index of the $SO(2N)$ Neumann boundary condition is given by
\begin{align}
\label{h_D_Neu}
&
\mathbb{II}_{\mathcal{N}}^{\textrm{4d $SO(2N)$}}(t;q)
\nonumber\\
&=\frac{1}{2^{N-1} N!}
\frac{(q)_{\infty}^N}{(q^{\frac12}t^{-2};q)_{\infty}^N}
\oint \prod_{i=1}^{N}
\frac{ds_i}{2\pi is_i}
\prod_{1\le i<j\le N}
\frac{(s_i^{\pm}s_j^{\mp};q)_{\infty} (s_i^{\pm}s_j^{\pm};q)_{\infty}}
{(q^{\frac12}t^{-2}s_i^{\pm}s_j^{\mp};q)_{\infty} (q^{\frac12}t^{-2}s_i^{\pm}s_j^{\pm};q)_{\infty}}. 
\end{align}

The Neumann half-index for the other connected component of the $SO(2N)$ gauge theory takes the form 
\begin{align}
\label{h_D_Neu2}
&
\mathbb{II}_{\mathcal{N}}^{\textrm{4d $SO(2N)^{-}$}}(t;q)
\nonumber\\
&=\frac{1}{2^{N-1}(N-1)!}
\frac{(q)_{\infty}^{N-1}(-q;q)_{\infty}}{(q^{\frac12}t^{-2};q)_{\infty}^{N-1}(-q^{\frac12}t^{-2};q)_{\infty}}
\oint \prod_{i=1}^{N-1}\frac{ds_i}{2\pi is_i}
\frac{(s_i^{\pm};q)_{\infty} (-s_i^{\pm};q)_{\infty}}{(q^{\frac12}t^{-2}s_i^{\pm};q)_{\infty}(-q^{\frac12}t^{-2}s_i^{\pm};q)_{\infty}}
\nonumber\\
&\times 
\prod_{1\le i<j\le N}
\frac{(s_i^{\pm}s_j^{\mp};q)_{\infty} (s_i^{\pm}s_j^{\pm};q)_{\infty}}
{(q^{\frac12}t^{-2}s_i^{\pm}s_j^{\mp};q)_{\infty} (q^{\frac12}t^{-2}s_i^{\pm}s_j^{\pm};q)_{\infty}}. 
\end{align}

Then the Neumann half-indices of $O(2N)$ gauge theory are given by
\begin{align}
\label{h_O2N_Neu}
\mathbb{II}_{\mathcal{N}}^{\textrm{4d $O(2N)^{\pm}$}}(t;q)
&=\frac12 \left[
\mathbb{II}_{\mathcal{N}}^{\textrm{4d $SO(2N)$}}(t;q)
\pm 
\mathbb{II}_{\mathcal{N}}^{\textrm{4d $SO(2N)^{-}$}}(t;q)
\right]. 
\end{align}
Here we have two choices for sign. 
For $+$ (resp. $-$), 
the half-index counts the $\mathbb{Z}_2$-even (resp. $\mathbb{Z}_2$-odd) boundary BPS local operators. 

Under the $S$ transformation the half NS5-brane becomes a half D5$'$-brane on which only $2N-1$ of $2N$ D3-branes terminate 
as the O3-plane on the other side of the half D5-brane is of the $\widetilde{\textrm{O3}}^-$ type (see Figure \ref{fig_o2NNeuNahm}). 
It gives rise to the S-dual Nahm pole boundary condition in $O(2N)$ gauge theory, 
however, it does not correspond to a $2N$-dimensional irreducible representation of $SU(2)$ 
but rather a decomposition $2N=(2N-1)+1$ \cite{Gaiotto:2008ak}. 

Let us derive the half-index for the S-dual Nahm pole boundary condition by means of the Higgsing procedure in \cite{Gaiotto:2019jvo}. 
We begin by the half-index of the Dirichlet boundary condition $\mathcal{D}'$ for $SO(2N)$ gauge theory. 
It is given by
\begin{align}
\label{h_D_Dir}
\mathbb{II}_{\mathcal{D}'}^{\textrm{4d $SO(2N)$}}(t,x_i;q)
&=\frac{(q)_{\infty}^{N}}{(q^{\frac12}t^{-2};q)_{\infty}^{N}}
\prod_{1\le i<j\le N}
\frac{
(qx_{i}^{\pm}x_{j}^{\mp};q)_{\infty} 
(qx_{i}^{\pm}x_{j}^{\pm};q)_{\infty} 
}{
(q^{\frac12}t^{-2}x_{i}^{\pm}x_{j}^{\mp};q)_{\infty} 
(q^{\frac12}t^{-2}x_{i}^{\pm}x_{j}^{\pm};q)_{\infty} 
}. 
\end{align}
By setting the fugacity $x_i$ for the boundary global symmetry to $q^{\frac{i-1}{2}}t^{-2(i-1)}$, 
the Dirichlet half-index is factorizes as 
\begin{align}
\label{h_D_Higgs1even}
&
\mathbb{II}_{\mathcal{D}'}^{\textrm{4d $SO(2N)$}}(t,x_{i}=q^{\frac{i-1}{2}}t^{-2(i-1)};q)
\nonumber\\
&=
\frac{
(q^{\frac12+\frac{N}{2}}t^{-2N+2};q)_{\infty}
}
{
(q^{\frac{N}{2}}t^{-2N};q)_{\infty}
}
\prod_{k=1}^{N-1}
\frac{
(q^{\frac12+k}t^{-4k+2};q)_{\infty}
}
{
(q^{k}t^{-4k};q)_{\infty}
}
\times 
\mathcal{I}^{\textrm{3d tHM}}(x=q^{\frac14}t^{-1})^{N}
\nonumber\\
&\times 
\prod_{l=\frac{N}{2}+1}^{N-1}
\mathcal{I}^{\textrm{3d tHM}}(x=q^{N-l-\frac14}t^{-4N+4l+1})^{l}
\times 
\mathcal{I}^{\textrm{3d tHM}}(x=q^{N-l+\frac14}t^{-4N+4l-1})^{l}
\nonumber\\
&\times 
\prod_{m=1}^{\frac{N}{2}-1}
\mathcal{I}^{\textrm{3d tHM}}(x=q^{N-m-\frac54}t^{-4N+4m+5})^{m}
\times 
\mathcal{I}^{\textrm{3d tHM}}(x=q^{N-m-\frac34}t^{-4N+4m+3})^{m}
\end{align}
for even $N$ and 
\begin{align}
\label{h_D_Higgs1odd}
&
\mathbb{II}_{\mathcal{D}'}^{\textrm{4d $SO(2N)$}}(t,x_{i}=q^{\frac{i-1}{2}}t^{-2(i-1)};q)
\nonumber\\
&=
\frac{
(q^{\frac12+\frac{N}{2}}t^{-2N+2};q)_{\infty}
}
{
(q^{\frac{N}{2}}t^{-2N};q)_{\infty}
}
\prod_{k=1}^{N-1}
\frac{
(q^{\frac12+k}t^{-4k+2};q)_{\infty}
}
{
(q^{k}t^{-4k};q)_{\infty}
}
\times 
\mathcal{I}^{\textrm{3d tHM}}(x=q^{\frac14}t^{-1})^{N}
\nonumber\\
&\times 
\prod_{l=\frac{N+3}{2}}^{N-1}
\mathcal{I}^{\textrm{3d tHM}}(x=q^{N-l-\frac14}t^{-4N+4l+1})^{l}
\times 
\mathcal{I}^{\textrm{3d tHM}}(x=q^{N-l+\frac14}t^{-4N+4l-1})^{l}
\nonumber\\
&\times 
\mathcal{I}^{\textrm{3d tHM}}(x=q^{\frac{N}{2}-\frac34}t^{-2N+3})^{\frac{N+1}{2}}
\times 
\mathcal{I}^{\textrm{3d tHM}}(x=q^{\frac{N}{2}-\frac14}t^{-2N+1})^{\frac{N-1}{2}}
\nonumber\\
&\times 
\prod_{m=1}^{\frac{N-3}{2}}
\mathcal{I}^{\textrm{3d tHM}}(x=q^{N-m-\frac54}t^{-4N+4m+5})^{m}
\times 
\mathcal{I}^{\textrm{3d tHM}}(x=q^{N-m-\frac34}t^{-4N+4m+3})^{m}
\end{align}
for odd $N$. 

From the expressions (\ref{h_D_Higgs1even}) and (\ref{h_D_Higgs1odd}) for the Higgsed Dirichlet half-indices, 
we get the Nahm pole half-index as the common factor 
\begin{align}
\label{h_D_Nahm}
\mathbb{II}_{\textrm{Nahm}'}^{\textrm{4d $SO(2N)$}}(t;q)
&=
\frac{
(q^{\frac12+\frac{N}{2}} t^{-2N+2};q)_{\infty}
}
{
(q^{\frac{N}{2}}t^{-2N};q)_{\infty}
}
\prod_{k=1}^{N-1}
\frac{
(q^{\frac12+k}t^{2-4k};q)_{\infty}
}
{
(q^{k}t^{-4k};q)_{\infty}
}. 
\end{align}
Indeed, we find that the $SO(2N)$ Neumann half-index (\ref{h_D_Neu}) precisely agrees with the Nahm pole half-index (\ref{h_D_Nahm}).  

Similarly, for the disconnected component we can perform the Higgsing procedure. 
We have the Dirichlet half-index of the form
\begin{align}
\label{h_D_Dir2}
&
\mathbb{II}_{\mathcal{D}'}^{\textrm{4d $SO(2N)^{-}$}}(t,x_i;q)
\nonumber\\
&=\frac{(q)_{\infty}^{N-1} (-q;q)_{\infty}}{(q^{\frac12}t^{-2};q)_{\infty}^{N-1}(-q^{\frac12}t^{-2};q)_{\infty}}
\prod_{i=1}^{N-1}
\frac{(qx_i^{\pm};q)_{\infty}(-qx_i^{\pm};q)_{\infty}}{(q^{\frac12}t^{-2}x_i^{\pm};q)_{\infty}(-q^{\frac12}t^{-2}x_i^{\pm};q)_{\infty}}
\nonumber\\
&\times 
\prod_{1\le i<j\le N-1}
\frac{(qx_i^{\pm}x_j^{\mp};q)_{\infty} (qx_i^{\pm}x_j^{\mp};q)_{\infty}}
{(q^{\frac12}t^{-2}x_i^{\pm}x_j^{\mp};q)_{\infty} (q^{\frac12}t^{-2}x_i^{\pm}x_j^{\pm};q)_{\infty}}. 
\end{align}
When the fugacities $x_i$ are specialized as $q^{\frac{i}{2}}t^{-2i}$, 
one finds 
\begin{align}
\label{h_D_Higgs2}
&
\mathbb{II}_{\mathcal{D}'}^{\textrm{4d $SO(2N)^{-}$}}(t,x_{i}=q^{\frac{i}{2}}t^{-2i};q)
\nonumber\\
&=\frac{(-q^{\frac12+\frac{N}{2}}t^{-2N+2};q)_{\infty}}
{(-q^{\frac{N}{2}}t^{-2N};q)_{\infty}}
\prod_{k=1}^{N-1}
\frac{(q^{\frac12+k}t^{2-4k};q)_{\infty}}
{(q^kt^{-4k};q)_{\infty}}
\nonumber\\
&\times 
\mathcal{I}^{\textrm{3d tHM}}(x=q^{\frac14}t^{-1})^{N-1}
\prod_{l=1}^{N-1}
\mathcal{I}^{\textrm{3d tHM}}(x=-q^{\frac{2l-1}{4}}t^{-(2l-1)})
\nonumber\\
&\times 
\prod_{m=1}^{N-1}
\mathcal{I}^{\textrm{3d tHM}}(x=q^{\frac{4m-1}{4}}t^{-(4m-1)})^{N-1-m}
\mathcal{I}^{\textrm{3d tHM}}(x=q^{\frac{4m+1}{4}}t^{-(4m+1)})^{N-1-m}. 
\end{align}
Taking away the contributions from the decoupled hypermultiplets, we obtain the half-index 
\begin{align}
\label{h_D_Nahm2}
\mathbb{II}_{\textrm{Nahm}'}^{\textrm{4d $SO(2N)^{-}$}}(t;q)
&=
\frac{(-q^{\frac12+\frac{N}{2}} t^{-2N+2};q)_{\infty}}
{(-q^{\frac{N}{2}}t^{-2N};q)_{\infty}}
\prod_{k=1}^{N-1}
\frac{(q^{\frac12+k}t^{2-4k};q)_{\infty}}
{(q^{k}t^{-4k};q)_{\infty}}. 
\end{align}
of the $SO(2N)$ Nahm pole boundary condition for the disconnected part. 

In fact, we find that 
the Neumann half-index (\ref{h_D_Neu2}) for the other disconnected component of $O(2N)$ gauge theory 
precisely matches with the Nahm pole half-index (\ref{h_D_Nahm2})! 
Eventually we obtain the Nahm pole half-iindices for $O(2N)$ gauge theory
\begin{align}
\label{h_O2N_Nahm}
&
\mathbb{II}_{\textrm{Nahm}'}^{\textrm{4d $O(2N)^{\pm}$}}(t;q)
\nonumber\\
&=
\frac12 
\left[
\frac{(q^{\frac12+\frac{N}{2}} t^{-2N+2};q)_{\infty}}
{(q^{\frac{N}{2}}t^{-2N};q)_{\infty}}
\pm 
\frac{(-q^{\frac12+\frac{N}{2}} t^{-2N+2};q)_{\infty}}
{(-q^{\frac{N}{2}}t^{-2N};q)_{\infty}}
\right]
\prod_{k=1}^{N-1}
\frac{(q^{\frac12+k}t^{2-4k};q)_{\infty}}
{(q^{k}t^{-4k};q)_{\infty}}, 
\end{align}
which match with the Neumann half-indices (\ref{h_O2N_Neu}) of $O(2N)$ gauge theory. 

To summarize, the equality between (\ref{h_D_Neu}) and (\ref{h_D_Nahm}) 
supports the following duality of boundary conditions: 
\begin{align}
&\textrm{Neumann b.c. for $\mathcal{N}=4$ $SO(2N)$ SYM}
\nonumber\\
&\Leftrightarrow
\textrm{Nahm pole b.c. for $\mathcal{N}=4$ $SO(2N)$ SYM}. 
\end{align}
Together with the agreement of (\ref{h_D_Neu2}) with (\ref{h_D_Nahm2}), 
we get the identity between (\ref{h_O2N_Neu}) and (\ref{h_O2N_Nahm}), 
which further indicates 
\begin{align}
&\textrm{Neumann b.c. for $\mathcal{N}=4$ $O(2N)$ SYM}
\nonumber\\
&\Leftrightarrow
\textrm{Nahm pole b.c. for $\mathcal{N}=4$ $O(2N)$ SYM}. 
\end{align}

In the Coulomb limit, the $SO(2N)$ half-indices (\ref{h_D_Neu}) and (\ref{h_D_Nahm}) become the half-BPS index 
\begin{align}
\label{1/2BPSindex_so2N}
\mathcal{I}^{SO(2N)}_{\textrm{$\frac12$BPS}}(\mathfrak{q})
&=
\frac{1}{1-\mathfrak{q}^{2N}}
\prod_{n=1}^{N-1}\frac{1}{1-\mathfrak{q}^{4n}}
\nonumber\\
&=
\prod_{n=1}^{N}\frac{1}{1-\mathfrak{q}^{4n}}
+\mathfrak{q}^{2N}
\prod_{n=1}^{N}\frac{1}{1-\mathfrak{q}^{4n}}. 
\end{align}
This agrees with the result in \cite{Caputa:2013vla}. 
The first term (reps. second term) in the second line correspond to the 
operators involving even number (resp. any odd number) 
of antisymmetric tensors $\epsilon^{i_1\cdots i_{N}}$. 
In particular, the factor $\mathfrak{q}^{2N}$ in the second term 
is contributed from the half-BPS Pfaffian operator consisting $N$ bosons. 

On the other hand, the Coulomb limits of the $O(2N)$ half-indices (\ref{h_O2N_Neu}) and (\ref{h_O2N_Nahm}) are given by
\begin{align}
\label{1/2BPSindex_O2N+}
\mathcal{I}^{O(2N)^+}_{\textrm{$\frac12$BPS}}(\mathfrak{q})
&=\prod_{n=1}^{N}\frac{1}{1-\mathfrak{q}^{4n}}, \\
\label{1/2BPSindex_O2N-}
\mathcal{I}^{O(2N)^-}_{\textrm{$\frac12$BPS}}(\mathfrak{q})
&=\mathfrak{q}^{2N}\prod_{n=1}^{N}\frac{1}{1-\mathfrak{q}^{4n}}. 
\end{align}
We see that the index (\ref{1/2BPSindex_O2N+}) for the $\mathbb{Z}_2$-even states 
agrees with the half-BPS index (\ref{1/2BPSindex_BC}) for $SO(2N+1)$ and $USp(2N)$ gauge theories. 

\subsection{$USp(2N)'$}
\begin{figure}
\centering
\scalebox{1.2}{
\begin{tikzpicture}
\draw[thick,solid,black] (-3.5,1) -- (-3.5,3) node[above] {\textrm{1/2 NS5}};
\draw[thick,dashed,black] (3.5,1) -- (3.5,3) node[above] {\textrm{1/2 D5$'$}};
\draw[thick,dotted,red] (-3.5,2) -- (-1.5,2) node[below] {$\widetilde{\textrm{O3}}^+$};
\draw[thick,dotted,red] (3.5,2) -- (5.5,2) node[below] {$\widetilde{\textrm{O3}}^+$};
\draw[thick,solid,yellow] (-3.5,2.3)  -- (-1.5,2.3) node[above] {\textcolor{black}{\textrm{D3}}};
\draw[thick,solid,yellow] (-3.5,2.2)  -- (-1.5,2.2);
\draw[thick,solid,yellow] (-3.5,2.1)  -- (-1.5,2.1);
\draw[thick,solid,yellow] (3.5,2.3)  -- (5.5,2.3) node[above] {\textcolor{black}{\textrm{D3}}};
\draw[thick,solid,yellow] (3.5,2.2)  -- (5.5,2.2);
\draw[thick,solid,yellow] (3.5,2.1)  -- (5.5,2.1);
\end{tikzpicture}
}
\caption{\textrm{
The S-dual brane configurations for the Neumann and Nahm pole boundary conditions in $USp(2N)'$ SYM theory. 
}}
\label{fig_usp2N'NeuNahm}
\end{figure}
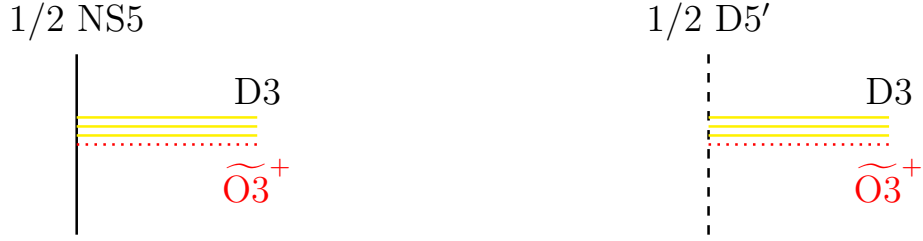

There is a variant of the basic Neumann boundary condition for symplectic gauge theory $USp(2N)'$, 
which is constructed by $N$ D3-branes in the background of $\widetilde{\textrm{O3}}^{+}$ ending on a half NS5-brane 
and a single half D3-brane on the other side (see Figure \ref{fig_usp2N'NeuNahm}).
It realizes the modified Neumann boundary condition with 
a coupling between the $USp(2N)'$ gauge field and a single half-hypermultiplet transforming in the fundamental representation 
that cancels the anomaly induced from the half-integral Chern-Simons term corresponding to the 4d theta angle \cite{Gaiotto:2008ak}. 

The half-index takes the form
\begin{align}
\label{hsp2N'_Neu}
\mathbb{II}_{\mathcal{N}+\textrm{hyp}}^{\textrm{4d $USp(2N)'$}}(t;q)
&=\frac{1}{2^N N!}
\frac{(q)_{\infty}^N}{(q^{\frac12}t^{-2};q)_{\infty}^N}
\oint \prod_{i=1}^N \frac{ds_i}{2\pi is_i}
\frac{(s_i^{\pm 2};q)_{\infty}}{(q^{\frac12}t^{-2}s_i^{\pm 2};q)_{\infty}}
\nonumber\\
&\times 
\prod_{1\le i<j\le N}
\frac{(s_i^{\pm} s_j^{\mp};q)_{\infty} (s_{i}^{\pm}s_{j}^{\pm};q)_{\infty}}
{(q^{\frac12}t^{-2}s_i^{\pm}s_j^{\mp};q)_{\infty} (q^{\frac12}t^{-2}s_i^{\pm}s_j^{\pm};q)_{\infty}}
\prod_{i=1}^{N}\frac{(q^{\frac34}t^{-1}s_i^{\pm};q)_{\infty}}{(q^{\frac14}ts_i^{\pm};q)_{\infty}}. 
\end{align}

S-duality gives rise to the dual configuration with the half NS5-brane being replaced with a half D5$'$-brane (see Figure \ref{fig_usp2N'NeuNahm}). 
It realizes the regular Nahm pole boundary condition for $USp(2N)'$ gauge theory as 
the S-dual of the modified Neumann boundary condition for the same theory \cite{Gaiotto:2008ak}. 

Applying the Higgsing manipulation to the Dirichlet half-index for the $USp(2N)'$ gauge theory will lead to the same result as that for the $USp(2N)$ gauge theory. 
In fact, we find that the half-index (\ref{hsp2N'_Neu}) for the modified Neumann boundary condition exactly coincides with the half-index (\ref{h_C_Nahm}) for the Nahm pole boundary condition. This indicates the duality  
\begin{align}
&\textrm{Neumann b.c. for $\mathcal{N}=4$ $USp(2N)'$ SYM $+$ a fund. half-hyper}
\nonumber\\
&\Leftrightarrow
\textrm{Nahm pole b.c. for $\mathcal{N}=4$ $USp(2N)'$ SYM}. 
\end{align}

\subsection{$SO(2N+1)|USp(2M)'$}
Now consider the half-BPS interfaces in $\mathcal{N}=4$ gauge theories of orthogonal and symplectic gauge groups 
which can be constructed in Type IIB string theory. 

A basic example of the NS5-type interface contains 
a half NS5-brane at $x^6=0$ and $N$ D3-branes in both sides, 
i.e. a stack of $N$ semi-infinite D3-branes in $x^6<0$ and the other stack of $N$ semi-infinite D3-branes in $x^6>0$. 
Since the half NS5-brane supports the $H$-field, the NS flux jumps across the NS5-brane 
and therefore different types of O3-planes appear across the NS5-brane. 
Accordingly, one should find a pair of orthogonal and symplectic gauge groups across the NS5-type interface. 

\begin{figure}
\centering
\scalebox{1.2}{
\begin{tikzpicture}
\draw[thick,solid,black] (-3.5,1) -- (-3.5,3) node[above] {\textrm{1/2 NS5}};
\draw[thick,dashed,black] (3.5,1) -- (3.5,3) node[above] {\textrm{1/2 D5$'$}};
\draw[thick,dotted,blue] (-5.5,2) node[below] {$\widetilde{\textrm{O3}}^-$} -- (-3.5,2);
\draw[thick,dotted,red] (-3.5,2) -- (-1.5,2) node[below] {$\widetilde{\textrm{O3}}^+$};
\draw[thick,dotted,red] (1.5,2) node[below] {$\textrm{O3}^+$} -- (3.5,2);
\draw[thick,dotted,red] (3.5,2) -- (5.5,2) node[below] {$\textrm{O3}^+$};
\draw[thick,solid,yellow] (-5.5,2.2) node[above] {\textcolor{black}{\textrm{D3}}} -- (-3.5,2.2);
\draw[thick,solid,yellow] (-5.5,2.1)  -- (-3.5,2.1);
\draw[thick,solid,yellow] (-3.5,2.3)  -- (-1.5,2.3) node[above] {\textcolor{black}{\textrm{D3}}};
\draw[thick,solid,yellow] (-3.5,2.2)  -- (-1.5,2.2);
\draw[thick,solid,yellow] (-3.5,2.1)  -- (-1.5,2.1);
\draw[thick,solid,yellow] (1.5,2.2) node[above] {\textcolor{black}{\textrm{D3}}} -- (3.5,2.2);
\draw[thick,solid,yellow] (1.5,2.1)  -- (3.5,2.1);
\draw[thick,solid,yellow] (3.5,2.3)  -- (5.5,2.3) node[above] {\textcolor{black}{\textrm{D3}}};
\draw[thick,solid,yellow] (3.5,2.2)  -- (5.5,2.2);
\draw[thick,solid,yellow] (3.5,2.1)  -- (5.5,2.1);
\end{tikzpicture}
}
\caption{\textrm{
The S-dual brane configurations for the NS5-type $SO(2N+1)|USp(2M)'$ interface and D5$'$-type $USp(2N)|USp(2M)'$ interface.  
}}
\label{fig_so2N+1|usp2M'}
\end{figure}
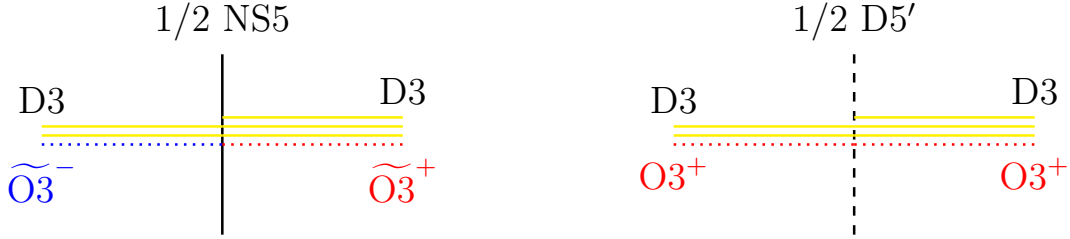

We begin with the NS5-type interface with both sides of $N$ D3-branes 
as well as $\widetilde{\textrm{O3}}^-$ in $x^6<0$ and $\widetilde{\textrm{O3}}^+$ in $x^6>0$ (see Figure \ref{fig_so2N+1|usp2M'}). 
The configuration realizes $\mathcal{N}=4$ $SO(2N+1)$ gauge theory for $x^6<0$ 
and $\mathcal{N}=4$ $USp(2N)'$ gauge theory for $x^6>0$. 
In addition, there exist a 3d $\mathcal{N}=4$ hypermultiplet transforming in the bifundamental representation 
under the $SO(2N+1)$ $\times$ $USp(2N)'$ gauge group corresponding to the fluctuation modes of open strings ending on D3-branes across the NS5-brane. 

The half-index of the NS5-type interface between $SO(2N+1)$ and $USp(2N)'$ gauge theories is given by
\begin{align}
\label{h_so2N+1|usp2N_N}
&
\mathbb{II}_{\mathcal{N}}^{\textrm{4d $SO(2N+1)|USp(2N)'$}}(t;q)
\nonumber\\
&
=\frac{1}{2^N N!}
\frac{(q)_{\infty}^N}{(q^{\frac12}t^{-2};q)_{\infty}^N}
\oint \prod_{i=1}^N \frac{ds_i^{(1)}}{2\pi is_i^{(1)}}
\frac{(s_i^{(1)\pm};q)_{\infty}}{(q^{\frac12}t^{-2}s_i^{(1)\pm};q)_{\infty}}
\nonumber\\
&\times 
\prod_{i<j}
\frac{(s_i^{(1)\pm}s_j^{(1)\mp};q)_{\infty}(s_i^{(1)\pm}s_j^{(1)\pm};q)_{\infty}}
{(q^{\frac12}t^{-2}s_i^{(1)\pm}s_j^{(1)\mp};q)_{\infty}(q^{\frac12}t^{-2}s_i^{(1)\pm}s_j^{(1)\pm};q)_{\infty}}
\nonumber\\
&\times 
\frac{1}{2^N N!}
\frac{(q)_{\infty}^N}{(q^{\frac12}t^{-2};q)_{\infty}^N}
\oint \prod_{i=1}^N \frac{ds_i^{(2)}}{2\pi is_i^{(2)}}
\frac{(s_i^{(2)\pm 2};q)_{\infty}}{(q^{\frac12}t^{-2}s_i^{(2)\pm 2};q)_{\infty}}
\nonumber\\
&\times 
\prod_{i< j}
\frac{(s_i^{(2)\pm}s_j^{(2)\mp};q)_{\infty}(s_i^{(2)\pm}s_j^{(2)\pm};q)_{\infty}}
{(q^{\frac12}t^{-2}s_i^{(2)\pm}s_j^{(2)\mp};q)_{\infty}(q^{\frac12}t^{-2}s_i^{(2)\pm}s_j^{(2)\pm};q)_{\infty}}
\nonumber\\
&\times 
\prod_{i,j=1}^N
\frac{(q^{\frac34}t^{-1}s_i^{(1)\pm}s_j^{(2)\mp};q)_{\infty}(q^{\frac34}t^{-1}s_i^{(1)\pm}s_j^{(2)\pm};q)_{\infty}(q^{\frac34}t^{-1}s_j^{(2)\pm};q)_{\infty}}
{(q^{\frac14}ts_i^{(1)\pm}s_{j}^{(2)\mp};q)_{\infty}(q^{\frac14}ts_i^{(1)\pm}s_{j}^{(2)\pm};q)_{\infty}(q^{\frac14}ts_{j}^{(2)\pm};q)_{\infty}}, 
\end{align}
where the first two lines of R.H.S. in (\ref{h_so2N+1|usp2N_N}) are the Neumann half-index of $SO(2N+1)$ gauge theory, 
the next two the Neumann half-index of $USp(2N)'$ gauge theory 
and the last line is a contribution from the bifundamental hypermultiplet. 

Upon the $S$ operation, we find the configuration which contains 
a half D5$'$-brane at $x^2=0$, $O3^+$ in $x^2<0$, $\widetilde{\textrm{O3}}^+$ in $x^2>0$ and both sides of $N$ D3-branes (see Figure \ref{fig_so2N+1|usp2M'}). 
It gives rise to the S-dual D5$'$-type interface between $USp(2N)$ and $USp(2N)'$ gauge theories \cite{Gaiotto:2008ak}. 
At the interface the $USp(2N)\times USp(2N)$ gauge group breaks down to its subgroup $USp(2N)$. 
There should exist a coupling of the 4d $USp(2N)$ gauge fields to the twisted half-hypermultiplet transforming in the fundamental representation 
which correspond to the fluctuation of open strings between D3- and D5$'$-branes. 
In other words, the D5$'$-type interface can be described by 
a whole 4d $\mathcal{N}=4$ $USp(2N)$ gauge theory with a coupling to a 3d $\mathcal{N}=4$ twisted half-hypermultiplet in the fundamental representation at the interface. 

The half-index of the $USp(2N)|USp(2N)'$ interface of the D5$'$-type is given by
\begin{align}
\label{h_usp2N|usp2N_D}
&
\mathbb{II}_{\mathcal{D}'}^{\textrm{4d $USp(2N)|USp(2N)'$}}(t;q)
\nonumber\\
&=\frac{1}{2^N N!} \frac{(q)_{\infty}^{2N}}{(q^{\frac12}t^{\pm};q)_{\infty}^N}
\oint \prod_{i=1}^{N} \frac{ds_i}{2\pi is_i}
\frac{(s_i^{\pm 2};q)_{\infty}(qs_i^{\pm 2};q)_{\infty}}
{(q^{\frac12}t^{2}s_i^{\pm 2};q)_{\infty}(q^{\frac12}t^{-2}s_i^{\pm 2};q)_{\infty}}
\nonumber\\
&\times \prod_{i<j}
\frac{(s_i^{\pm}s_j^{\mp};q)_{\infty}(s_i^{\pm}s_j^{\pm};q)_{\infty}(qs_i^{\pm}s_j^{\mp};q)_{\infty}(qs_i^{\pm}s_j^{\pm};q)_{\infty}}
{(q^{\frac12}t^{2}s_i^{\pm}s_j^{\mp};q)_{\infty}(q^{\frac12}t^{2}s_i^{\pm}s_j^{\pm};q)_{\infty}
(q^{\frac12}t^{-2}s_i^{\pm}s_j^{\mp};q)_{\infty}(q^{\frac12}t^{-2}s_i^{\pm}s_j^{\pm};q)_{\infty}}
\nonumber\\
&\times 
\prod_{i=1}^{N}
\frac{(q^{\frac34}t s_i^{\pm};q)_{\infty}}{(q^{\frac14}t^{-1} s_i^{\pm};q)_{\infty}}. 
\end{align}
The first two lines of R.H.S. are the full index of $\mathcal{N}=4$ $USp(2N)$ gauge theory 
and the last line is the contribution from the half-hyper living at the interface. 

In fact, we find that the half-indices (\ref{h_so2N+1|usp2N_N}) and (\ref{h_usp2N|usp2N_D}) beautifully agree with each other! 
The equality between (\ref{h_so2N+1|usp2N_N}) and (\ref{h_usp2N|usp2N_D}) indicates the following duality of the interfaces: 
\begin{align}
\label{dual_so2n+1|usp2n'}
&\textrm{$SO(2N+1)|USp(2N)'$ NS5-type interface}
\nonumber\\
&\Leftrightarrow
\textrm{$USp(2N)|USp(2N)'$ D5$'$-type interface}. 
\end{align}

More generally, when $N$ D3-branes in $x^6<0$ end on a half NS5-brane from one side 
and $M$ D3-branes in $x^6>0$ end on it from the other side with $M\neq N$, 
we get the NS5-type interface between $SO(2N+1)$ and $USp(2M)'$ gauge theories (see Figure \ref{fig_so2N+1|usp2M'}).  
The domain wall supports a hypermultiplet transforming in the bifundamental representation of the $SO(2N+1)$ $\times$ $USp(2M)'$ gauge group. 

The half-index is given by
\begin{align}
\label{h_so2N+1|usp2M_N}
&
\mathbb{II}_{\mathcal{N}}^{\textrm{4d $SO(2N+1)|USp(2M)'$}}(t;q)
\nonumber\\
&
=\frac{1}{2^N N!}
\frac{(q)_{\infty}^N}{(q^{\frac12}t^{-2};q)_{\infty}^N}
\oint \prod_{i=1}^N \frac{ds_i^{(1)}}{2\pi is_i^{(1)}}
\frac{(s_i^{(1)\pm};q)_{\infty}}{(q^{\frac12}t^{-2}s_i^{(1)\pm};q)_{\infty}}
\nonumber\\
&\times 
\prod_{i<j}
\frac{(s_i^{(1)\pm}s_j^{(1)\mp};q)_{\infty}(s_i^{(1)\pm}s_j^{(1)\pm};q)_{\infty}}
{(q^{\frac12}t^{-2}s_i^{(1)\pm}s_j^{(1)\mp};q)_{\infty}(q^{\frac12}t^{-2}s_i^{(1)\pm}s_j^{(1)\pm};q)_{\infty}}
\nonumber\\
&\times 
\frac{1}{2^M M!}
\frac{(q)_{\infty}^M}{(q^{\frac12}t^{-2};q)_{\infty}^M}
\oint \prod_{i=1}^M \frac{ds_i^{(2)}}{2\pi is_i^{(2)}}
\frac{(s_i^{(2)\pm 2};q)_{\infty}}{(q^{\frac12}t^{-2}s_i^{(2)\pm 2};q)_{\infty}}
\nonumber\\
&\times 
\prod_{i< j}
\frac{(s_i^{(2)\pm}s_j^{(2)\mp};q)_{\infty}(s_i^{(2)\pm}s_j^{(2)\pm};q)_{\infty}}
{(q^{\frac12}t^{-2}s_i^{(2)\pm}s_j^{(2)\mp};q)_{\infty}(q^{\frac12}t^{-2}s_i^{(2)\pm}s_j^{(2)\pm};q)_{\infty}}
\nonumber\\
&\times 
\prod_{i=1}^N\prod_{j=1}^M
\frac{(q^{\frac34}t^{-1}s_i^{(1)\pm}s_j^{(2)\mp};q)_{\infty}(q^{\frac34}t^{-1}s_i^{(1)\pm}s_j^{(2)\pm};q)_{\infty}(q^{\frac34}t^{-1}s_j^{(2)\pm};q)_{\infty}}
{(q^{\frac14}ts_i^{(1)\pm}s_{j}^{(2)\mp};q)_{\infty}(q^{\frac14}ts_i^{(1)\pm}s_{j}^{(2)\pm};q)_{\infty}(q^{\frac14}ts_{j}^{(2)\pm};q)_{\infty}}, 
\end{align}
where the first two lines and the next ones of R.H.S. describe the contributions 
from the Neumann half-indices for $SO(2N+1)$ and $USp(2M)'$ gauge theories. 
The last line describes the contributions from the bifundamental hyper. 

Similarly, the S-dual is the D5$'$-type $USp(N)|USp(M)$ domain wall (see Figure \ref{fig_so2N+1|usp2M'}). 
The configuration has a half D5$'$-brane on which $N$ D3-branes in $x^2<0$ and $M$ D3-branes in $x^2>0$ terminate 
as well as O3$^+$ in $x^2<0$ and $\widetilde{\textrm{O3}}^+$ in $x^2>0$. 
Unlike the case with $N=M$, the gauge group jumps from $USp(\max (N,M))$ to $USp(\min (N,M))$ so that there is a gauge group $USp(\min (N,M))$ 
together with a Nahm pole of rank $|N-M|$. 
While there is no twisted hypermultiplet, the broken $USp(|N-M|)$ part survives as a global symmetry at the domain wall. 

The half-index of the dual D5$'$-type domain wall takes the form 
\begin{align}
\label{h_usp2N|usp2M_D}
&
\mathbb{II}_{\mathcal{D}'}^{\textrm{4d $USp(2N)|USp(2M)'$}}(t;q)
\nonumber\\
&=\frac{1}{2^{\min (N,M)} (\min (N,M))!} \frac{(q)_{\infty}^{2 \min (N,M)}}{(q^{\frac12}t^{\pm};q)_{\infty}^{\min (N,M)}}
\oint \prod_{i=1}^{\min (N,M)} \frac{ds_i}{2\pi is_i}
\frac{(s_i^{\pm 2};q)_{\infty}(qs_i^{\pm 2};q)_{\infty}}
{(q^{\frac12}t^{2}s_i^{\pm 2};q)_{\infty}(q^{\frac12}t^{-2}s_i^{\pm 2};q)_{\infty}}
\nonumber\\
&\times \prod_{i<j}
\frac{(s_i^{\pm}s_j^{\mp};q)_{\infty}(s_i^{\pm}s_j^{\pm};q)_{\infty}(qs_i^{\pm}s_j^{\mp};q)_{\infty}(qs_i^{\pm}s_j^{\pm};q)_{\infty}}
{(q^{\frac12}t^{2}s_i^{\pm}s_j^{\mp};q)_{\infty}(q^{\frac12}t^{2}s_i^{\pm}s_j^{\pm};q)_{\infty}
(q^{\frac12}t^{-2}s_i^{\pm}s_j^{\mp};q)_{\infty}(q^{\frac12}t^{-2}s_i^{\pm}s_j^{\pm};q)_{\infty}}
\nonumber\\
&\times 
\prod_{k=1}^{|N-M|}
\frac{(q^{k+\frac12}t^{-4k+2};q)_{\infty}}
{(q^{k}t^{-4k};q)_{\infty}}
\prod_{i=1}^{\min (N,M)}
\frac{(q^{\frac34+\frac{|N-M|}{2}}t^{1-2|N-M|} s_i^{\pm};q)_{\infty}}
{(q^{\frac14+\frac{|N-M|}{2}}t^{-1-2|N-M|} s_i^{\pm};q)_{\infty}}. 
\end{align}
Here the first two lines of R.H.S. are the full index of $\mathcal{N}=4$ $USp(\min (N,M))$ gauge theory 
and the last line involves the half-index for the regular Nahm pole of $USp(|N-M|)$ gauge theory 
and the contributions from the broken $USp(|N-M|)$ part remaining at the domain wall. 

Again, we find that the half-index (\ref{h_so2N+1|usp2M_N}) exactly coincides with the half-index (\ref{h_usp2N|usp2M_D})! 
The matching of the half-indices supports the duality 
\begin{align}
\label{dual_so2n+1|usp2m'}
&\textrm{$SO(2N+1)|USp(2M)'$ NS5-type interface}
\nonumber\\
&\Leftrightarrow
\textrm{$USp(2N)|USp(2M)'$ D5$'$-type interface}, 
\end{align}
which generalizes the duality (\ref{dual_so2n+1|usp2n'}). 

In the Coulomb limit and Higgs limit we obtain
\begin{align}
\label{h_usp2N|usp2M_C}
\mathbb{II}_{\mathcal{N}}^{\textrm{4d $SO(2N+1)|USp(2M)'$}(C)}(\mathfrak{q})
&=\mathbb{II}_{\mathcal{D}'}^{\textrm{4d $USp(2N)|USp(2M)'$}(C)}(\mathfrak{q})
\nonumber\\
&=\frac{1}{(\mathfrak{q}^4;\mathfrak{q}^4)_{N}(\mathfrak{q}^4;\mathfrak{q}^4)_{M}}, \\
\label{h_usp2N|usp2M_H}
\mathbb{II}_{\mathcal{N}}^{\textrm{4d $SO(2N+1)|USp(2M)'$}(H)}(\mathfrak{q})
&=\mathbb{II}_{\mathcal{D}'}^{\textrm{4d $USp(2N)|USp(2M)'$}(H)}(\mathfrak{q})
\nonumber\\
&=\frac{1}{(\mathfrak{q}^4;\mathfrak{q}^4)_{\min (N,M)}}. 
\end{align}
We see that the Coulomb index (\ref{h_usp2N|usp2M_C}) is simply factorized into 
a pair of the $SO(2N+1)$ half-BPS index and the $USp(2M)'$ half-BPS index, 
each of which is given by (\ref{1/2BPSindex_BC}). 
On the other hand, the Higgs index (\ref{h_usp2N|usp2M_H}) 
can be viewed as the twisted version of the ``Coulomb limit'' of the D5$'$-type interface. 
It is equal to the half-BPS index of the theory with lower rank preserved at the D5$'$-type interface. 

\subsection{$O(2N)|USp(2M)$}
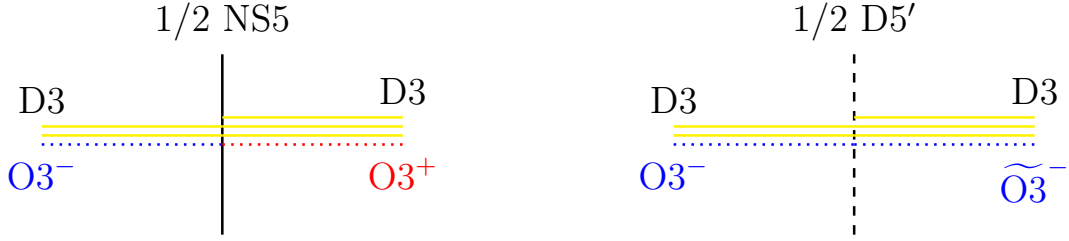
\begin{figure}
\centering
\scalebox{1.2}{
\begin{tikzpicture}
\draw[thick,solid,black] (-3.5,1) -- (-3.5,3) node[above] {\textrm{1/2 NS5}};
\draw[thick,dashed,black] (3.5,1) -- (3.5,3) node[above] {\textrm{1/2 D5$'$}};
\draw[thick,dotted,blue] (-5.5,2) node[below] {$\textrm{O3}^-$} -- (-3.5,2);
\draw[thick,dotted,red] (-3.5,2) -- (-1.5,2) node[below] {$\textrm{O3}^+$};
\draw[thick,dotted,blue] (1.5,2) node[below] {$\textrm{O3}^-$} -- (3.5,2);
\draw[thick,dotted,blue] (3.5,2) -- (5.5,2) node[below] {$\widetilde{\textrm{O3}}^-$};
\draw[thick,solid,yellow] (-5.5,2.2) node[above] {\textcolor{black}{\textrm{D3}}} -- (-3.5,2.2);
\draw[thick,solid,yellow] (-5.5,2.1)  -- (-3.5,2.1);
\draw[thick,solid,yellow] (-3.5,2.3)  -- (-1.5,2.3) node[above] {\textcolor{black}{\textrm{D3}}};
\draw[thick,solid,yellow] (-3.5,2.2)  -- (-1.5,2.2);
\draw[thick,solid,yellow] (-3.5,2.1)  -- (-1.5,2.1);
\draw[thick,solid,yellow] (1.5,2.2) node[above] {\textcolor{black}{\textrm{D3}}} -- (3.5,2.2);
\draw[thick,solid,yellow] (1.5,2.1)  -- (3.5,2.1);
\draw[thick,solid,yellow] (3.5,2.3)  -- (5.5,2.3) node[above] {\textcolor{black}{\textrm{D3}}};
\draw[thick,solid,yellow] (3.5,2.2)  -- (5.5,2.2);
\draw[thick,solid,yellow] (3.5,2.1)  -- (5.5,2.1);
\end{tikzpicture}
}
\caption{\textrm{
The S-dual brane configurations for the NS5-type $O(2N)|USp(2M)$ interface and D5$'$-type $O(2N)|SO(2M+1)'$ interface.  
}}
\label{fig_o2N|usp2M}
\end{figure}

Next consider the NS5-type interface 
with both sides of $N$ D3-branes as well as O3$^-$ in $x^6<0$ and O3$^+$ in $x^6>0$ (see Figure \ref{fig_o2N|usp2M}). 
In this case, we have $\mathcal{N}=4$ $O(2N)$ gauge theory in $x^6<0$ 
and $\mathcal{N}=4$ $USp(2N)$ gauge theory in $x^6>0$. 
At the interface there also exists a 3d $\mathcal{N}=4$ bifundamental hypermultiplet. 

Let us first consider the NS5-type interface between $SO(2N)$ and $USp(2N)$ gauge theories. 
The half-index is evaluated by the following matrix integral
\begin{align}
\label{h_so2N|usp2N_N}
&
\mathbb{II}_{\mathcal{N}}^{\textrm{4d $SO(2N)|USp(2N)$}}(t;q)
\nonumber\\
&
=\frac{1}{2^{N-1} N!}
\frac{(q)_{\infty}^N}{(q^{\frac12}t^{-2};q)_{\infty}^N}
\oint \prod_{i=1}^N \frac{ds_i^{(1)}}{2\pi is_i^{(1)}}
\nonumber\\
&\times 
\prod_{i<j}
\frac{(s_i^{(1)\pm}s_j^{(1)\mp};q)_{\infty}(s_i^{(1)\pm}s_j^{(1)\pm};q)_{\infty}}
{(q^{\frac12}t^{-2}s_i^{(1)\pm}s_j^{(1)\mp};q)_{\infty}(q^{\frac12}t^{-2}s_i^{(1)\pm}s_j^{(1)\pm};q)_{\infty}}
\nonumber\\
&\times 
\frac{1}{2^N N!}
\frac{(q)_{\infty}^N}{(q^{\frac12}t^{-2};q)_{\infty}^N}
\oint \prod_{i=1}^N \frac{ds_i^{(2)}}{2\pi is_i^{(2)}}
\frac{(s_i^{(2)\pm 2};q)_{\infty}}{(q^{\frac12}t^{-2}s_i^{(2)\pm 2};q)_{\infty}}
\nonumber\\
&\times 
\prod_{i< j}
\frac{(s_i^{(2)\pm}s_j^{(2)\mp};q)_{\infty}(s_i^{(2)\pm}s_j^{(2)\pm};q)_{\infty}}
{(q^{\frac12}t^{-2}s_i^{(2)\pm}s_j^{(2)\mp};q)_{\infty}(q^{\frac12}t^{-2}s_i^{(2)\pm}s_j^{(2)\pm};q)_{\infty}}
\nonumber\\
&\times 
\prod_{i,j=1}^N
\frac{(q^{\frac34}t^{-1}s_i^{(1)\pm}s_j^{(2)\mp};q)_{\infty}(q^{\frac34}t^{-1}s_i^{(1)\pm}s_j^{(2)\pm};q)_{\infty}}
{(q^{\frac14}ts_i^{(1)\pm}s_{j}^{(2)\mp};q)_{\infty}(q^{\frac14}ts_i^{(1)\pm}s_{j}^{(2)\pm};q)_{\infty}}. 
\end{align}
Here the first two lines of R.H.S. in (\ref{h_so2N|usp2N_N}) are the Neumann half-index of $SO(2N)$ gauge theory, 
the next two the Neumann half-index of $USp(2N)$ gauge theory. 
The last line is the index of the bifundamental hypermultiplet. 
Similarly, for the other disconnected component of $O(2N)$ gauge group 
we have the half-index 
\begin{align}
\label{h_so2N|usp2N_N2}
&
\mathbb{II}_{\mathcal{N}}^{\textrm{4d $SO(2N)^{-}|USp(2N)$}}(t;q)
\nonumber\\
&=\frac{1}{2^{N-1}(N-1)!}\frac{(q)_{\infty}^{N-1} (-q;q)_{\infty}}
{(q^{\frac12}t^{-2};q)_{\infty}^{N-1} (-q^{\frac12}t^{-2};q)_{\infty}}
\oint \prod_{i=1}^{N-1}
\frac{(s_i^{(1)\pm};q)_{\infty}(-s_i^{(1)\pm};q)_{\infty}}
{(q^{\frac12}t^{-2}s_i^{(1)\pm};q)_{\infty}(-q^{\frac12}t^{-2}s_i^{(1)\pm};q)_{\infty}}
\nonumber\\
&\times 
\prod_{i<j}
\frac{(s_i^{(1)\pm}s_j^{(1)\mp};q)_{\infty}(s_i^{(1)\pm}s_j^{(1)\pm};q)_{\infty}}
{(q^{\frac12}t^{-2}s_i^{(1)\pm}s_j^{(1)\mp};q)_{\infty}(q^{\frac12}t^{-2}s_i^{(1)\pm}s_j^{(1)\pm};q)_{\infty}}
\nonumber\\
&\times 
\frac{1}{2^N N!}
\frac{(q)_{\infty}^N}{(q^{\frac12}t^{-2};q)_{\infty}^N}
\oint \prod_{i=1}^N \frac{ds_i^{(2)}}{2\pi is_i^{(2)}}
\frac{(s_i^{(2)\pm 2};q)_{\infty}}{(q^{\frac12}t^{-2}s_i^{(2)\pm 2};q)_{\infty}}
\nonumber\\
&\times 
\prod_{i< j}
\frac{(s_i^{(2)\pm}s_j^{(2)\mp};q)_{\infty}(s_i^{(2)\pm}s_j^{(2)\pm};q)_{\infty}}
{(q^{\frac12}t^{-2}s_i^{(2)\pm}s_j^{(2)\mp};q)_{\infty}(q^{\frac12}t^{-2}s_i^{(2)\pm}s_j^{(2)\pm};q)_{\infty}}
\nonumber\\
&\times 
\prod_{i=1}^{N-1}
\prod_{j=1}^{N}
\frac{
(q^{\frac34}t^{-1}s_i^{(1)\pm}s_j^{(2)\mp};q)_{\infty}(q^{\frac34}t^{-1}s_i^{(1)\pm}s_j^{(2)\pm};q)_{\infty}
(q^{\frac34}t^{-1}s_j^{(2)\pm};q)_{\infty}(-q^{\frac34}t^{-1}s_j^{(2)\pm};q)_{\infty}
}
{
(q^{\frac14}ts_i^{(1)\pm}s_{j}^{(2)\mp};q)_{\infty}(q^{\frac14}ts_i^{(1)\pm}s_{j}^{(2)\pm};q)_{\infty}
(q^{\frac14}ts_j^{(2)\pm};q)_{\infty}(-q^{\frac14}ts_j^{(2)\pm};q)_{\infty}
}. 
\end{align}
After gauging the discrete $\mathbb{Z}_2$ global symmetry of $SO(2N)$ gauge theory, 
we get the half-index of the NS5-type interface between $O(2N)$ and $USp(2N)$ gauge theories
\begin{align}
 \label{h_o2N|usp2N_N}
\mathbb{II}_{\mathcal{N}}^{\textrm{4d $O(2N)^{\pm}|USp(2N)$}}(t;q)
&=\frac12\left[
\mathbb{II}_{\mathcal{N}}^{\textrm{4d $SO(2N)|USp(2N)$}}(t;q)
\pm 
\mathbb{II}_{\mathcal{N}}^{\textrm{4d $SO(2N)^{-}|USp(2N)$}}(t;q)
\right], 
\end{align}
where $+$ (resp. $-$) corresponds to the $O(2N)$ theory that contains the $\mathbb{Z}_2$ even (resp. odd) BPS local operators. 

In the S-dual configuration there exists a half D5$'$-brane at $x^2=0$, 
O3$^-$ in $x^2<0$, $\widetilde{\textrm{O3}}^-$ in $x^2>0$ and both sides of $N$ D3-branes (see Figure \ref{fig_o2N|usp2M}). 
This is identified with the S-dual D5$'$-type interface between $O(2N)$ and $SO(2N+1)$ gauge theories \cite{Gaiotto:2008ak}. 
This can be realized as the orthogonal type orientifold projection of the D5$'$-type interface between $U(N)$ and $U(N)$ gauge theories. 

The half-index of the $SO(2N)|SO(2N+1)$ interface of the D5$'$-type takes the form 
\begin{align}
\label{h_so2N|so2N+1_D}
&
\mathbb{II}_{\mathcal{D}'}^{\textrm{4d $SO(2N)|SO(2N+1)$}}(t;q)
\nonumber\\
&=\frac{1}{2^{N-1} N!} \frac{(q)_{\infty}^{2N}}{(q^{\frac12}t^{\pm 2};q)_{\infty}^N}
\oint \prod_{i=1}^{N} \frac{ds_i}{2\pi is_i}
\nonumber\\
&\times \prod_{i< j}
\frac{(s_i^{\pm}s_j^{\mp};q)_{\infty}(s_i^{\pm}s_j^{\pm};q)_{\infty}(qs_i^{\pm}s_j^{\mp};q)_{\infty}(qs_i^{\pm}s_j^{\pm};q)_{\infty}}
{(q^{\frac12}t^{2}s_i^{\pm}s_j^{\mp};q)_{\infty}(q^{\frac12}t^{2}s_i^{\pm}s_j^{\pm};q)_{\infty}
(q^{\frac12}t^{-2}s_i^{\pm}s_j^{\mp};q)_{\infty}(q^{\frac12}t^{-2}s_i^{\pm}s_j^{\pm};q)_{\infty}}
\nonumber\\
&\times 
\prod_{i=1}^{N}
\frac{(q s_i^{\pm};q)_{\infty}}{(q^{\frac12}t^{-2} s_i^{\pm};q)_{\infty}}. 
\end{align}
The integrand contains the expected contributions from the 4d $SO(2N+1)$ gauginos and scalar fields for $x^2>0$ which are not part of the $SO(2N)$ gauge theory 
as well as the 4d $SO(2N)$ gauge theory fields contributions.

In fact, we have checked that the half-indices (\ref{h_so2N|usp2N_N}) and (\ref{h_so2N|so2N+1_D}) precisely coincide with each other! 
The equality between (\ref{h_so2N|usp2N_N}) and (\ref{h_so2N|so2N+1_D}) should imply the following duality of the interfaces: 
\begin{align}
\label{dual_so2n|usp2n}
&\textrm{$SO(2N)|USp(2N)$ NS5-type interface}
\nonumber\\
&\Leftrightarrow
\textrm{$SO(2N)|SO(2N+1)$ D5$'$-type interface}. 
\end{align}

Furthermore, we can evaluate the half-index of the $SO(2N)^{-}|SO(2N+1)$ interface of the D5$'$-type 
associated with the disconnected part of $O(2N)$ as
\begin{align}
\label{h_so2N|so2N+1_D2}
&
\mathbb{II}_{\mathcal{D}'}^{\textrm{4d $SO(2N)^{-}|SO(2N+1)$}}(t;q)
\nonumber\\
&=\frac{1}{2^{N-1} (N-1)!} \frac{(q)_{\infty}^{2N-2}(-q;q)_{\infty}^2}
{(q^{\frac12}t^{\pm2 };q)_{\infty}^{N-1}(-q^{\frac12}t^{\pm2};q)_{\infty}}
\oint \prod_{i=1}^{N-1} \frac{ds_i}{2\pi is_i}
\nonumber\\
&\times 
\prod_{i=1}^{N-1}
\frac{(s_i^{\pm};q)_{\infty} (-s_i^{\pm};q)_{\infty} (qs_i^{\pm};q)_{\infty} (-qs_i^{\pm};q)_{\infty}}
{(q^{\frac12}t^2s_i^{\pm};q)_{\infty} (-q^{\frac12}t^2s_i^{\pm};q)_{\infty} (q^{\frac12}t^{-2}s_i^{\pm};q)_{\infty} (-q^{\frac	12}t^{-2}s_i^{\pm};q)_{\infty}}
\nonumber\\
&\times \prod_{i<j}
\frac{(s_i^{\pm}s_j^{\mp};q)_{\infty}(s_i^{\pm}s_j^{\pm};q)_{\infty}(qs_i^{\pm}s_j^{\mp};q)_{\infty}(qs_i^{\pm}s_j^{\pm};q)_{\infty}}
{(q^{\frac12}t^{2}s_i^{\pm}s_j^{\mp};q)_{\infty}(q^{\frac12}t^{2}s_i^{\pm}s_j^{\pm};q)_{\infty}
(q^{\frac12}t^{-2}s_i^{\pm}s_j^{\mp};q)_{\infty}(q^{\frac12}t^{-2}s_i^{\pm}s_j^{\pm};q)_{\infty}}
\nonumber\\
&\times 
\prod_{i=1}^{N-1}
\frac{(qs_i^{\pm};q)_{\infty}}{(q^{\frac12}t^{-2}s_i^{\pm};q)_{\infty}}
\frac{(\pm q;q)_{\infty}}{(\pm q^{\frac12}t^{-2};q)_{\infty}}. 
\end{align}

Again the half-index of the D5$'$-type interface between 
$O(2N)$ and $SO(2N+1)$ gauge theories is obtained by gauging the $\mathbb{Z}_2$ global symmetry as
 \begin{align}
 \label{h_o2N|so2N+1_D}
\mathbb{II}_{\mathcal{D}'}^{\textrm{4d $O(2N)^{\pm}|SO(2N+1)$}}(t;q)
&=\frac12\left[
\mathbb{II}_{\mathcal{D}'}^{\textrm{4d $SO(2N)|SO(2N+1)$}}(t;q)
\pm 
\mathbb{II}_{\mathcal{D}'}^{\textrm{4d $SO(2N)^{-}|SO(2N+1)$}}(t;q)
\right]. 
\end{align}
From the formal power series expansion of the indices, 
we find that the half-indices (\ref{h_so2N|usp2N_N2}) and (\ref{h_so2N|so2N+1_D2}) give rise to the same result! 
So the half-indices (\ref{h_o2N|usp2N_N}) and (\ref{h_o2N|so2N+1_D}) are equivalent. 
This implies the duality of the interfaces involving the disconnected $O(2N)$ gauge group
\begin{align}
\label{dual_o2n|usp2n}
&\textrm{$O(2N)|USp(2N)$ NS5-type interface}
\nonumber\\
&\Leftrightarrow
\textrm{$O(2N)|SO(2N+1)$ D5$'$-type interface}. 
\end{align}

We can also consider the setups with unequal numbers of D3-branes on the two sides of the $5$-branes. 
When we have $N$ D3-branes and O3$^-$ in $x^6<0$, 
$M$ D3-branes and O3$^+$ in $x^6>0$ 
and a half NS5-brane at $x^6=0$, one obtains the NS5-type interface between $O(2N)$ and $USp(2M)$ gauge theories 
that has a bifundamental hypermultiplet of the $O(2N)\times USp(2M)$ gauge group (see Figure \ref{fig_o2N|usp2M}). 

First consider the case with $SO(2N)$ gauge theory. 
The half-index of the NS5-type $SO(2N)|USp(M)$ interface is given by
\begin{align}
\label{h_so2N|usp2M_N}
&
\mathbb{II}_{\mathcal{N}}^{\textrm{4d $SO(2N)|USp(2M)$}}(t;q)
\nonumber\\
&
=\frac{1}{2^{N-1} N!}
\frac{(q)_{\infty}^N}{(q^{\frac12}t^{-2};q)_{\infty}^N}
\oint \prod_{i=1}^N \frac{ds_i^{(1)}}{2\pi is_i^{(1)}}
\nonumber\\
&\times 
\prod_{i<j}
\frac{(s_i^{(1)\pm}s_j^{(1)\mp};q)_{\infty}(s_i^{(1)\pm}s_j^{(1)\pm};q)_{\infty}}
{(q^{\frac12}t^{-2}s_i^{(1)\pm}s_j^{(1)\mp};q)_{\infty}(q^{\frac12}t^{-2}s_i^{(1)\pm}s_j^{(1)\pm};q)_{\infty}}
\nonumber\\
&\times 
\frac{1}{2^M M!}
\frac{(q)_{\infty}^M}{(q^{\frac12}t^{-2};q)_{\infty}^M}
\oint \prod_{i=1}^M \frac{ds_i^{(2)}}{2\pi is_i^{(2)}}
\frac{(s_i^{(2)\pm 2};q)_{\infty}}{(q^{\frac12}t^{-2}s_i^{(2)\pm 2};q)_{\infty}}
\nonumber\\
&\times 
\prod_{i< j}
\frac{(s_i^{(2)\pm}s_j^{(2)\mp};q)_{\infty}(s_i^{(2)\pm}s_j^{(2)\pm};q)_{\infty}}
{(q^{\frac12}t^{-2}s_i^{(2)\pm}s_j^{(2)\mp};q)_{\infty}(q^{\frac12}t^{-2}s_i^{(2)\pm}s_j^{(2)\pm};q)_{\infty}}
\nonumber\\
&\times 
\prod_{i=1}^N
\prod_{j=1}^M
\frac{(q^{\frac34}t^{-1}s_i^{(1)\pm}s_j^{(2)\mp};q)_{\infty}(q^{\frac34}t^{-1}s_i^{(1)\pm}s_j^{(2)\pm};q)_{\infty}
}
{(q^{\frac14}ts_i^{(1)\pm}s_{j}^{(2)\mp};q)_{\infty}(q^{\frac14}ts_i^{(1)\pm}s_{j}^{(2)\pm};q)_{\infty}
}. 
\end{align}
Here the first two lines and the next ones of R.H.S. are the Neumann half-indices 
for $SO(2N)$ and $USp(2M)$ gauge theories respectively. 
The last line is contributed from the bifundamental hyper living at the interface. 

For the interface involving the other disconnected part of $O(2N)$ gauge group 
the half-index is given by the following matrix integral
\begin{align}
\label{h_so2N|usp2M_N2}
&
\mathbb{II}_{\mathcal{N}}^{\textrm{4d $SO(2N)^{-}|USp(2M)$}}(t;q)
\nonumber\\
&
=\frac{1}{2^{N-1} (N-1)!}
\frac{(q)_{\infty}^{N-1}(-q;q)_{\infty}}
{(q^{\frac12}t^{-2};q)_{\infty}^{N-1}(-q^{\frac12}t^{-2};q)_{\infty}}
\oint \prod_{i=1}^{N-1} \frac{ds_i^{(1)}}{2\pi is_i^{(1)}}
\frac{(s_i^{(1)\pm};q)_{\infty} (-s_i^{(1)\pm};q)_{\infty}}
{(q^{\frac12}t^{-2}s_i^{(1)\pm};q)_{\infty}(-q^{\frac12}t^{-2}s_i^{(1)\pm};q)_{\infty}}
\nonumber\\
&\times 
\prod_{i<j}
\frac{(s_i^{(1)\pm}s_j^{(1)\mp};q)_{\infty}(s_i^{(1)\pm}s_j^{(1)\pm};q)_{\infty}}
{(q^{\frac12}t^{-2}s_i^{(1)\pm}s_j^{(1)\mp};q)_{\infty}(q^{\frac12}t^{-2}s_i^{(1)\pm}s_j^{(1)\pm};q)_{\infty}}
\nonumber\\
&\times 
\frac{1}{2^M M!}
\frac{(q)_{\infty}^M}{(q^{\frac12}t^{-2};q)_{\infty}^M}
\oint \prod_{i=1}^M \frac{ds_i^{(2)}}{2\pi is_i^{(2)}}
\frac{(s_i^{(2)\pm 2};q)_{\infty}}{(q^{\frac12}t^{-2}s_i^{(2)\pm 2};q)_{\infty}}
\nonumber\\
&\times 
\prod_{i< j}
\frac{(s_i^{(2)\pm}s_j^{(2)\mp};q)_{\infty}(s_i^{(2)\pm}s_j^{(2)\pm};q)_{\infty}}
{(q^{\frac12}t^{-2}s_i^{(2)\pm}s_j^{(2)\mp};q)_{\infty}(q^{\frac12}t^{-2}s_i^{(2)\pm}s_j^{(2)\pm};q)_{\infty}}
\nonumber\\
&\times 
\prod_{i=1}^{N-1}
\prod_{j=1}^M
\frac{(q^{\frac34}t^{-1}s_i^{(1)\pm}s_j^{(2)\mp};q)_{\infty}(q^{\frac34}t^{-1}s_i^{(1)\pm}s_j^{(2)\pm};q)_{\infty}
(q^{\frac34}t^{-1}s_j^{(2)\pm};q)_{\infty}(-q^{\frac34}t^{-1}s_j^{(2)\pm};q)_{\infty}}
{(q^{\frac14}ts_i^{(1)\pm}s_{j}^{(2)\mp};q)_{\infty}(q^{\frac14}ts_i^{(1)\pm}s_{j}^{(2)\pm};q)_{\infty}
(q^{\frac14}ts_{j}^{(2)\pm};q)_{\infty}(-q^{\frac14}ts_{j}^{(2)\pm};q)_{\infty}}. 
\end{align}
Again, the half-index of the NS5-type interface between $O(2N)$ and $USp(2M)$ gauge theories can be obtained from (\ref{h_so2N|usp2M_N}) and (\ref{h_so2N|usp2M_N2}) 
by gauging the $\mathbb{Z}_2$ global symmetry
\begin{align}
 \label{h_o2N|usp2M_N}
\mathbb{II}_{\mathcal{N}}^{\textrm{4d $O(2N)^{\pm}|USp(2M)$}}(t;q)
&=\frac12\left[
\mathbb{II}_{\mathcal{N}}^{\textrm{4d $SO(2N)|USp(2M)$}}(t;q)
\pm 
\mathbb{II}_{\mathcal{N}}^{\textrm{4d $SO(2N)^{-}|USp(2M)$}}(t;q)
\right]. 
\end{align}

The dual interface is obtained as the D5$'$-type between $O(2N)$ and $SO(2M+1)$ gauge theories under S-duality in Type IIB string theory. 
We consider the $O(2N)$ by taking $SO(2N)$ and gauging the $\mathbb{Z}_2$ symmetry. 

For $N>M$ the $SO(2N)$ $\times$ $SO(2M+1)$ gauge group breaks down to $SO(2M+1)$ and there is a Nahm pole for $SO(2(N-M))$ gauge theory. 
A broken part of the gauge group $SO(2N)$ remains as a global symmetry at the domain wall 
whereas there is no hypermultiplet at the interface. 
The half-index of the D5$'$-type $SO(2N)$ and $SO(2M+1)$ interface with $N>M$ takes the form 
\begin{align}
\label{h_so2N|so2M+1_D1}
&
\mathbb{II}_{\mathcal{D}'}^{\textrm{4d $SO(2N)|SO(2M+1)$}}(t;q)
\nonumber\\
&=\frac{1}{2^M M!} \frac{(q)_{\infty}^{2M}}{(q^{\frac12}t^{\pm2};q)_{\infty}^M}
\oint \prod_{i=1}^{M} \frac{ds_i}{2\pi is_i}
\frac{(s_i^{\pm};q)_{\infty} (qs_i^{\pm};q)_{\infty}}{(q^{\frac12}t^2 s_i^{\pm};q)_{\infty}(q^{\frac12}t^{-2}s_i^{\pm};q)_{\infty}}
\nonumber\\
&\times \prod_{i<j}
\frac{(s_i^{\pm}s_j^{\mp};q)_{\infty}(s_i^{\pm}s_j^{\pm};q)_{\infty}(qs_i^{\pm}s_j^{\mp};q)_{\infty}(qs_i^{\pm}s_j^{\pm};q)_{\infty}}
{(q^{\frac12}t^{2}s_i^{\pm}s_j^{\mp};q)_{\infty}(q^{\frac12}t^{2}s_i^{\pm}s_j^{\pm};q)_{\infty}
(q^{\frac12}t^{-2}s_i^{\pm}s_j^{\mp};q)_{\infty}(q^{\frac12}t^{-2}s_i^{\pm}s_j^{\pm};q)_{\infty}}
\nonumber\\
&\times 
\frac{(q^{\frac12+\frac{N-M}{2}} t^{-2(N-M)+2};q)_{\infty}}
{(q^{\frac{N-M}{2}}t^{-2(N-M)};q)_{\infty}}
\prod_{k=1}^{N-M-1}
\frac{(q^{\frac12+k}t^{2-4k};q)_{\infty}}
{(q^k t^{-4k};q)_{\infty}}
\nonumber\\
&\times 
\prod_{i=1}^{M}
\frac{(q^{1+\frac{N-M-1}{2}} t^{-2(N-M-1)} s_i^{\pm};q)_{\infty}}{(q^{\frac12+\frac{N-M-1}{2}}t^{-2-2(N-M-1)} s_i^{\pm};q)_{\infty}}. 
\end{align}
The matrix integral is associated with the surviving $SO(2M+1)$ gauge group. 
The last two lines of R.H.S. contain the $SO(2(N-M))$ Nahm pole half-index 
and the contributions from the broken part of $SO(2(N-M))$ gauge theory. 
For the disconnected part we have
\begin{align}
\label{h_so2N|so2M+1_D2}
&
\mathbb{II}_{\mathcal{D}'}^{\textrm{4d $SO(2N)^{-}|SO(2M+1)$}}(t;q)
\nonumber\\
&=\frac{1}{2^M M!} \frac{(q)_{\infty}^{2M}}{(q^{\frac12}t^{\pm2};q)_{\infty}^M}
\oint \prod_{i=1}^{M} \frac{ds_i}{2\pi is_i}
\frac{(s_i^{\pm};q)_{\infty} (qs_i^{\pm};q)_{\infty}}{(q^{\frac12}t^2 s_i^{\pm};q)_{\infty}(q^{\frac12}t^{-2}s_i^{\pm};q)_{\infty}}
\nonumber\\
&\times \prod_{i<j}
\frac{(s_i^{\pm}s_j^{\mp};q)_{\infty}(s_i^{\pm}s_j^{\pm};q)_{\infty}(qs_i^{\pm}s_j^{\mp};q)_{\infty}(qs_i^{\pm}s_j^{\pm};q)_{\infty}}
{(q^{\frac12}t^{2}s_i^{\pm}s_j^{\mp};q)_{\infty}(q^{\frac12}t^{2}s_i^{\pm}s_j^{\pm};q)_{\infty}
(q^{\frac12}t^{-2}s_i^{\pm}s_j^{\mp};q)_{\infty}(q^{\frac12}t^{-2}s_i^{\pm}s_j^{\pm};q)_{\infty}}
\nonumber\\
&\times 
\frac{(-q^{\frac12+\frac{N-M}{2}} t^{-2(N-M)+2};q)_{\infty}}
{(-q^{\frac{N-M}{2}}t^{-2(N-M)};q)_{\infty}}
\prod_{k=1}^{N-M-1}
\frac{(q^{\frac12+k}t^{2-4k};q)_{\infty}}
{(q^k t^{-4k};q)_{\infty}}
\nonumber\\
&\times 
\prod_{i=1}^{M}
\frac{(-q^{1+\frac{N-M-1}{2}} t^{-2(N-M-1)} s_i^{\pm};q)_{\infty}}{(-q^{\frac12+\frac{N-M-1}{2}}t^{-2-2(N-M-1)} s_i^{\pm};q)_{\infty}}. 
\end{align}

For $N<M$ the half D5$'$-brane breaks the $SO(2N)$ $\times$ $SO(2M+1)$ gauge group down to $SO(2N)$. 
The interface involves a Nahm pole for $SO(2(M-N)+1)$ gauge theory 
and contains a degrees of freedom from the broken part of the gauge group $SO(2M+1)$. 
In this case, we have the half-index 
\begin{align}
\label{h_so2N|so2M+1_D3}
&
\mathbb{II}_{\mathcal{D}'}^{\textrm{4d $SO(2N)|SO(2M+1)$}}(t;q)
\nonumber\\
&=\frac{1}{2^{N-1} N!} \frac{(q)_{\infty}^{2N}}{(q^{\frac12}t^{\pm2};q)_{\infty}^N}
\oint \prod_{i=1}^{N} \frac{ds_i}{2\pi is_i}
\nonumber\\
&\times \prod_{i<j}
\frac{(s_i^{\pm}s_j^{\mp};q)_{\infty}(s_i^{\pm}s_j^{\pm};q)_{\infty}(qs_i^{\pm}s_j^{\mp};q)_{\infty}(qs_i^{\pm}s_j^{\pm};q)_{\infty}}
{(q^{\frac12}t^{2}s_i^{\pm}s_j^{\mp};q)_{\infty}(q^{\frac12}t^{2}s_i^{\pm}s_j^{\pm};q)_{\infty}
(q^{\frac12}t^{-2}s_i^{\pm}s_j^{\mp};q)_{\infty}(q^{\frac12}t^{-2}s_i^{\pm}s_j^{\pm};q)_{\infty}}
\nonumber\\
&\times 
\prod_{k=1}^{M-N}
\frac{(q^{\frac12+k}t^{2-4k};q)_{\infty}}
{(q^k t^{-4k};q)_{\infty}}
\nonumber\\
&\times 
\prod_{i=1}^{N}
\frac{(q^{1+\frac{M-N}{2}} t^{-2(M-N)} s_i^{\pm};q)_{\infty}}{(q^{\frac12+\frac{M-N}{2}}t^{-2-2(M-N)} s_i^{\pm};q)_{\infty}}. 
\end{align}
Also the half-index for the disconnected component is 
\begin{align}
\label{h_so2N|so2M+1_D4}
&
\mathbb{II}_{\mathcal{D}'}^{\textrm{4d $SO(2N)^{-}|SO(2M+1)$}}(t;q)
\nonumber\\
&=\frac{1}{2^{N-1} (N-1)!} 
\frac{(q)_{\infty}^{2N-2} (-q;q)_{\infty}^2}
{(q^{\frac12}t^{\pm2};q)_{\infty}^{N-1}(-q^{\frac12}t^{\pm2};q)_{\infty}}
\oint \prod_{i=1}^{N} \frac{ds_i}{2\pi is_i}
\nonumber\\
&\times
\prod_{i=1}^{N-1} 
\frac{(s_i^{\pm};q)_{\infty}(-s_i^{\pm};q)_{\infty} (qs_i^{\pm};q)_{\infty}(-qs_i^{\pm};q)_{\infty}}
{(q^{\frac12}t^{2}s_i^{\pm};q)_{\infty}(-q^{\frac12}t^{2}s_i^{\pm};q)_{\infty}(q^{\frac12}t^{-2}s_i^{\pm};q)_{\infty}(-q^{\frac12}t^{-2}s_i^{\pm};q)_{\infty}}
\nonumber\\
&\times \prod_{i<j}
\frac{(s_i^{\pm}s_j^{\mp};q)_{\infty}(s_i^{\pm}s_j^{\pm};q)_{\infty}(qs_i^{\pm}s_j^{\mp};q)_{\infty}(qs_i^{\pm}s_j^{\pm};q)_{\infty}}
{(q^{\frac12}t^{2}s_i^{\pm}s_j^{\mp};q)_{\infty}(q^{\frac12}t^{2}s_i^{\pm}s_j^{\pm};q)_{\infty}
(q^{\frac12}t^{-2}s_i^{\pm}s_j^{\mp};q)_{\infty}(q^{\frac12}t^{-2}s_i^{\pm}s_j^{\pm};q)_{\infty}}
\nonumber\\
&\times 
\prod_{k=1}^{M-N}
\frac{(q^{\frac12+k}t^{2-4k};q)_{\infty}}
{(q^k t^{-4k};q)_{\infty}}
\nonumber\\
&\times 
\prod_{i=1}^{N-1}
\frac{(q^{1+\frac{M-N}{2}} t^{-2(M-N)} s_i^{\pm};q)_{\infty}}{(q^{\frac12+\frac{M-N}{2}}t^{-2-2(M-N)} s_i^{\pm};q)_{\infty}}
\frac{(\pm q^{1+\frac{M-N}{2}} t^{-2(M-N)})}{(\pm q^{\frac12+\frac{M-N}{2}}t^{-2-2(M-N)};q)_{\infty}}. 
\end{align}
Note that the expressions (\ref{h_so2N|so2M+1_D3}) and (\ref{h_so2N|so2M+1_D4}) become 
(\ref{h_so2N|so2N+1_D}) and (\ref{h_so2N|so2N+1_D2}) when $N=M$ respectively. 

It turns out that 
the half-index (\ref{h_so2N|usp2M_N}) for the NS5$'$-type $SO(2N)|USp(2M)$ interface excellently matches with 
the half-index (\ref{h_so2N|so2M+1_D1}) for $N>M$ and the half-index (\ref{h_so2N|so2M+1_D3}) for $N<M$! 
These equalities imply the following duality: 
\begin{align}
\label{dual_so2n|usp2m}
&\textrm{$SO(2N)|USp(2M)$ NS5-type interface}
\nonumber\\
&\Leftrightarrow
\textrm{$SO(2N)|SO(2M+1)$ D5$'$-type interface}, 
\end{align}
which generalizes (\ref{dual_so2n|usp2n}) with gauge groups of equal rank. 
Moreover, the half-index (\ref{h_so2N|usp2M_N}) for the NS5$'$-type $SO(2N)^{-}|USp(2M)$ 
coincides with the half-index (\ref{h_so2N|so2M+1_D2}) for $N>M$ and the half-index (\ref{h_so2N|so2M+1_D4}) for $N<M$! 
This supports the duality of interfaces involving the disconnected $O(2N)$ gauge group
\begin{align}
\label{dual_o2n|usp2m}
&\textrm{$O(2N)|USp(2M)$ NS5-type interface}
\nonumber\\
&\Leftrightarrow
\textrm{$O(2N)|SO(2M+1)$ D5$'$-type interface}. 
\end{align}

In the Coulomb limit and Higgs limit the half-indices (\ref{h_so2N|so2M+1_D1}) and (\ref{h_so2N|so2M+1_D2}) become
\begin{align}
\label{h_so2N|so2M+1_C}
\mathbb{II}_{\mathcal{N}}^{\textrm{4d $SO(2N)|USp(2M)$}(C)}(\mathfrak{q})
&=\mathbb{II}_{\mathcal{D}'}^{\textrm{4d $SO(2N)|SO(2M+1)$}(C)}(\mathfrak{q})
\nonumber\\
&=\frac{1}{1-\mathfrak{q}^{2N}}\frac{1}{(\mathfrak{q}^4;\mathfrak{q}^4)_{N-1}(\mathfrak{q}^4;\mathfrak{q}^4)_{M}}, \\
\label{h_so2N|so2M+1_H}
\mathbb{II}_{\mathcal{N}}^{\textrm{4d $SO(2N)|USp(2M)$}(H)}(\mathfrak{q})
&=\mathbb{II}_{\mathcal{D}'}^{\textrm{4d $SO(2N)|SO(2M+1)$}(H)}(\mathfrak{q})
\nonumber\\
&=\begin{cases}
\frac{1}{(\mathfrak{q}^4;\mathfrak{q}^4)_{M}}&\textrm{for $N>M$}\cr
\frac{1}{1-\mathfrak{q}^{2N}}\frac{1}{(\mathfrak{q}^4;\mathfrak{q}^4)_{N-1}}&\textrm{for $N\le M$}\cr
\end{cases}
\end{align}
and the half-indices (\ref{h_so2N|so2M+1_D1}) and (\ref{h_so2N|so2M+1_D2}) associated with the disconnected group $O(2N)$ become 
\begin{align}
\label{h_o2N|so2M+1_C+}
\mathbb{II}_{\mathcal{N}}^{\textrm{4d $O(2N)^+|USp(2M)$}(C)}(\mathfrak{q})
&=\mathbb{II}_{\mathcal{D}'}^{\textrm{4d $O(2N)^+|SO(2M+1)$}(C)}(\mathfrak{q})
\nonumber\\
&=\frac{1}{(\mathfrak{q}^4;\mathfrak{q}^4)_{N}(\mathfrak{q}^4;\mathfrak{q}^4)_{M}}, \\
\label{h_o2N|so2M+1_H+}
\mathbb{II}_{\mathcal{N}}^{\textrm{4d $O(2N)^+|USp(2M)$}(H)}(\mathfrak{q})
&=\mathbb{II}_{\mathcal{D}'}^{\textrm{4d $O(2N)^+|SO(2M+1)$}(H)}(\mathfrak{q})
\nonumber\\
&=
\frac{1}{(\mathfrak{q}^4;\mathfrak{q}^4)_{\min (N,M)}}, 
\end{align}
\begin{align}
\label{h_o2N|so2M+1_C-}
\mathbb{II}_{\mathcal{N}}^{\textrm{4d $O(2N)^-|USp(2M)$}(C)}(\mathfrak{q})
&=\mathbb{II}_{\mathcal{D}'}^{\textrm{4d $O(2N)^-|SO(2M+1)$}(C)}(\mathfrak{q})
\nonumber\\
&=\frac{\mathfrak{q}^{2N}}{(\mathfrak{q}^4;\mathfrak{q}^4)_{N}(\mathfrak{q}^4;\mathfrak{q}^4)_{M}}, \\
\label{h_o2N|so2M+1_H-}
\mathbb{II}_{\mathcal{N}}^{\textrm{4d $O(2N)^-|USp(2M)$}(H)}(\mathfrak{q})
&=\mathbb{II}_{\mathcal{D}'}^{\textrm{4d $O(2N)^-|SO(2M+1)$}(H)}(\mathfrak{q})
\nonumber\\
&=\begin{cases}
\frac{1}{(\mathfrak{q}^4;\mathfrak{q}^4)_{M}}&\textrm{for $N>M$}\cr
\frac{\mathfrak{q}^{2N}}{(\mathfrak{q}^4;\mathfrak{q}^4)_{N-1}}&\textrm{for $N\le M$}\cr
\end{cases}. 
\end{align}
The Coulomb indices (\ref{h_so2N|so2M+1_C}), (\ref{h_o2N|so2M+1_C+}) and (\ref{h_o2N|so2M+1_C-}) are factorized into 
pairs of the half-BPS indices of the gauge theories in both sides of the interfaces. 
The Higgs indices (\ref{h_so2N|so2M+1_H}), (\ref{h_o2N|so2M+1_H+}) and (\ref{h_o2N|so2M+1_H-}) are given by the 
half-BPS indices of the gauge theory with lower rank gauge group on either side of the interfaces. 
They can be thought of as the ``Coulomb limit'' of the D5$'$-type interface 
in such a way that they count the remaining gauge group invariant local operators at the domain walls.  

\section{Giant graviton expansions}
\label{sec_ggexp}

\subsection{Large $N$ limits}
In the large $N$ limit, the half-indices capture the spectra of 
the KK modes on the holographic dual $AdS_4$ bagpipe geometries involving the orbifold ETW brane 
as well as the asymptotic $AdS_5\times \mathbb{RP}^5$ regions. 

For the Neumann and Nahm pole boundary conditions of $SO(2N+1)$, $USp(2N)$, $SO(2N)$, $O(2N)^+$ and $USp(2N)'$ gauge theories 
the half-indices coincide in the large $N$ limit. 
From the exact closed-form expressions for the Nahm pole half-indices, (\ref{h_C_Nahm}), (\ref{h_B_Nahm}) and (\ref{h_O2N_Nahm}) which are also equal to the half-indices for the S-dual Neumann boundary conditions, we find
\begin{align}
\label{large_BCD}
&\mathbb{II}_{\mathcal{N}}^{SO(\infty)}
=\mathbb{II}_{\mathcal{N}}^{USp(\infty)}
=\mathbb{II}_{\mathcal{N}}^{O(\infty)^+}
=\mathbb{II}_{\mathcal{N}}^{USp(\infty)'}
\nonumber\\
=&\mathbb{II}_{\textrm{Nahm}'}^{SO(\infty)}
=\mathbb{II}_{\textrm{Nahm}'}^{USp(\infty)}
=\mathbb{II}_{\textrm{Nahm}'}^{O(\infty)^+}
=\mathbb{II}_{\textrm{Nahm}'}^{USp(\infty)'}
=\prod_{n=0}^{\infty}
\prod_{k=0}^{\infty} \frac{1-q^{n+k+\frac32}t^{-4k-2}}
{1-q^{n+k+1} t^{-4k-4}}. 
\end{align}
The single particle gravity index for the orbifold ETW brane 
is obtained by the plethystic logarithm \cite{MR1601666} of the large $N$ index (\ref{large_BCD}). 
We get
\begin{align}
\label{orbETW_sind}
i^{\textrm{$\mathbb{Z}_2$ ETW}}
&=-\frac{q^{\frac32}t^{-2}}{(1-q)(1-qt^{-4})}+\frac{qt^{-4}}{(1-q)(1-qt^{-4})}. 
\end{align}
In the unflavored limit, we get
\begin{align}
\label{large_BCDunf}
&\mathbb{II}_{\mathcal{N}}^{SO(\infty)}
=\mathbb{II}_{\mathcal{N}}^{USp(\infty)}
=\mathbb{II}_{\mathcal{N}}^{O(\infty)^+}
=\mathbb{II}_{\mathcal{N}}^{USp(\infty)'}
\nonumber\\
=&\mathbb{II}_{\textrm{Nahm}'}^{SO(\infty)}
=\mathbb{II}_{\textrm{Nahm}'}^{USp(\infty)}
=\mathbb{II}_{\textrm{Nahm}'}^{O(\infty)^+}
=\mathbb{II}_{\textrm{Nahm}'}^{USp(\infty)'}
=\prod_{n=1}^{\infty}
\frac{(1-q^{n+\frac12})^n}{(1-q^n)^n}. 
\end{align}
In the half-BPS limit, the half-index becomes the large $N$ half-BPS index
\begin{align}
\label{large_BCD_half}
&\mathbb{II}_{\textrm{$\frac12$BPS}}^{SO(\infty)}
=\mathbb{II}_{\textrm{$\frac12$BPS}}^{USp(\infty)}
=\mathbb{II}_{\textrm{$\frac12$BPS}}^{O(\infty)^+}
=\mathbb{II}_{\textrm{$\frac12$BPS}}^{USp(\infty)'}
=\prod_{n=1}^{\infty}\frac{1}{1-\mathfrak{q}^{4n}}. 
\end{align}

Next consider the large gauge rank limits of the interface half-indices. 
It follows that when either of the gauge ranks is taken large, 
the interface half-index is factorized into the large $N$ half-indices and the full indices, i.e. Schur indices. 
We find that 
\begin{align}
\mathbb{II}_{\mathcal{N}}^{\textrm{4d $SO(2N+1)|USp(\infty)'$}}(t;q)
&=\mathcal{I}^{SO(2N+1)}(t;q)\times \mathbb{II}_{\mathcal{N}}^{USp(\infty)'}(t;q), \\
\mathbb{II}_{\mathcal{N}}^{\textrm{4d $SO(\infty)|USp(2M)'$}}(t;q)
&=\mathcal{I}^{USp(2M)'}(t;q)\times \mathbb{II}_{\mathcal{N}}^{SO(\infty)}(t;q), \\
\mathbb{II}_{\mathcal{N}}^{\textrm{4d $SO(2N)|USp(\infty)$}}(t;q)
&=\mathcal{I}^{SO(2N)}(t;q)\times \mathbb{II}_{\mathcal{N}}^{USp(\infty)}(t;q), \\
\mathbb{II}_{\mathcal{N}}^{\textrm{4d $SO(\infty)|USp(2M)$}}(t;q)
&=\mathcal{I}^{USp(2M)}(t;q)\times \mathbb{II}_{\mathcal{N}}^{SO(\infty)}(t;q), \\
\mathbb{II}_{\mathcal{N}}^{\textrm{4d $O(2N)^+|USp(\infty)$}}(t;q)
&=\mathcal{I}^{O(2N)^+}(t;q)\times \mathbb{II}_{\mathcal{N}}^{USp(\infty)}(t;q), \\
\mathbb{II}_{\mathcal{N}}^{\textrm{4d $O(\infty)^+|USp(2M)$}}(t;q)
&=\mathcal{I}^{USp(2M)}(t;q)\times \mathbb{II}_{\mathcal{N}}^{O(\infty)^+}(t;q), 
\end{align}
where $\mathcal{I}^{G}(t;q)$ is the flavored Schur index of $\mathcal{N}=4$ SYM theory 
(or equivalently $\mathcal{N}=2^*$ Schur index) of gauge group $G$. 
Such factorizations also appear in the half-indices of the interfaces between unitary gauge theories \cite{Hatsuda:2024uwt}. 
In particular, when both gauge ranks are taken large, we obtain
\begin{align}
\label{large_BCDinterface}
&\mathbb{II}_{\mathcal{N}}^{\textrm{4d $SO(\infty)|USp(\infty)'$}}(t;q)
=\mathbb{II}_{\mathcal{N}}^{\textrm{4d $SO(\infty)|USp(\infty)$}}(t;q)
=\mathbb{II}_{\mathcal{N}}^{\textrm{4d $O(\infty)^+|USp(\infty)$}}(t;q)
\nonumber\\
&=\mathcal{I}^{SO(\infty)}(t;q)\times \mathbb{II}_{\mathcal{N}}^{USp(\infty)'}(t;q)
=\mathcal{I}^{USp(\infty)'}(t;q)\times \mathbb{II}_{\mathcal{N}}^{SO(\infty)}(t;q)
\nonumber\\
&=\mathcal{I}^{SO(\infty)}(t;q)\times \mathbb{II}_{\mathcal{N}}^{USp(\infty)}(t;q)
=\mathcal{I}^{USp(\infty)}(t;q)\times \mathbb{II}_{\mathcal{N}}^{SO(\infty)}(t;q)
\nonumber\\
&=\mathcal{I}^{O(\infty)^+}(t;q)\times \mathbb{II}_{\mathcal{N}}^{USp(\infty)}(t;q)
=\mathcal{I}^{USp(\infty)}(t;q)\times \mathbb{II}_{\mathcal{N}}^{O(\infty)^+}(t;q), 
\end{align}
where 
\begin{align}
\label{large_BCDSchur}
&
\mathcal{I}^{SO(\infty)}(t;q)
=\mathcal{I}^{USp(\infty)}(t;q)
=\mathcal{I}^{O(\infty)^+}(t;q)
=\mathcal{I}^{USp(\infty)'}(t;q)
\nonumber\\
=&
\prod_{n,m,l=0}^{\infty}
\prod_{\pm}
\frac{(1-q^{n+m+l+\frac32} t^{-4m+4l\pm 2})^2}
{(1-q^{n+m+l+1} t^{-4m+4l\pm 4})(1-q^{n+m+l+1} t^{-4m+4l})(1-q^{n+m+l+3} t^{-4m+4l})}
\end{align}
is the large $N$ limit of the flavored Schur index of the orthogonal or symplectic gauge theory. 
According to the factorized forms of the interface half-indices (\ref{large_BCDinterface}), 
the single particle gravity index is simply obtained as a sum of the orbifold ETW brane index (\ref{orbETW_sind}) 
and the single particle gravity index 
\begin{align}
i^{AdS_5\times \mathbb{RP}^5}
&=-\frac{q^{\frac32}t^{-2}}{(1-q)(1-qt^{-4})(1-qt^{4})}-\frac{q^{\frac32}t^2}{(1-q)(1-qt^{-4})(1-qt^{4})}
\nonumber\\
&+\frac{qt^{-4}}{(1-q)(1-qt^{-4})(1-qt^{4})}+\frac{qt^{4}}{(1-q)(1-qt^{-4})(1-qt^{4})}
\nonumber\\
&+\frac{q}{(1-q)(1-qt^{-4})(1-qt^{4})}+\frac{q^3}{(1-q)(1-qt^{-4})(1-qt^{4})}
\end{align}
for the spectrum of the KK modes on the $AdS_5\times \mathbb{RP}^5$ geometry,  
which is obtained by taking the plethystic logarithm of the large $N$ full index (\ref{large_BCDSchur}). 
In the unflavored limit the large $N$ Schur index (\ref{large_BCDSchur}) becomes 
\begin{align}
\label{large_BCDSchur_unf}
&
\mathcal{I}^{SO(\infty)}(q)
=\mathcal{I}^{USp(\infty)}(q)
=\mathcal{I}^{O(\infty)^+}(q)
=\mathcal{I}^{USp(\infty)'}(q)
\nonumber\\
=&\prod_{n=1}^{\infty}
\frac{(1-q^{n+\frac12})^{2n^2+2n}}{(1-q^{n})^{2n^2+1}}. 
\end{align}
In the Coulomb and Higgs limits, the large gauge rank interface half-index (\ref{large_BCDinterface}) becomes 
\begin{align}
&\mathbb{II}_{\mathcal{N}}^{\textrm{4d $SO(\infty)|USp(\infty)' (C)$}}(\mathfrak{q})
=\mathbb{II}_{\mathcal{N}}^{\textrm{4d $SO(\infty)|USp(\infty) (C)$}}(\mathfrak{q})
=\mathbb{II}_{\mathcal{N}}^{\textrm{4d $O(\infty)^+|USp(\infty) (C)$}}(\mathfrak{q})
\nonumber\\
&=\prod_{n=1}^{\infty}\frac{1}{(1-\mathfrak{q}^{4n})^2}, \\
&\mathbb{II}_{\mathcal{N}}^{\textrm{4d $SO(\infty)|USp(\infty)' (H)$}}(\mathfrak{q})
=\mathbb{II}_{\mathcal{N}}^{\textrm{4d $SO(\infty)|USp(\infty) (H)$}}(\mathfrak{q})
=\mathbb{II}_{\mathcal{N}}^{\textrm{4d $O(\infty)^+|USp(\infty) (H)$}}(\mathfrak{q})
\nonumber\\
&=\prod_{n=1}^{\infty}\frac{1}{(1-\mathfrak{q}^{4n})}. 
\end{align}

\subsection{Giant graviton expansions}
When we expand the half-indices for finite $N$, the rank of gauge groups, with respect to $q$, 
we encounter corrections to the large $N$ half-indices. 
From the gravity dual point of view, they can be considered as the effect of orbifold ETW giant gravitons. 

Consider the half-indices of Neumann or Nahm pole boundary conditions for $SO(2N+1)$ or $USp(2N)$ gauge theories. 
Since the equalities of the half-indices, it is suffice to consider the $USp(2N)$ Neumann half-indices. 
In the unflavored limit, the half-indices for $N$ $=$ $1,\cdots, 5$ can be expanded as follows: 
\begin{align}
\mathbb{II}_{\mathcal{N}}^{\textrm{4d $USp(2)$}}
&=1+q-q^{3/2}+2q^2-2q^{5/2}+3q^3-4q^{7/2}+6q^4-7q^{9/2}+9q^5+\cdots, \\
\mathbb{II}_{\mathcal{N}}^{\textrm{4d $USp(4)$}}
&=1+q-q^{3/2}+3q^2-3q^{5/2}+5q^3-7q^{7/2}+12q^4-15q^{9/2}+21q^5+\cdots, \\
\mathbb{II}_{\mathcal{N}}^{\textrm{4d $USp(6)$}}
&=1+q-q^{3/2}+3q^2-3q^{5/2}+6q^3-8q^{7/2}+14q^4-18q^{9/2}+27q^5+\cdots, \\
\mathbb{II}_{\mathcal{N}}^{\textrm{4d $USp(8)$}}
&=1+q-q^{3/2}+3q^2-3q^{5/2}+6q^3-8q^{7/2}+15q^4-19q^{9/2}+29q^5+\cdots, \\
\mathbb{II}_{\mathcal{N}}^{\textrm{4d $USp(10)$}}
&=1+q-q^{3/2}+3q^2-3q^{5/2}+6q^3-8q^{7/2}+15q^4-19q^{9/2}+30q^5+\cdots. 
\end{align}
We see that the finite $N$ correction starts from the term with $q^{N+1}$. 

Let us investigate the giant graviton expansions \cite{Arai:2019xmp,Arai:2020qaj,Gaiotto:2021xce} of the half-indices, 
that is the expansions of the ratios of the finite $N$ half-indices to the large $N$ half-index. 
We first observe that the ratio
\begin{align}
\frac{\mathbb{II}_{\mathcal{N}}^{\textrm{4d $USp(2N)$}}}
{\mathbb{II}_{\mathcal{N}}^{\textrm{4d $USp(2N-2)$}}}=\frac{(q^{N+\frac12}t^{-4N+2};q)_{\infty}}{(q^N t^{-4N};q)_{\infty}}
=\frac{(x^{2N}y;q)_{\infty}}{(x^{2N};q)_{\infty}}
=\sum_{k=0}^\infty x^{2kN}\frac{(y;q)_k}{(q;q)_k}
\end{align}
has a simple large $N$ expansion. Here $x:=q^{1/2}t^{-2}$ and $y:=q^{1/2}t^2$. 
Let us consider the giant graviton expansion
\begin{align}
\frac{\mathbb{II}_{\mathcal{N}}^{\textrm{4d $USp(2N)$}}}{\mathbb{II}_{\mathcal{N}}^{\textrm{4d $USp(\infty)$}}}
=\sum_{k=0}^\infty x^{2kN}\hat{f}_k(x,y;q), 
\end{align}
where the expansion variable $x^{2kN}$ corresponds to the orbifold ETW giant gravitons of wrapping number $2k$. 
Then the above equation leads to
\begin{align}
\frac{\mathbb{II}_{\mathcal{N}}^{\textrm{4d $USo(2N)$}}}
{\mathbb{II}_{\mathcal{N}}^{\textrm{4d $USp(\infty)$}}}
=\frac{\mathbb{II}_{\mathcal{N}}^{\textrm{4d $USp(2N-2)$}}}{\mathbb{II}_{\mathcal{N}}^{\textrm{4d $USp(\infty)$}}}\frac{(x^{2N}y;q)_{\infty}}{(x^{2N};q)_{\infty}},
\end{align}
and substituting the giant graviton expansion, we obtain
\begin{align}
\sum_{k=0}^\infty z^k \hat{f}_k(x,y;q)&=\biggl( \sum_{\ell=0}^\infty z^\ell x^{-2\ell}\hat{f}_\ell(x,y;q) \biggr)\biggl(\sum_{m=0}^\infty z^m\frac{(y;q)_m}{(q;q)_m}\biggr) \\
&=\sum_{k=0}^\infty z^k \sum_{\ell=0}^k \frac{(y;q)_{k-\ell}}{(q;q)_{k-\ell}}\frac{\hat{f}_\ell(x,y;q)}{x^{2\ell}},
\end{align}
where we have defined $z:=x^{2N}$. 
Therefore 
\begin{align}
\hat{f}_k(x,y;q)=\sum_{\ell=0}^k \frac{(y;q)_{k-\ell}}{(q;q)_{k-\ell}}\frac{\hat{f}_\ell(x,y;q)}{x^{2\ell}}.
\end{align}
Solving this equation with respect to $\hat{f}_k(x,y;q)$, we get 
the giant graviton indices 
\begin{align}
\label{GGind_NeuNahm_bcd}
\hat{f}_k(x,y;q)=\frac{1}{1-x^{-2k}} \sum_{\ell=0}^{k-1} \frac{(y;q)_{k-\ell}}{(q;q)_{k-\ell}}\frac{\hat{f}_\ell(x,y;q)}{x^{2\ell}}.
\end{align}
This recursively determines all the coefficients $\hat{f}_k(x,y;q)$ with $\hat{f}_0(x,y;q)=1$. For instance,
\begin{align}
\hat{f}_1(x,y;q)&=-\frac{x^2(1-y)}{(1-q)(1-x^2)}, \\
\hat{f}_2(x,y;q)&=\frac{x^4(1-y)(q+x^2-y-qx^2y)}{(1-q)(1-q^2)(1-x^2)(1-x^4)}.
\end{align}

We can also derive another expression more directly. We start with an infinite product form
\begin{align}
\frac{\mathbb{II}_{\mathcal{N}}^{\textrm{4d $USp(2N)$}}}
{\mathbb{II}_{\mathcal{N}}^{\textrm{4d $USp(\infty)$}}}&=\prod_{k=N+1}^\infty \frac{(q^k t^{-4k};q)_{\infty}}{(q^{k+\frac12}t^{-4k+2};q)_{\infty}}
=\prod_{m=1}^\infty \frac{(x^{2N+2m};q)_\infty}{(x^{2N+2m}y;q)_\infty} \\
&=\prod_{m=1}^\infty \prod_{\ell=0}^\infty \frac{1-zx^{2m}q^\ell}{1-zx^{2m}yq^\ell}.
\end{align}
This form itself is not useful to compute the giant graviton expansion. We rewrite it as follows.
\begin{align}
\log \left(\frac{\mathbb{II}_{\mathcal{N}}^{\textrm{4d $USp(2N)$}}}{\mathbb{II}_{\mathcal{N}}^{\textrm{4d $USp(\infty)$}}}\right)
&=\sum_{m=1}^\infty \sum_{\ell=0}^\infty \log \frac{1-zx^{2m}q^\ell}{1-zx^{2m}yq^\ell}\\
&=\sum_{m=1}^\infty \sum_{\ell=0}^\infty \sum_{n=1}^\infty \frac{z^n}{n}(x^{2mn}y^nq^{\ell n}-x^{2mn}q^{\ell n} ) \\
&=-\sum_{n=1}^\infty \frac{z^n x^{2n}(1-y^n)}{n(1-x^{2n})(1-q^n)}. 
\end{align}
Therefore we find
\begin{align}
\sum_{k=0}^\infty z^k \hat{f}_k(x,y;q)&= \exp \left[ -\sum_{n=1}^\infty \frac{z^n x^{2n}(1-y^n)}{n(1-x^{2n})(1-q^n)} \right]
\nonumber\\
&=\prod_{n=0}^{\infty}\prod_{m=0}^{\infty}
\frac{1-x^{2n+2} q^{m}z}{1-yx^{2n+2}q^{m}z}. 
\end{align}

This expression allows us to find another recursion relation
\begin{equation}
\begin{aligned}
\hat{f}_k(x,y;q)=\frac{1}{k}\sum_{r=1}^k p_r \hat{f}_{k-r}(x,y;q),\qquad p_r:=-\frac{x^{2r}(1-y^r)}{(1-x^{2r})(1-q^r)}. 
\end{aligned}
\end{equation}
For  $k=1,2,3$, we have
\begin{align}
\hat{f}_1&=p_1, \\
\hat{f}_2&=\frac{1}{2}(p_1^2+p_2), \\
\hat{f}_3&=\frac{1}{6}(p_1^3+3p_1p_2+2p_3).
\end{align}
Also an explicit form of the giant graviton index can be written as a sum over partitions of integers:
\begin{align}
\label{GGind_NeuNahm_bcd2}
\hat{f}_k(x,y;q)=\sum_{|\lambda|=k} \frac{p_\lambda}{z_\lambda}, 
\end{align}
where $\lambda=(1^{m_1}, 2^{m_2},3^{m_3}, \dots)$ is a partition of $k$ satisfiying
\begin{align}
\sum_{i=1}^\infty im_i=k
\end{align}
and
\begin{align}
p_\lambda&:=\prod_{i=1}^\infty p_i^{m_i}=p_1^{m_1} p_2^{m_2} p_3^{m_3}\cdots \\
z_\lambda&:=\prod_{i=1}^\infty i^{m_i}m_i!=(1^{m_1}m_1!)(2^{m_2}m_2!)(3^{m_3}m_3!)\cdots.
\end{align}
From the point of view of the effective theory on the orbifold ETW giant gravitons, 
the index $f_k(x,y;q)$ can be obtained from the giant graviton index $\hat{f}_k(x,y;q)$ by redefining the fugacities \cite{Arai:2019xmp,Gaiotto:2021xce}
\begin{align}
\label{x_change}
\sigma_x: (x,y,q)&\rightarrow (x^{-1},q,y). 
\end{align}
Again it is evaluated from the grand canonical index 
\begin{align}
\sum_{k=0}^\infty z^k f_k(x,y;q)
&= \exp \left[ \sum_{n=1}^\infty \frac{z^n (1-q^n)}
{n(1-x^{2n})(1-y^n)} \right]
\nonumber\\
&=\prod_{n=0}^{\infty}\prod_{m=0}^{\infty}
\frac{1-q x^{2n} y^{m}z}{1-x^{2n}y^{m}z}. 
\end{align}
In the large $k$ limit, the index $f_k(x,y;q)$ becomes 
\begin{align}
f_{\infty}(x,y;q)&=
\prod_{n=1}^{\infty}\frac{1}{(1-x^{2n})(1-y^n)}
\prod_{m=1}^{\infty}\frac{1-x^{2n-1}y^m}{1-x^{2n}y^m}. 
\end{align}
When we expand the index $f_k(x,y;q)$ with respect to $q$, 
we see that the finite $k$ correction appears at the term with $q^{\frac{k+1}{2}}$. 
Quite interestingly, we find that it admits the inverse giant graviton expansion with the form
\begin{align}
\frac{f_k(x,y;q)}{f_{\infty}(x,y;q)}
&=\sum_{N=0}^{\infty} x^{2Nk} \sigma_x \mathbb{II}_{\mathcal{N}}^{\textrm{4d $USp(2N)$}}. 
\end{align}

It would be intriguing to reproduce our exact form (\ref{GGind_NeuNahm_bcd}) or (\ref{GGind_NeuNahm_bcd2}) 
for the orbifold ETW giant graviton index by means of other methods 
and to figure out the giant graviton expansions for other half-indices including the interface half-indices. 
For the half-index of Neumann or Nahm pole boundary conditions for $U(N)$ gauge theory, 
it is shown in \cite{Hatsuda:2024uwt} that 
the giant graviton index with wrapping number $k$ can be identified with the large $N$ normalized two-point function 
of the Wilson lines in the rank-$k$ antisymmetric representation in $U(N)$ gauge theories \cite{Hatsuda:2023iwi,Hatsuda:2023imp} 
upon changes of variables. 
More generally, for the half-indices of the interface between $U(N)$ and $U(M)$ gauge theories, 
the giant graviton index depends on a pair $(m,k)$ of the wrapping numbers for the giants in the ETW region and those in the bulk. 
It is found in \cite{Hatsuda:2024uwt} that the interface half-index agrees with 
the two-point function of the Wilson lines in the rank-$k$ antisymmetric representation in $U(m)$ gauge theories \cite{Hatsuda:2023iwi,Hatsuda:2023imp}. 
We hope to report on the detailed study of the giant graviton expansions and the Schur line defect correlators in the near future. 

Recall that the half-indices become the half-BPS indices in the half-BPS limit. 
For $SO(2N+1)$, $USp(2N)$, $O(2N)^+$ and $USp(2N)'$ they have the same expression. 
For simplicity, we take the $USp(2N)$ half-BPS index (\ref{1/2BPSindex_BC})
\begin{align}
\mathcal{I}_{\text{$\frac{1}{2}$BPS}}^{USp(2N)}(\mathfrak{q})
&=\prod_{n=1}^{N}\frac{1}{1-\mathfrak{q}^{4n}}.
\end{align}
Since this is obtained from the half-BPS index of the $U(N)$ theory 
by replacing $\mathfrak{q}^2$ with $\mathfrak{q}^4$, 
it admits the giant graviton expansion with respect to the $(\mathfrak{q}^2)^{2k}$ 
associated with wrapping numbers $2k$, $k=0,1,\cdots$
\begin{align}
\label{gg_1/2BPS_usp2N}
\frac{\mathcal{I}_{\text{$\frac{1}{2}$BPS}}^{USp(2N)}(\mathfrak{q})}
{\mathcal{I}_{\text{$\frac{1}{2}$BPS}}^{USp(\infty)}(\mathfrak{q})}
&=\sum_{k=0}^{\infty} \mathfrak{q}^{4kN}
\mathcal{I}_{\text{$\frac{1}{2}$BPS}}^{USp(2k)}(\mathfrak{q}^{-1}). 
\end{align}

As pointed out recently in e.g. \cite{Deddo:2024liu,Chang:2024zqi}, giant graviton expansions for the half-BPS giants 
can also be derived by analyzing the half-BPS geometries, from supergravity bubbling geometries. 
Here we give a geometric derivation of giant graviton expansions  
for the $\mathbb{Z}_{2}$ projected cases, i.e. the orientifold cases, from the geometric backgrounds 
where the half-BPS giant gravitons have backreacted for the dual backreacted geometry. 
See \cite{Mukhi:2005cv} by Mukhi and Smedback and \cite{Fiol:2014fla} by Fiol, Garolera and Torrents for detailed analyses. 
The family of solutions have a configuration space, equipped with a family of symplectic structures. 
The symplectic structures and their associated Poisson brackets are evaluated to commutators. 
The family of solutions and their orbifold cousins are quantized in this formalism, 
using covariant canonical quantization methods, see \cite{Maoz:2005nk} and detailed analyses
e.g. \cite{Chang:2024zqi,Gieres:2021ekc,Deddo:2024liu,Berenstein:2017rrx,Fiol:2014fla} and their related references.

The half-BPS bubbling geometries in Type IIB string theory are families of solutions 
of Type IIB supergravity whose metrics contain two three-spheres $S_{(1)}^3$ and $S_{(2)}^3$ \cite{Lin:2004nb}. 
They are determined by the harmonic function $z$ $=$ $z(x_1,x_2,y)$ of three coordinates $x_1$, $x_2$ and $y$. 
They are non-singular if the function $z$ obeys the boundary condition $z=\pm \frac12$ at $y=0$. 
For example, for $z=\frac12$ one of the three-spheres shrinks to zero size while the other remains finite. 
For $z=-\frac{1}{2}$ one encounters the shrinking of the other type of three-spheres.
Accordingly, the bubbling geometries can be determined by two-colorings of the $(x_1,x_2)$-plane 
with regions of $z=-\frac12$ (black) and those of $z=\frac12$ (white). 
One can view these regions as the fermion configurations 
where the black regions are occupied and white regions are empty. 
A black disk with a circular boundary describes the $AdS_5\times S^5$ solution. 
A deformation of this profile by adding thin concentric shells describes giant gravitons (see Figure \ref{fig_bubbling1}(a)). 
For the orientifold cases, the profiles are invariant under $(x_1,x_2)$ $\rightarrow$ $(-x_1,-x_2)$ 
so that they can be represented by the upper half-plane $x_2\ge 0$. 
The giant graviton corresponds to the configuration in Figure \ref{fig_bubbling1}(b). 
\usetikzlibrary{decorations.pathmorphing}
\tikzset{snake it/.style={decorate, decoration=snake}}
\begin{figure}
\centering
\scalebox{1.2}{
\begin{tikzpicture}
\draw [decorate,opacity=0.6, fill=black, decorate,decoration={snake,amplitude=.4mm,segment length=2.5mm,post length=0mm}] (-3,0) arc (0:360:2cm);
\draw [decorate,opacity=1, fill=white, decorate,decoration={snake,amplitude=.2mm,segment length=1mm,post length=0mm}] (-4.5,0) arc (0:360:0.5cm);
\draw (-5,-3) node [scale=0.75, black] {\textrm{(a)}};
\draw [decorate,opacity=0.6, fill=black, decorate,decoration={snake,amplitude=.4mm,segment length=2.5mm,post length=0mm}] (2,0) arc (0:180:2cm);
\draw [decorate,opacity=1, fill=white, decorate,decoration={snake,amplitude=.2mm,segment length=1mm,post length=0mm}] (0.5,0) arc (0:180:0.5cm);
\draw (0,-3) node [scale=0.75, black] {\textrm{(b)}};
\draw [very thick,white, ->] (1,0) arc (0:45:1cm);
\draw (1.25,0.25) node [scale=0.75, white] {$\phi$};
\draw[thick,red] (0,0)--(1.4,1.4); 
\draw (0.4,1) node [scale=0.75, red] {$R(\phi)$};
\draw[thick,blue] (0.05,0)--(0.4,0.35);
\draw (0.2,-0.2) node [scale=0.75, blue] {$r(\phi)$}; 
\end{tikzpicture}
}
\caption{\textrm{
(a) The maximal giant gravitons in $AdS_5\times S^5$. 
(b) The maximal giant gravitons in $AdS_5\times \mathbb{RP}^5$. 
}}
\label{fig_bubbling1}
\end{figure}
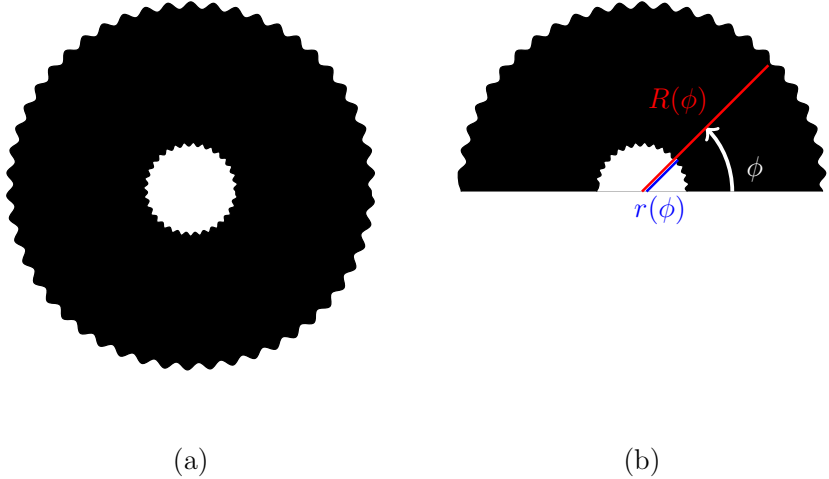

We also use polar coordinates $x_{1}+ix_{2}=\rho e^{i\phi }$, and the $\mathbb{Z}_{2}$ identification is $e^{i\phi }\rightarrow -e^{i\phi }$, 
which is $\phi \rightarrow \phi +\pi $. After the quotient, $\phi \in \lbrack0,\pi ]$. 
Due to the smooth $\mathbb{Z}_{2}$ identification, the $\phi =0$ is identified with $\phi =\pi $, 
and hence $\phi $ parametrizes a smooth $S^{1}/\mathbb{Z}_{2}$ and 
there is no quotient singularity at $\phi =0$ and $\pi $, since they are smoothly identified with each other.

We thus have, inside a black disk, the configurations of extra excitations
of white droplets corresponding to the excitations of maximal giants. 
Hence the black region will have a outer boundary and a inner boundary, see Figure \ref{fig_bubbling1}(b) for an illustration. 
The maximal giants are hence dual to the configuration of the inner boundary. The outer boundary of the
black disk is parameterized by a curve $R(\phi )$ corresponding to the fluctuations of gravitons 
and the inner boundary of the black disk is parameterized by a curve $r(\phi )$ 
corresponding to the fluctuations of the maximal giant gravitons. We have the boundary constraints 
$R(\phi =0)=R(\phi=\pi )$ and $r(\phi =0)=r(\phi =\pi )$. Hence we expand them in modes with
\begin{align}
\label{Rr_def}
R(\phi )^{2}& =\sum_{n\in \mathbb{Z}}\alpha _{n}e^{2in\phi },\qquad \alpha
_{0}=R^{2},\qquad \alpha _{-n}=\alpha _{n}^{\ast },  \notag \\
r(\phi )^{2}& =\sum_{j\in \mathbb{Z}}\beta _{j}e^{2ij\phi },\qquad \beta
_{0}=r^{2},\qquad \beta _{-j}=\beta _{j}^{\ast }, 
\end{align}
where $\alpha_{n}$, $\beta_{j}$ can be interpreted as wave modes on the boundary of droplets. 
Due to the smooth $\mathbb{Z}_{2}$ identification, $e^{2in\pi }=1$. 
According to the $\mathbb{Z}_{2}$ projection, the flux value, before the quotient, is increased to twice.

The configuration space of the above geometric configurations have a symplectic structure \cite{Maoz:2005nk,Deddo:2024liu,Mukhi:2005cv,Fiol:2014fla}. 
We have the droplet boundary, $\partial \mathcal{D}$, which has components $\bigcup_{b}\partial\mathcal{D}^{(b)}$, 
where the superscript $b$ denotes disjoint droplet boundaries. 
$\partial \mathcal{D}^{(b)}$ is represented by a closed curve component $\gamma ^{(b)}(s)$ where $s\in \partial \mathcal{D}^{(b)}$. 
We denote $\delta \gamma _{\perp }^{(b)}(s)$ as the variation of $\partial\mathcal{D}^{(b)}$ in the normal direction, 
which is related to the above curve $\delta r(\phi )$, e.g. (\ref{Rr_def}), 
by $\delta \gamma _{\perp}^{(b)}=\frac{\mathrm{d}\phi}{\mathrm{d}s}r(\phi )\delta r$. 
The configuration space has a family of symplectic structures \cite{Maoz:2005nk,Deddo:2024liu}
\begin{align}
\omega =\frac{1}{8\pi \hbar }\sum_{b}\int_{\gamma ^{(b)}}\mathrm{d}%
s\int_{\gamma ^{(b)}}\mathrm{d}\tilde{s} \ \mathrm{Sign}(s-\tilde{s}%
)\,\delta \gamma _{\perp }^{(b)}(s)\wedge \delta \gamma _{\perp }^{(b)}(%
\tilde{s}),
\end{align}
and they have the Poisson brackets
\begin{align}
\{\delta \gamma _{\perp }^{(b)}(s),\delta \gamma _{\perp }^{(\tilde{b})}(%
\tilde{s})\}=2\pi \hbar \delta ^{\prime }(s-\tilde{s})\delta _{b\tilde{b}}.
\end{align}

We use the covariant canonical quantization formalism, and the Poisson
brackets become commutators for the above modes,
\begin{equation}
\lbrack a_{n},a_{l}]=2n\delta _{n+l},\qquad \lbrack b_{n},b_{j}]=2n\delta
_{n+j},
\end{equation}%
where $a_{n}=\alpha _{n}/(2\hbar )$ and $b_{j}=\beta _{j}/(2\hbar )$ denote
the normalized operators. Note that $a_{n}$, $b_{j}$ denote even modes $2n$,
$2j$, respectively.

Thereby, the energy $E=\Delta $ and angular momentum $J$ of the maximal
giant graviton excitations are given by
\begin{equation}
\Delta =J=2mN+2\sum_{n\geq 0}a_{-n}a_{n}-2\sum_{j\geq 0}b_{-j}b_{j}.
\label{eqn 05}
\end{equation}%
After the smooth $\mathbb{Z}_{2}~$quotient, the $m$ is the number of flux
quanta corresponding to the giant gravitons.

The index computed from the gravitational data (\ref{eqn 05}) hence takes the form
\begin{align}
 \label{eqn 07}
\mathcal{I}_{\text{$\frac{1}{2}$BPS}}^{USp(2N)}(\mathfrak{q})
=\sum_{m=0}^{\infty }\mathfrak{q}^{4mN}\left( \prod_{n\geq 1}\sum_{N_{n}\geq 0}
{\mathfrak{q}}^{4nN_{n}}\right) \left( \prod_{j\geq 1}\sum_{n_{j}\geq 0}{\mathfrak{q}}^{-4jn_{j}}\right).
\end{align}
Here $N_{n}$ and $n_{j}$ denote occupation numbers for mode $n$ and mode $j$. 
The inner radius is
\begin{align}
r(\phi )^{2}=4\hbar \left( m+\sum_{j}n_{j}e^{2ij\phi }\right).
\end{align}
The inner radius squared is non-negative and is bounded by $0$ from below 
so that we have the constraint on the total sum $\sum_{j}n_{j}\leq m$.
Therefore, we have the giant graviton index 
\begin{align}
\hat{f}_k(\mathfrak{q})=\prod_{j=1}^{k}\frac{1}{1-\mathfrak{q}^{-4j}}
=\mathcal{I}_{\text{$\frac{1}{2}$BPS}}^{USp(2k)}({\mathfrak{q}}^{-1}). 
\end{align}
Here $m$ of (\ref{eqn 07}) is identified with $k$ in the gauge theory notation. 
Hence (\ref{eqn 07}) gives the full index, in the form of giant graviton expansion (\ref{gg_1/2BPS_usp2N}). 

Next consider the $SO(2N)$ case, which is distinguished from the $USp(2N)$ case. 
The half-BPS index is given by (\ref{1/2BPSindex_so2N})
\begin{align}
\mathcal{I}^{SO(2N)}_{\textrm{$\frac12$BPS}}(\mathfrak{q})
&=(1+\mathfrak{q}^{2N})
\prod_{n=1}^{N}\frac{1}{1-\mathfrak{q}^{4n}}. 
\end{align}
The giant graviton expansion is given by
\begin{align}
\label{gg_1/2BPS_so2N}
\frac{\mathcal{I}_{\text{$\frac{1}{2}$BPS}}^{SO(2N)}(\mathfrak{q})}
{\mathcal{I}_{\text{$\frac{1}{2}$BPS}}^{SO(\infty)}(\mathfrak{q})}
&=(1+\mathfrak{q}^{2N})
\sum_{k=0}^{\infty} \mathfrak{q}^{4kN}\hat{f}_k(\mathfrak{q}). 
\end{align}

In this case the dual geometries have additional $S^{3}/\mathbb{Z}_{2}\cong \mathbb{RP}^{3}$ cycles \cite{Witten:1998xy}
and the wrapping number on the $\mathbb{Z}_2$-torsion $\mathbb{RP}^3$ cycle is $\mathbb{Z}_{2}$-valued, instead of $\mathbb{Z}$-valued. 
As an illustration of the example in Figure \ref{fig_bubbling1}(b), near the origin of $\rho =0$, 
inside the white half-disk, there is a $\mathbb{RP}^{3}$ cycle. 
This $\mathbb{RP}^{3}$ cycle is formed as follows.
The union of this cycle and its image, in the $1:2$ covering map, forms a full $S^{3}$ in the covering space. 
Hence the smooth $\mathbb{Z}_{2}\ $quotient of the full $S^{3}~$near the origin of $\rho =0$, 
is always a $\mathbb{RP}^{3}$ after the smooth quotient. 
Hence there is always a $\mathbb{RP}^{3}$ cycle at the origin ($\rho =0$). 
We can wrap either one or zero D3-brane on this $\mathbb{RP}^{3}$ due to $\mathbb{Z}_{2}$-valued cohomology group.
Hence, the angular momentum of this wrapped D3-brane is denoted by $lN$,
with $l=0$ for the $\mathbb{Z}_{2}$-even generator, and $l=1$ for the $\mathbb{Z}_{2}$-odd generator. 
Hence the energy and angular momentum of the configuration are
\begin{align}
\Delta =J=2mN+2\sum_{n\geq 0}a_{-n}a_{n}-2\sum_{j\geq 0}b_{-j}b_{j}+lN,
\label{eqn 13}
\end{align}
where $l$ is $0$ or $1$. 
The index computed from the gravitational data (\ref{eqn 13}) takes the form
\begin{align}
\mathcal{I}_{\text{$\frac{1}{2}$BPS}}^{SO(2N)}(\mathfrak{q}%
)=\sum_{m=0}^{\infty }\mathfrak{q}^{4mN}\left( \prod_{n\geq
1}\sum_{N_{n}\geq 0}{\mathfrak{q}}^{4nN_{n}}\right) \left( \prod_{j\geq
1}\sum_{n_{j}\geq 0}{\mathfrak{q}}^{-4jn_{j}}\right) \sum_{l=0}^{1}\mathfrak{%
q}^{2lN}.
\label{eqn_15}
\end{align}%
The index derived from above (\ref{eqn 13}) is hence
\begin{equation}
\mathcal{I}_{\text{$\frac{1}{2}$BPS}}^{SO(2N)}(\mathfrak{q})=\mathcal{I}_{%
\text{$\frac{1}{2}$BPS}}^{SO(\infty )}(\mathfrak{q})\cdot (1+\mathfrak{q}%
^{2N})\cdot \sum_{k=0}^{\infty }\frac{(-1)^{k}\mathfrak{q}^{2k(k+1)}}{(%
\mathfrak{q}^{4};\mathfrak{q}^{4})_{k}}\mathfrak{q}^{4kN}.  \label{eqn 21}
\end{equation}%
This is the same as the gauge theory analysis (\ref{gg_1/2BPS_so2N}). 
The factor $(1+\mathfrak{q}%
^{2N})$ in (\ref{eqn 21}) can be understood as the contribution from the
Pfaffian D3-brane wrapping $\mathbb{RP}^{3}\cong S^{3}/\mathbb{Z}_{2}$, 
where $1$ and $\mathfrak{q}^{2N}$ correspond to the $\mathbb{Z}_2$ even and odd generators. 
They also correspond to $l=0,1$ in (\ref{eqn 13}), (\ref{eqn_15}) respectively. 

To summarize, we show that the half-BPS indices for $\mathcal{N}=4$ orthogonal and symplectic gauge theories have the following giant graviton expansions: 
\begin{align}
\label{gg1/2_so2N+1}
\frac{\mathcal{I}_{\text{$\frac{1}{2}$BPS}}^{SO(2N+1)}(\mathfrak{q})}
{\mathcal{I}_{\text{$\frac{1}{2}$BPS}}^{SO(\infty)}(\mathfrak{q})}
&=\sum_{m=0}^{\infty}\mathfrak{q}^{2(2N+1)m}
\mathcal{I}_{\text{$\frac{1}{2}$BPS}}^{O(2m)^{-}}(\mathfrak{q}^{-1}), \\
\label{gg1/2_usp2N}
\frac{\mathcal{I}_{\text{$\frac{1}{2}$BPS}}^{USp(2N)}(\mathfrak{q})}
{\mathcal{I}_{\text{$\frac{1}{2}$BPS}}^{USp(\infty)}(\mathfrak{q})}
&=\sum_{m=0}^{\infty}\mathfrak{q}^{4mN}
\mathcal{I}_{\text{$\frac{1}{2}$BPS}}^{USp(2m)}(\mathfrak{q}^{-1}), \\
\label{gg1/2_so2N}
\frac{\mathcal{I}_{\text{$\frac{1}{2}$BPS}}^{SO(2N)}(\mathfrak{q})}
{\mathcal{I}_{\text{$\frac{1}{2}$BPS}}^{SO(\infty)}(\mathfrak{q})}
&=\sum_{m=0}^{\infty}\mathfrak{q}^{2mN}
\mathcal{I}_{\text{$\frac{1}{2}$BPS}}^{O(m)^{+}}(\mathfrak{q}^{-1}), \\
\label{gg1/2_o2N+}
\frac{\mathcal{I}_{\text{$\frac{1}{2}$BPS}}^{O(2N)^+}(\mathfrak{q})}
{\mathcal{I}_{\text{$\frac{1}{2}$BPS}}^{O(\infty)^+}(\mathfrak{q})}
&=\sum_{m=0}^{\infty}\mathfrak{q}^{4mN}
\mathcal{I}_{\text{$\frac{1}{2}$BPS}}^{O(2m)^{+}}(\mathfrak{q}^{-1}), \\
\label{gg1/2_o2N-}
\frac{\mathcal{I}_{\text{$\frac{1}{2}$BPS}}^{O(2N)^{-}}(\mathfrak{q})}
{\mathcal{I}_{\text{$\frac{1}{2}$BPS}}^{O(\infty)^{+}}(\mathfrak{q})}
&=\sum_{m=0}^{\infty}\mathfrak{q}^{2(2m+1)N}
\mathcal{I}_{\text{$\frac{1}{2}$BPS}}^{SO(2m+1)}(\mathfrak{q}^{-1}), \\
\label{gg1/2_usp'}
\frac{\mathcal{I}_{\text{$\frac{1}{2}$BPS}}^{USp(2N)'}(\mathfrak{q})}
{\mathcal{I}_{\text{$\frac{1}{2}$BPS}}^{USp(\infty)'}(\mathfrak{q})}
&=\sum_{m=0}^{\infty}\mathfrak{q}^{4mN}
\mathcal{I}_{\text{$\frac{1}{2}$BPS}}^{USp(2m)'}(\mathfrak{q}^{-1}). 
\end{align}
The half-BPS index for $O(2N+1)^+$ gauge theory is the same as that for $SO(2N+1)$ gauge theory. 
Our giant graviton expansions are all compatible with those conjectured for full supersymmetric indices in \cite{Fujiwara:2023bdc}. 
\footnote{
Our expressions (\ref{gg1/2_usp2N}) and  (\ref{gg1/2_usp'}) correspond to eq.(87) in \cite{Fujiwara:2023bdc}, 
while (\ref{gg1/2_so2N+1}), (\ref{gg1/2_so2N}), (\ref{gg1/2_o2N+}) and (\ref{gg1/2_o2N-}) to  eq.(93), (92), (78) and (91) in \cite{Fujiwara:2023bdc}. 
}

\subsection*{Acknowledgements}
We thank Masatoshi Noumi for useful discussion on the associated $q$-hypergeometric integrals. 
The work of Y.H. was supported in part by JSPS KAKENHI Grant Nos. 22K03641 and 23H01093. 
The work of H.L. was supported in part by National Key R\&D Program of China grant 2020YFA0713000, 
by Overseas high-level talents program, by Fundamental Research Funds for the Central Universities of China, and by Grant No. 3207012204. 
The work of T.O. was supported by the Startup Funding no.\ 4007012317 of the Southeast University. 

\appendix

\section{Series expansions}
\label{app_expansion}
We show first several terms in the $q$-series expansions. 
The indices agree at least up to the terms with $q^5$ unless the order is not indicated in the table. 
\subsection{$SO(2N+1)$ and $USp(2N)$}

\begin{align}
\begin{array}{c|c}
\textrm{half-indices}&\textrm{expansions}\\ \hline
\mathbb{II}_{\mathcal{N}}^{SO(3)}
=\mathbb{II}_{\textrm{Nahm}'}^{USp(2)}
&
1+t^{-4}q-t^{-2}q^{\frac32}+(t^{-8}+t^{-4})q^2-(t^{-6}+t^{-2})q^{\frac52}\\ 
&+(t^{-12}+t^{-8}+t^{-4})q^3-(t^{-10}+2t^{-6}+t^{-2})q^{\frac72}+\cdots\\ \hline 
\mathbb{II}_{\mathcal{N}}^{SO(5)}
=\mathbb{II}_{\textrm{Nahm}'}^{USp(4)}&
1+t^{-4}q-t^{-2}q^{\frac32}+(2t^{-8}+t^{-4})q^2-(2t^{-6}+t^{-2})q^{\frac52}\\ 
&+(2t^{-12}+2t^{-8}+t^{-4})q^3-(3t^{-10}+3t^{-6}+t^{-2})q^{\frac72}+\cdots\\ \hline 
\mathbb{II}_{\mathcal{N}}^{SO(7)}
=\mathbb{II}_{\textrm{Nahm}'}^{USp(6)}&
1+t^{-4}q-t^{-2}q^{\frac32}+(2t^{-8}+t^{-4})q^2-(2t^{-6}+t^{-2})q^{\frac52}\\ 
&+(3t^{-12}+2t^{-8}+t^{-4})q^3-(4t^{-10}+3t^{-6}+t^{-2})q^{\frac72}+\cdots\\ \hline 
\end{array}
\end{align}

\begin{align}
\begin{array}{c|c}
\textrm{half-indices}&\textrm{expansions}\\ \hline
\mathbb{II}_{\mathcal{N}}^{USp(2)}
=\mathbb{II}_{\textrm{Nahm}'}^{SO(3)}
&
1+t^{-4}q-t^{-2}q^{\frac32}+(t^{-8}+t^{-4})q^2-(t^{-6}+t^{-2})q^{\frac52}\\ 
&+(t^{-12}+t^{-8}+t^{-4})q^3-(t^{-10}+2t^{-6}+t^{-2})q^{\frac72}+\cdots\\ \hline 
\mathbb{II}_{\mathcal{N}}^{USp(4)}
=\mathbb{II}_{\textrm{Nahm}'}^{SO(5)}&
1+t^{-4}q-t^{-2}q^{\frac32}+(2t^{-8}+t^{-4})q^2-(2t^{-6}+t^{-2})q^{\frac52}\\ 
&+(2t^{-12}+2t^{-8}+t^{-4})q^3-(3t^{-10}+3t^{-6}+t^{-2})q^{\frac72}+\cdots\\ \hline 
\mathbb{II}_{\mathcal{N}}^{USp(6)}
=\mathbb{II}_{\textrm{Nahm}'}^{SO(7)}&
1+t^{-4}q-t^{-2}q^{\frac32}+(2t^{-8}+t^{-4})q^2-(2t^{-6}+t^{-2})q^{\frac52}\\ 
&+(3t^{-12}+2t^{-8}+t^{-4})q^3-(4t^{-10}+3t^{-6}+t^{-2})q^{\frac72}+\cdots\\ \hline 
\end{array}
\end{align}

\subsection{$SO(2N)$ and $O(2N)$}

\begin{align}
\begin{array}{c|c}
\textrm{half-indices}&\textrm{expansions}\\ \hline
\mathbb{II}_{\mathcal{N}}^{SO(2)}
=\mathbb{II}_{\textrm{Nahm}'}^{SO(2)}
&
1+t^{-2}q^{\frac12}+(-1+t^{-4})q+t^{-6}q^{\frac32}+(t^{-10}-t^{-2})q^{\frac52}\\ 
&+t^{-12}q^3+(t^{-14}-t^{-2})q^{\frac72}+\cdots\\ \hline 
\mathbb{II}_{\mathcal{N}}^{SO(4)}
=\mathbb{II}_{\textrm{Nahm}'}^{SO(4)}&
1+2t^{-4}q-2t^{-2}q^{\frac32}+(3t^{-8}+2t^{-4})q^2-(4t^{-6}+2t^{-2})q^{\frac52}\\ 
&+(4t^{-12}+4t^{-8}+3t^{-4})q^3+(6t^{-10}+8t^{-6}+2t^{-2})q^{\frac72}+\cdots\\ \hline 
\mathbb{II}_{\mathcal{N}}^{SO(6)}
=\mathbb{II}_{\textrm{Nahm}'}^{SO(6)}&
1+t^{-4}q+(t^{-6}-t^{-2})q^{\frac32}+2t^{-8}q^2-(-t^{-10}+t^{-6}+t^{-2})q^{\frac52}\\ 
&+3t^{-12}q^3-(-2t^{-14}+2t^{-10}+t^{-6}+t^{-2})q^{\frac72}+\cdots\\ \hline 
\end{array}
\end{align}

\begin{align}
\begin{array}{c|c}
\textrm{half-indices}&\textrm{expansions}\\ \hline
\mathbb{II}_{\mathcal{N}}^{O(2)}
=\mathbb{II}_{\textrm{Nahm}'}^{O(2)}
&
1+t^{-4}q-t^{-2}q^{\frac32}+(t^{-8}+t^{-4})q^2-(t^{-6}+2t^{-2})q^{\frac52} \\ 
&+(1+t^{-12}+t^{-8}+2t^{-4})q^3-(t^{-10}+2t^{-6}+3t^{-2})q^{\frac72}+\cdots \\ \hline 
\mathbb{II}_{\mathcal{N}}^{O(4)}
=\mathbb{II}_{\textrm{Nahm}'}^{O(4)}&
1+t^{-4}q-t^{-2}q^{\frac32}+(2t^{-8}+t^{-4})q^2-(2t^{-6}+t^{-2})q^{\frac52} \\ 
&+(2t^{-12}+2t^{-8}+t^{-4})q^3-(3t^{-10}+4t^{-6}+t^{-2})q^{\frac72}+\cdots \\ \hline 
\mathbb{II}_{\mathcal{N}}^{O(6)}
=\mathbb{II}_{\textrm{Nahm}'}^{O(6)}&
1+t^{-4}q-t^{-2}q^{\frac32}+(2t^{-8}+t^{-4})q^2-(2t^{-6}+t^{-2})q^{\frac52} \\ 
&+(3t^{-12}+2t^{-8}+t^{-4})q^3-(4t^{-10}+3t^{-6}+t^{-2})q^{\frac72}+\cdots\\ \hline 
\end{array}
\end{align}

\subsection{$SO(2N+1)|USp(2M)'$ and $USp(2N)|USp(2M)'$}

\begin{align}
\begin{array}{c|c}
\textrm{half-indices}&\textrm{expansions}\\ \hline
\mathbb{II}_{\mathcal{N}}^{SO(3)|USp(2)'}
=\mathbb{II}_{\mathcal{D}'}^{USp(2)|USp(2)'}
&
1+(1+2t^{-4}+t^4)q-(3t^{-2}+2t^2)q^{\frac32} \\ 
&+(3+3t^{-8}+3t^{-4}+2t^4+t^8)q^2+\cdots \\ \hline 
\mathbb{II}_{\mathcal{N}}^{SO(3)|USp(4)'}
=\mathbb{II}_{\mathcal{D}'}^{USp(2)|USp(4)'}&
1+(1+2t^{-4}+t^4)q-(3t^{-2}+2t^2)q^{\frac32} \\ 
&+(4+4t^{-8}+4t^{-4}+2t^4+t^8)q^2+\cdots \\ \hline 
\mathbb{II}_{\mathcal{N}}^{SO(3)|USp(6)'}
=\mathbb{II}_{\mathcal{D}'}^{USp(2)|USp(6)'}&
1+(1+2t^{-4}+t^4)q-(3t^{-2}+2t^2)q^{\frac32} \\ 
&+(4+4t^{-8}+4t^{-4}+2t^4+t^8)q^2+\cdots+\mathcal{O}(q^3) \\ \hline 
\mathbb{II}_{\mathcal{N}}^{SO(5)|USp(2)'}
=\mathbb{II}_{\mathcal{D}'}^{USp(4)|USp(2)'}&
1+(1+2t^{-4}+t^4)q-(3t^{-2}+2t^2)q^{\frac32} \\ 
&+(4+4t^{-8}+4t^{-4}+2t^4+t^8)q^2+\cdots \\ \hline 
\mathbb{II}_{\mathcal{N}}^{SO(5)|USp(4)'}
=\mathbb{II}_{\mathcal{D}'}^{USp(4)|USp(4)'}&
1+(1+2t^{-4}+t^4)q-(3t^{-2}+2t^2)q^{\frac32} \\ 
&+(6+5t^{-8}+5t^{-4}+3t^4+2t^8)q^2+\cdots+\mathcal{O}(q^3) \\ \hline 
\mathbb{II}_{\mathcal{N}}^{SO(5)|USp(6)'}
=\mathbb{II}_{\mathcal{D}'}^{USp(4)|USp(6)'}&
1+(1+2t^{-4}+t^4)q-(3t^{-2}+2t^2)q^{\frac32} \\ 
&+(6+4t^{-8}+5t^{-4}+4t^4+2t^8)q^2+\cdots+\mathcal{O}(q^3) \\ \hline 
\mathbb{II}_{\mathcal{N}}^{SO(7)|USp(2)'}
=\mathbb{II}_{\mathcal{D}'}^{USp(6)|USp(2)'}&
1+(1+2t^{-4}+t^4)q-(3t^{-2}+2t^2)q^{\frac32} \\ 
&+(4+4t^{-8}+4t^{-4}+2t^4+t^8)q^2+\cdots+\mathcal{O}(q^3) \\ \hline 
\mathbb{II}_{\mathcal{N}}^{SO(7)|USp(4)'}
=\mathbb{II}_{\mathcal{D}'}^{USp(6)|USp(4)'}&
1+(1+2t^{-4}+t^4)q-(3t^{-2}+2t^2)q^{\frac32} \\ 
&+(6+5t^{-8}+5t^{-4}+3t^4+2t^8)q^2+\cdots+\mathcal{O}(q^3) \\ \hline 
\mathbb{II}_{\mathcal{N}}^{SO(7)|USp(6)'}
=\mathbb{II}_{\mathcal{D}'}^{USp(6)|USp(6)'}&
1+(1+2t^{-4}+t^4)q-(3t^{-2}+2t^2)q^{\frac32}+\cdots+\mathcal{O}(q^2)  \\ \hline
\end{array}
\end{align}

\subsection{$SO(2N)|USp(2M)$ and $SO(2N)|SO(2M+1)$}

\begin{align}
\begin{array}{c|c}
\textrm{half-indices}&\textrm{expansions}\\ \hline
\mathbb{II}_{\mathcal{N}}^{SO(2)|USp(2)}
=\mathbb{II}_{\mathcal{D}'}^{SO(2)|SO(3)}
&
1+(t^{-2}+t^2)q^{\frac12}+(-1+2t^{-4}+t^4)q+(2t^{-6}-t^{-2}+t^{6})q^{\frac32} \\ 
&+(3t^{-8}-t^{-4}+t^8)q^2+(3t^{-10}-t^{-6}-t^{2}+t^{10})q^{\frac52}+\cdots \\ \hline 
\mathbb{II}_{\mathcal{N}}^{SO(2)|USp(4)}
=\mathbb{II}_{\mathcal{D}'}^{SO(2)|SO(5)}
&
1+(t^{-2}+t^2)q^{\frac12}+(-1+2t^{-4}+t^4)q+(2t^{-6}+t^{6})q^{\frac32} \\ 
&+(4t^{-8}-t^{-4}+t^8)q^2+(4t^{-10}-t^{-6}-t^{2}+t^{10})q^{\frac52}+\cdots \\ \hline 
\mathbb{II}_{\mathcal{N}}^{SO(2)|USp(6)}
=\mathbb{II}_{\mathcal{D}'}^{SO(2)|SO(7)}
&
1+(t^{-2}+t^2)q^{\frac12}+(-1+2t^{-4}+t^4)q+(2t^{-6}+t^{6})q^{\frac32} \\ 
&+(4t^{-8}-t^{-4}+t^8)q^2+(4t^{-10}-t^{-6}-t^{2}+t^{10})q^{\frac52}+\cdots+\mathcal{O}(q^3) \\ \hline 
\mathbb{II}_{\mathcal{N}}^{SO(4)|USp(2)}
=\mathbb{II}_{\mathcal{D}'}^{SO(4)|SO(3)}
&
1+(2+3t^{-4}+t^4)q-(5t^{-2}+3t^{2})q^{\frac32} \\ 
&+(7+6t^{-8}+7t^{-4}+3t^4+t^8)q^2+\cdots \\ \hline 
\mathbb{II}_{\mathcal{N}}^{SO(4)|USp(4)}
=\mathbb{II}_{\mathcal{D}'}^{SO(4)|SO(5)}
&
1+(2+3t^{-4}+2t^4)q-(5t^{-2}+4t^{2})q^{\frac32} \\ 
&+(10+7t^{-8}+9t^{-4}+6t^4+3t^8)q^2+\cdots+\mathcal{O}(q^3) \\ \hline 
\mathbb{II}_{\mathcal{N}}^{SO(4)|USp(6)}
=\mathbb{II}_{\mathcal{D}'}^{SO(4)|SO(7)}
&
1+(2+3t^{-4}+2t^4)q-(5t^{-2}+4t^{2})q^{\frac32} \\ 
&+(11+7t^{-8}+9t^{-4}+6t^4+3t^8)q^2+\cdots+\mathcal{O}(q^3) \\ \hline 
\mathbb{II}_{\mathcal{N}}^{SO(6)|USp(2)}
=\mathbb{II}_{\mathcal{D}'}^{SO(6)|SO(3)}
&
1+(1+2t^{-4}+t^4)q+(t^{-6}-2t^{-2}-2t^2)q^{\frac32} \\ 
&+(3+4t^{-8}+2t^{-4}+2t^4+t^8)q^2+\cdots+\mathcal{O}(q^3) \\ \hline 
\mathbb{II}_{\mathcal{N}}^{SO(6)|USp(4)}
=\mathbb{II}_{\mathcal{D}'}^{SO(6)|SO(5)}
&
1+(1+2t^{-4}+t^4)q-(-t^{-6}+2t^{-2}+t^2)q^{\frac32} \\ 
&+(4+5t^{-8}+3t^{-4}+2t^4+2t^8)q^2+\cdots+\mathcal{O}(q^3) \\ \hline 
\mathbb{II}_{\mathcal{N}}^{SO(6)|USp(6)}
=\mathbb{II}_{\mathcal{D}'}^{SO(6)|SO(7)}
&
1+(1+2t^{-4}+t^4)q+(t^{-6}-2t^{-2}-t^2+t^6)q^{\frac32}+\mathcal{O}(q^2) \\ \hline
\end{array}
\end{align}

\subsection{$O(2N)^+|USp(2M)$ and $O(2N)^+|SO(2M+1)$}

\begin{align}
\begin{array}{c|c}
\textrm{half-indices}&\textrm{expansions}\\ \hline
\mathbb{II}_{\mathcal{N}}^{O(2)^{+}|USp(2)}
=\mathbb{II}_{\mathcal{D}'}^{O(2)^{+}|SO(3)}
&
1+(1+2t^{-4}+t^4)q-(3t^{-2}+2t^2)q^{\frac32} \\ 
&+(4+3t^{-8}+3t^{-4}+2t^4+t^8)q^2
+\cdots \\ \hline 
\mathbb{II}_{\mathcal{N}}^{O(2)^+|USp(4)}
=\mathbb{II}_{\mathcal{D}'}^{O(2)^+|SO(5)}
&
1+(1+2t^{-4}+t^4)q-(3t^{-2}+2t^2)q^{\frac32} \\ 
&+(5+4t^{-8}+4t^{-4}+2t^4+t^8)q^2+\cdots \\ \hline 
\mathbb{II}_{\mathcal{N}}^{O(2)^+|USp(6)}
=\mathbb{II}_{\mathcal{D}'}^{O(2)^+|SO(7)}
&
1+(1+2t^{-4}+t^4)q-(3t^{-2}+2t^2)q^{\frac32} \\ 
&+(5+4t^{-8}+4t^{-4}+2t^4+t^8)q^2+\cdots+\mathcal{O}(q^3) \\ \hline 
\mathbb{II}_{\mathcal{N}}^{O(4)^+|USp(2)}
=\mathbb{II}_{\mathcal{D}'}^{O(4)^+|SO(3)}
&
1+(1+2t^{-4}+t^4)q-(3t^{-2}+2t^2)q^{\frac32} \\ 
&+(4+4t^{-8}+4t^{-4}+2t^4+t^8)q^2+\cdots\\ \hline 
\mathbb{II}_{\mathcal{N}}^{O(4)^+|USp(4)}
=\mathbb{II}_{\mathcal{D}'}^{O(4)^+|SO(5)}
&
1+(1+2t^{-4}+t^4)q-(3t^{-2}+2t^2)q^{\frac32} \\ 
&+(6+5t^{-8}+5t^{-4}+3t^4+2t^8)q^2+\cdots+\mathcal{O}(q^3) \\ \hline 
\mathbb{II}_{\mathcal{N}}^{O(4)^+|USp(6)}
=\mathbb{II}_{\mathcal{D}'}^{O(4)^+|SO(5)}
&
1+(1+2t^{-4}+t^4)q-(3t^{-2}+2t^2)q^{\frac32} \\ 
&+(6+5t^{-8}+5t^{-4}+3t^4+2t^8)q^2+\cdots+\mathcal{O}(q^3) \\ \hline 
\mathbb{II}_{\mathcal{N}}^{O(6)^+|USp(2)}
=\mathbb{II}_{\mathcal{D}'}^{O(6)^+|SO(3)}
&
1+(1+2t^{-4}+t^4)q-(3t^{-2}+2t^2)q^{\frac32} \\ 
&+(4+4t^{-8}+4t^{-4}+2t^4+t^8)q^2+\cdots+\mathcal{O}(q^3) \\ \hline 
\mathbb{II}_{\mathcal{N}}^{O(6)^+|USp(4)}
=\mathbb{II}_{\mathcal{D}'}^{O(6)^+|SO(5)}
&
1+(1+2t^{-4}+t^4)q-(3t^{-2}+2t^2)q^{\frac32} \\ 
&+(6+5t^{-8}+5t^{-4}+3t^4+2t^8)q^2+\cdots+\mathcal{O}(q^3) \\ \hline 
\mathbb{II}_{\mathcal{N}}^{O(6)^+|USp(6)}
=\mathbb{II}_{\mathcal{D}'}^{O(6)^+|SO(7)}
&
1+(1+2t^{-4}+t^4)q-(3t^{-2}+2t^2)q^{\frac32}+\mathcal{O}(q^2) \\ \hline 
\end{array}
\end{align}

\bibliographystyle{utphys}
\bibliography{ref}

\end{document}